\documentclass[a4paper,11pt]{article}
\usepackage{jheppub}
\usepackage[utf8]{inputenc}
\usepackage{multirow}

\newcommand{\cC}{\mathcal{C}}
\newcommand{\cO}{\mathcal{O}}
\newcommand{\cJ}{{\cal J}}

\title{\boldmath Violations of Quark-Hadron Duality in Low-Energy Determinations of $\alpha_s$}
\author[a]{Antonio Pich} 
\author[b]{and Antonio Rodríguez-Sánchez}
\affiliation[a]{Departament de Física Teòrica, IFIC, Universitat de València – CSIC,\\
Parque Científico, Catedrático José Beltrán 2, E-46980 Paterna, Spain}
\affiliation[b]{Université Paris-Saclay, CNRS/IN2P3, IJCLab, 91405 Orsay, France}

\emailAdd{Antonio.Pich@ific.uv.es}
\emailAdd{arodriguez@ijclab.in2p3.fr}

\begin{document}

\abstract{
Using the spectral functions measured in $\tau$ decays, we investigate the actual numerical impact of duality violations on the extraction of the strong coupling.
These effects are tiny in the standard $\alpha_s(m_\tau^2)$ determinations from integrated distributions of the hadronic spectrum with pinched weights, or from the total $\tau$ hadronic width. 
The pinched-weight factors suppress very efficiently the violations of duality, making their numerical effects negligible in comparison with the larger perturbative uncertainties.
However, combined fits of $\alpha_s$ and duality-violation parameters, performed with non-protected weights, are subject to large systematic errors associated with the assumed modelling of duality-violation effects. These  uncertainties have not been taken into account in the published analyses,
based on specific models of quark-hadron duality. 
}


\maketitle

\section{Introduction}

Confinement implies a dual description of QCD observables. First-principles theoretical calculations are made in terms of the fundamental quark and gluon degrees of freedom appearing in the Lagrangian, while experimental measurements rely on the detected hadronic spectrum. Both descriptions should agree, provided confinement is exact, but there is always some degree of ambiguity at the observable level, which introduces
unavoidable theoretical uncertainties.

In order to perform precise tests of the perturbative QCD predictions, one usually studies inclusive or semi-inclusive observables. The inclusive production of hadrons in processes that do not contain strongly-interacting particles in the initial state is particularly well suited for this purpose \cite{Pich:2020gzz}. Since the total probability that quarks and gluons hadronize is just one and the separate identities of the produced hadrons are not specified, the two dual descriptions are indeed equivalent in this case. Nevertheless, the different infrared sensitivity of both approaches still generates some ambiguities. Even at very high energies, perturbation theory predicts the appearance of multiple thresholds, corresponding to the production of additional gluons and quark-antiquark pairs,
while nature only exhibits multi-hadron production. The infrared problems associated with the binding of quarks and gluons in physical colour-singlet particles can be minimized, smearing the observable cross sections over a suitable energy range, which washes out the threshold sensitivity \cite{Poggio:1975af}. Similarly, in jet physics, one tries to minimize the sensitivity to the final hadronization. A clean jet observable should be infrared safe, {\it i.e.}, free of collinear and soft singularities \cite{Moretti:1998qx,Ellis:2007ib,Salam:2010nqg}.
Fully inclusive observables such as the hadronic decay widths of the $Z$, $W^\pm$ and $H$ bosons are defined at a specific energy point given by the boson mass and, therefore, are subject to the threshold ambiguity. They are precisely known in perturbation theory, including all possible gluon emissions up to $\cO(\alpha_s^4)$, but it remains an uncontrolled uncertainty associated with the nearby thresholds for multi-hadron
production. Fortunately, the numerical size of this effect is strongly suppressed by the heavy boson mass because $\Lambda_{\mathrm{QCD}}/M_Z\sim 2\times 10^{-3}$.

A similar argument can be applied to $\sigma(e^+e^-\to\mathrm{hadrons})$ at very high energies. However, at low and intermediate values of $s$ the resonance structure of the hadronic spectrum shows up. In order to smear the violations of duality, one then considers integrals of the hadronic invariant-mass distribution over the full energy range, from threshold up to a given value $s_{\mathrm{max}}$, high enough that perturbative methods are reliable. These finite-energy sum rules are double-inclusive observables and, using the operator product expansion (OPE) \cite{Wilson:1969zs,Shifman:1978bx,Shifman:1978by,Shifman:1978bw,Novikov:1980uj}, can be computed with a much higher precision than the production cross section at fixed values of the hadronic invariant mass \cite{Eidelman:1978xy,Bertlmann:1984ih,Narison:1993sx,Bodenstein:2011hm,Boito:2018yvl}. 
The experimental determination of the distribution of the final hadrons in $e^+e^-$ annihilation has been considerably improved in recent years 
\cite{Jegerlehner:2017lbd,Davier:2017zfy,Keshavarzi:2018mgv,Colangelo:2018mtw,Hoferichter:2019mqg,Davier:2019can,Keshavarzi:2019abf},
with the goal to refine the dispersive Standard Model prediction of the muon anomalous magnetic moment \cite{Aoyama:2020ynm} and the running of the electromagnetic coupling up to $M_Z$. Thus, there exists an interesting data set which could be used to perform precision QCD tests. Unfortunately, the achievable accuracy is still limited by significant discrepancies among different experiments which are not yet fully resolved.

A very special role has been played by the inclusive $\tau$ hadronic width \cite{Braaten:1988hc,Braaten:1988ea,Narison:1988ni,Braaten:1991qm,LeDiberder:1992jjr,LeDiberder:1992zhd}, which provides a very clean observable from both the experimental and theoretical points of view \cite{Pich:2013lsa}. 
The tau mass is high enough to safely apply the OPE, non-perturbative corrections can be shown to be suppressed \cite{Braaten:1991qm} and the perturbative contribution, which is known to $\cO(\alpha_s^4)$ \cite{Baikov:2008jh}, is very sizeable (dominant) because $\alpha_s(m_\tau^2)$ is large. Furthermore, violations of quark-hadron duality are heavily suppressed because this inclusive observable is given by an integral over the full hadronic invariant-mass distribution that, moreover, it is weighted by a kinematic factor with a double zero at the upper end of the integration range
\cite{Braaten:1991qm,LeDiberder:1992zhd}. The small size of non-perturbative effects can be assessed through the study of additional weighted integrals of this spectral distribution 
\cite{LeDiberder:1992zhd}. The detailed experimental analyses performed by the ALEPH \cite{ALEPH:1993qhw,ALEPH:1998rgl,ALEPH:2005qgp}, CLEO \cite{CLEO:1995nlc} and OPAL \cite{OPAL:1998rrm} collaborations corroborated a long time ago the predicted suppression of non-perturbative contributions and established a quite precise determination of $\alpha_s(m_\tau^2)$, which has been later updated with the $\cO(\alpha_s^4)$ QCD corrections and improved experimental information \cite{Davier:2005xq,Davier:2008sk,Davier:2013sfa,Pich:2016bdg}. 

A quite different strategy has been advocated in Refs.~\cite{Cata:2008ye,Cata:2008ru,Boito:2011qt,Boito:2012cr,Boito:2014sta}.
Instead of suppressing the unwanted violations of quark-hadron duality, these references analyse observables that are very sensitive to such effects with the aim of measuring the size of the duality violations. Analyses of this type could help to better understand the complicated infrared dynamics responsible for the observed differences between the low-energy hadronic world and its partonic description. However, what it is actually done is a rough phenomenological estimate of the duality-violation  (DV) contribution to the chosen observable, which is then subtracted from its measured value in order to determine $\alpha_s$, assuming that perturbation theory gives a good description of the reminder. While this procedure is obviously not more precise
than the actual theoretical control we have over the subtracted DV contribution, a surprisingly
accurate determination of $\alpha_s$ has been claimed.
It was already demonstrated in Refs.~\cite{Pich:2016bdg,Pich:2016mgv,Pich:2018jiy} that the numerical value of the strong coupling obtained in this way is model dependent because it is fully correlated with the adopted functional form of the DV correction. Small changes on the assumed DV ansatz result in large variations of $\alpha_s$, which gets then converted into one additional model parameter.

Some arguments concerning the applicability of the OPE at the $\tau$ mass scale and the theoretically-admissible functional form of the DV ansatz have been put forward~\cite{Boito:2016oam,Boito:2019iwh,Boito:2020xli}, trying to evade the conclusions of Refs.~\cite{Pich:2016bdg,Pich:2016mgv,Pich:2018jiy}. 
In this work we provide a much more detailed analysis that exhibits the intrinsic inconsistency of these arguments. We aim to clarify the subject by making as transparent as possible the implicit assumptions of the DV approach to the strong coupling. The numerical correlation between the fitted value of $\alpha_s$ and the assumed DV ansatz can be easily understood. 
The DV algorithm turns out to determine $\alpha_s$ at a quite low energy scale, $\hat s_0\sim (1.2~\mathrm{GeV})^2$, from a theoretically subtracted integral of the $\tau$ decay distribution up to $\hat s_0$. The $\tau$ data in the energy bins above $\hat s_0$ must be used to fit the ansatz parameters and calculate the DV subtraction,
but the resulting value of this subtraction changes in a quite significant way with slight modifications of the DV ansatz, generating an uncontrolled systematic uncertainty on $\alpha_s$. Once the strong coupling and the DV parameters get fixed with a given ansatz, all perturbative and DV deformations introduced by the chosen model can only be reabsorbed into the power corrections. An incorrect value of $\alpha_s$ needs to be compensated with unphysical values of the vacuum condensates (as many as observables) in order to reproduce the experimental moments of the $\tau$ hadronic distribution. As a result, the spread of $\alpha_s$ values enforces a much larger spread of fitted OPE corrections and a significant loss of theoretical control, which in some cases can even induce pathological behaviours not required by any data. Those DV ansatzs that do not display such pathologies turn out to generate condensates of smaller size and values of $\alpha_s(m_\tau^2)$ in agreement with the standard determination with pinched weights~\cite{Pich:2016bdg}.

Violations of quark-hadron duality are interesting phenomena per se \cite{Shifman:1995qj,Chibisov:1996wf,Blok:1997hs,Shifman:2000jv,Bigi:2001ys,Golterman:2001pj,Cata:2005zj,Beneke:2009az,Gonzalez-Alonso:2010kpl,Gonzalez-Alonso:2010lvh,Beylich:2011aq,Jamin:2011vd,Caprini:2014qda,Boito:2017cnp}, so it is worthwhile to investigate their effects through quantitative tests. In the absence of a better understanding of confinement,
achieving a rigorous description of DV corrections is a very difficult (may be hopeless) enterprise,  but nevertheless, it is important to assess their phenomenological impact in low-energy determinations of the strong coupling. This is in fact the main motivation of the analysis that will be presented next, which attempts to provide a quantitative estimate of the uncertainties associated with DV effects.

The manuscript is organised as follows. In section~\ref{sec:formalism} we briefly review the well-known analyticity properties of current correlators that make possible to rigorously analyse weighted integrals of the measured hadronic distributions with the short-distance OPE. The main results of the exhaustive analysis of $\alpha_s(m_\tau^2)$ determinations, performed in Ref.~\cite{Pich:2016bdg} with a broad variety of methodologies, are summarized in section~\ref{sec:alphasResults}, which collects different pieces of phenomenological evidence that will be used in the subsequent discussion. Section~\ref{sec:DVmethod} anatomizes the DV method employed in \cite{Cata:2008ye,Cata:2008ru,Boito:2011qt,Boito:2012cr,Boito:2014sta}, clarifying its assumptions and the adopted computational algorithm, and reproduces the numerical results of Ref.~\cite{Boito:2014sta}. The sensitivity of this approach to the assumed functional form of the DV ansatz is studied in detail, exhibiting the very large (unaccounted) systematic uncertainties associated with our poor control of DV phenomena.
In addition, this section discusses the applicability region of the inverse power expansion and points out the formal inconsistencies implicit in recent arguments against the truncation of the OPE, showing that those criticisms are inherently flawed.
All these results are then used in section~\ref{sec:assessment} to quantitative assess the actual impact of DV effects in the more standard determinations of the strong coupling presented in section~\ref{sec:alphasResults}. The estimated DV corrections are in this case well below the perturbative and non-perturbative uncertainties already considered in~\cite{Pich:2016bdg}, demonstrating the robustness of the final extraction of $\alpha_s(m_\tau^2)$. Some summarizing comments are finally given in section~\ref{sec:Summary}  that concludes giving our estimated value of $\alpha_s(m_\tau^2)$ from the available $\tau$ data. We relegate to the appendix some complementary results, which are not crucial for the central discussion but expose
the tautological nature of several tests within the DV approach.

\section{Theoretical formalism}
\label{sec:formalism}

For the inclusive observables we are interested in, the  QCD dynamics is encoded in the two-point correlation functions
\begin{eqnarray}\label{eq:correlators}
\Pi^{\mu \nu}_{ij,\cJ}(q)& \equiv &
 i \int d^4x \;\, \mathrm{e}^{iqx}\,
\langle 0|T(\cJ^{\mu}_{ij}(x)\, \cJ^{\nu}_{ij}(0)^\dagger)|0\rangle
\nonumber\\
& = &
\left( -g^{\mu\nu} q^2 + q^{\mu} q^{\nu}\right) \, \Pi_{ij,\,\cJ}^{(1)}(q^2)
 +   q^{\mu} q^{\nu}\, \Pi_{ij,\,\cJ}^{(0)}(q^2) \, ,
\end{eqnarray}
where $\cJ=V,A$ are the colour-singlet vector $\, V^{\mu}_{ij} = \bar{q}_j \gamma^{\mu} q_i \, $
or axial-vector
$\, A^{\mu}_{ij} = \bar{q}_j \gamma^{\mu} \gamma_5 q_i \,$
quark currents ($i,j=u,d,s\ldots$),
and the superscripts denote the corresponding angular momentum $J=1$ and $J=0$
in the hadronic rest frame ($\vec q = \vec 0$). For values of $s\equiv q^2\le m_\tau^2$, the spectral functions (absorptive parts) of these correlators are directly measured by the invariant-mass distribution of the final hadrons in $\tau$ decay \cite{Braaten:1991qm}:
\begin{eqnarray}
\label{eq:R_tau}
R_{\tau} & \equiv &
\frac{ \Gamma [\tau^- \to \nu_\tau +\mathrm{hadrons}]}{ \Gamma [\tau^- \to \nu_\tau e^- {\bar \nu}_e]}
\nonumber\\
& =& 12 \pi\, S_{\mathrm{EW}} \int^{m_\tau^2}_{s_{\mathrm{th}}} \frac{ds}{m_\tau^2 } \,
 \left(1-\frac{s}{m_\tau^2}\right)^2
\biggl[ \left(1 + 2 \frac{s}{m_\tau^2}\right)\,
 \mathrm{Im}\, \Pi^{(1)}_\tau(s)
 \, +\, \mathrm{Im}\, \Pi^{(0)}_\tau(s) \biggr]  ,\quad
\end{eqnarray}
where $s_{\mathrm{th}}$ is the hadronic mass-squared threshold,
\begin{equation}\label{eq:piTau}
\Pi^{(J)}_\tau(s)  \; \equiv  \;
  |V_{ud}|^2 \, \left( \Pi^{(J)}_{ud,V}(s) + \Pi^{(J)}_{ud,A}(s) \right)
\, + \,
|V_{us}|^2 \, \left( \Pi^{(J)}_{us,V}(s) + \Pi^{(J)}_{us,A}(s) \right)
\end{equation}
and the global factor $S_{\mathrm{EW}}=1.0201\pm 0.0003$ accounts for the (renormalization-group improved) electroweak radiative corrections \cite{Marciano:1988vm,Braaten:1990ef,Erler:2002mv}.

We will restrict our discussion to the Cabibbo-allowed hadronic distribution.  Neglecting the tiny up and down quark masses,\footnote{Quark mass corrections are numerically negligible. The dominant residual contribution due to the non-zero pion mass is taken into account.} 
$s\, \Pi^{(0)}_{ud,\cJ}(s) = 0$.
Therefore, the relevant dynamical quantities are the scalar correlators
\begin{equation}\label{eq:pi}
\Pi_\cJ(s) \,\equiv\, \Pi^{(0+1)}_{ud,\cJ}(s)\, . 
\end{equation}
These correlators are analytic functions in the whole complex plane, except along the positive real $s$ axis where their imaginary parts have discontinuities. Using a closed complex contour circumventing the physical cut, one gets the following mathematical identity for any weighted integral of the hadronic spectral functions 
$\rho_\cJ (s)\equiv\frac{1}{\pi}\,\mathrm{Im}\,\Pi_{\cJ}(s)$
\cite{Pich:1989pq,Braaten:1991qm,LeDiberder:1992zhd}:
\begin{equation}\label{eq:weighted_integrals}
A_\cJ^\omega (s_0)\,\equiv\,\pi
\int_{s_{\mathrm{th}}}^{s_0} \frac{ds}{s_0}\; \omega (s)\; \rho_\cJ (s)
\; =\;
\frac{i}{2}\; \oint_{|s|=s_0} \frac{ds}{s_0}\; \omega (s) \;\Pi_{\cJ}(s)\, ,
\end{equation}
with $\omega (s)$ an arbitrary weight function without singularities in the
region $|s|\leq s_0$. 
The integral on the left-hand-side is directly determined by the experimental data, while for sufficiently large $s_0$ values the OPE 
\begin{equation}\label{eq:OPE}
\Pi_{\cJ}^{\mathrm{OPE}}(s)\, =\, \sum_{D=2n}\,
\frac{1}{(-s)^{D/2}}\,\sum_{\mathrm{dim}\, O =D}
\cC_{D,\,\cJ}(-s,\mu^2) \;\langle 0| \cO(\mu^2) |0\rangle
\,=\, \sum_{D=2n}\,\frac{\cO_{D,\cJ}}{(-s)^{D/2}}
\end{equation}
can be used to calculate the contour integral along the circle $|s|=s_0$, 
as an expansion in inverse powers of $s_0$. 

The observable $R_\tau$ corresponds to the particular weight $\omega_\tau(x)=(1-x)^2 (1+2x) = 1-3x^2+2x^3$, with $x\equiv s/s_0$ and $s_0 = m_\tau^2$ that is expected to be large enough to safely apply the OPE. Neglecting the logarithmic running of the Wilson coefficients, Cauchy's theorem implies that the contour integral is only sensitive to OPE corrections with dimensions $D=6$ and $8$, which are strongly suppressed by the corresponding powers of $m_\tau$. In the total $V+A$ distribution, there is in addition a strong cancellation between the vector and axial-vector power corrections, which have opposite signs \cite{Braaten:1991qm,Davier:2013sfa,Pich:2016bdg}. The QCD contribution to $R_\tau$ is dominated by the perturbative correction, which amounts to a large 20\% effect because $\alpha_s(m_\tau^2)\sim 0.3$ is sizeable. This explains the high sensitivity of this observable to the strong coupling.

In order to better analyze the different OPE contributions, it is convenient to particularize\footnote{This can be trivially generalized, taking into account that from $\omega(x)=\sum_{n}c_{n}x^n$ one has $A^{\omega}_\cJ(x)=\sum_{n}c_{n} A^{(n)}_\cJ(x)$.}
Eq.~(\ref{eq:weighted_integrals}) 
with a monomial weight $\omega_n(x) = (s/s_0)^n$.
Integrating by parts,
\begin{align}\nonumber
A^{(n)}_\cJ (s_0)\,&\equiv\,\pi
\int_{s_{\mathrm{th}}}^{s_0} \frac{ds}{s_0}\; \left(\frac{s}{s_0}\right)^n\; \rho_\cJ (s)
\; =\;
\frac{i}{2}\; \oint_{|s|=s_0} \frac{ds}{s_0}\; \left(\frac{s}{s_0}\right)^n \;\Pi_\cJ(s)\\&=\frac{1}{n+1}\,\mathrm{Im}\,\Pi_\cJ(s_0)+\frac{i}{2(n+1)}\oint_{|s|=s_0} \frac{ds}{s_0} \left(\frac{s}{s_0}\right)^n \tilde{D}_\cJ(s)\, , \label{eq:A(n)}
\end{align}
where
\begin{equation}
\tilde{D}_\cJ (s)\equiv-s\frac{d\Pi_\cJ(s)}{ds} \, .
\end{equation}
From the first line, it follows that, for any $n \geq 0$,
\begin{equation}\label{eq:ImPifrommom}
\mathrm{Im}\,\Pi_\cJ(s_0)= \frac{1}{s_0^n}\,\frac{d\left[s_0^{n+1} A^{(n)}_\cJ(s_0)\right]}{ds_0} \, .
\end{equation}

The dominant perturbative $D=0$ contribution to the different integrals is encoded in the associated Adler function, which is known up to four loops:
\begin{equation}
D(s)\,\equiv\, - s\,\frac{d}{ds}\,\Pi_V^{\mathrm{pert}}(s)\, =\,\frac{1}{4\pi^2}\,\sum_{n=0} K_n\,\left(\frac{\alpha_s(s)}{\pi}\right)^n  ,
\end{equation}
where $K_0 = K_1 = 1$, while for $n_f=3$ quark flavours $K_2=1.63982$, $K_3=6.37101$ and $K_4=49.0757$ ($\overline{\mathrm{MS}}$ scheme)  \cite{Baikov:2008jh}.
One easily finds:
\begin{align}\label{eq:pertimpi}
\mathrm{Im}\,\Pi_V^{\mathrm{pert}}(s_0)&=\frac{1}{8\pi^2}\sum_{m}K_{m}\int^{\pi}_{-\pi}d\varphi\; 
\left(\frac{\alpha_{s}( s_{0}\, e^{i\varphi})}{\pi}\right)^{m}  ,
\\ \label{eq:pertan}
A^{(n)}_{\mathrm{pert}}(s_0)&=\frac{1}{8\pi^2(n+1)}\sum_m K_m \int_{-\pi}^{\pi} d\varphi \;\left( 1-(-1)^{n+1}e^{i\varphi (n+1)} \right)  
\left(\frac{\alpha_{s}( s_{0}\, e^{i\varphi})}{\pi}\right)^{m} .
\end{align}
Thus, the perturbative spectral function itself can be rewritten as an integral over complex angles. For the more inclusive moments $A_{\mathrm{pert}}^{(n)}(s_0)$, the integrand is zero at $\varphi=-\pi,\pi$, which are the (dangerous) angular values associated with the physical axis. 

The perturbative integrals in Eqs.~(\ref{eq:pertimpi}) and (\ref{eq:pertan}) can be computed in two different ways.
One can either perform the contour integrations with a running coupling $\alpha_s(-s)$, by solving numerically the five-loop $\beta$-function equation (contour-improved perturbation theory, CIPT) \cite{LeDiberder:1992jjr,Pivovarov:1991rh}, or naively expand them in powers of $\alpha_s(s_0)$ (fixed-order perturbation theory, FOPT). The CIPT prescription makes a re-summation of large higher-order corrections, generated by the long running of $\alpha_s$ along the complex circle, which results in a slightly smaller perturbative contribution to $A^{(n)}_\cJ(s_0)$
than FOPT, for a given value of $\alpha_s(m_\tau^2)$. Therefore, when solving the equality (\ref{eq:weighted_integrals}), CIPT leads to a slightly larger fitted value of $\alpha_s$. Strong efforts are currently being made aimed to improve our understanding of the perturbative series, e.g. see~\cite{Beneke:2008ad,Beneke:2012vb,Boito:2018rwt,Wu:2019mky,Caprini:2019kwp,Caprini:2020lff,Hoang:2020mkw,Hoang:2021nlz,Ayala:2021mwc,Ayala:2021yct,Goriachuk:2021ayq,Benitez-Rathgeb:2022yqb,Ayala:2022cxo}.

Weighting the spectral distribution with different functional dependences on $s$, one becomes sensitive to different power corrections in the OPE \cite{Pich:1989pq,Braaten:1991qm,LeDiberder:1992zhd}. At LO in $\alpha_s$, the power correction $\cO_{D,\cJ}$ is independent on the energy. QCD loops, which a priori cannot be ignored, spoil this behaviour and, at NLO, one has

\begin{equation}\label{eq:opeeuc}
\left.\Pi_\cJ^{\mathrm{OPE}}(s)\right|_{D>0}\, =\,\sum_{D>0}\,\frac{\mathcal{O}_{D,\cJ}(\mu)+\mathcal{P}_{D,\cJ}\,\ln{(-s/\mu^2)} }{(-s)^{D/2}} \, .
\end{equation}
The factors $\mathcal{P}_{D,\cJ}$ determine the QCD running of the coefficients $\mathcal{O}_{D,\cJ}(\mu)$. Their values cannot however be inferred from the $\mathcal{O}_{D,\cJ}(\mu)$ evaluated at a single scale (in general they involve different nonperturbative vacuum matrix elements), but are suppressed with respect to $\mathcal{O}_{D,\cJ}(\mu)$ by a power of $\alpha_s$. 
At this order, up to tiny light-quark mass corrections, one has (e.g. see \cite{Pich:1999hc})
\begin{equation}
\mathcal{O}_{2,\cJ}(\mu)\, =\, \mathcal{P}_{2,\cJ}\, =\, \mathcal{P}_{4,\cJ}\, =\, 0 \, .
\end{equation}
Performing the needed integrals, one finds

\begin{equation}\label{eq:suleadingPowers}
\left. A^{(n)}_\cJ(s_0)\right|_{D>0}\, =\, -\pi\,\sum_{p=2} \frac{d_{p,\cJ}^{(n)}}{(-s_0)^p}\, ,
\end{equation}
where
\begin{equation}
d_{p,\cJ}^{(n)}\, =\, \left\{\begin{array}{lr}
\mathcal{O}_{2p,\cJ}(s_0), & \quad\text{if } p= n+ 1 \, ,\\[4pt]
\frac{\displaystyle\mathcal{P}_{2p,\cJ}}{\displaystyle n-p+1}, &\quad\text{if } p\neq n+ 1 \, .\\
\end{array}\right.
\end{equation}

The OPE is valid in the complex plane, away from the physical cut, which justifies its application in the contour integration except for the region near $s_0$, the point where the circle touches the real axis. The so-called duality violations originate precisely from this small integration range where the OPE description is not precise.
Fortunately, the $R_\tau$ weight contains a double zero at the upper end of the integration range that strongly suppresses the numerical contribution from this dangerous region and, therefore, the corresponding violations of quark-hadron duality.

A quantitative definition of duality violations is provided by the differences between the physical values of the integrals $A_\cJ^\omega (s_0)$ and their OPE approximations. Using again the analyticity properties of the correlators $\Pi_{\cJ}(s)$, the  size of these effects can be expressed in the form \cite{Cata:2008ye,Chibisov:1996wf,Gonzalez-Alonso:2010kpl,Gonzalez-Alonso:2016ndl}
\begin{equation}
\label{eq:DV_A}
\Delta A_\cJ^\omega (s_0)\,\equiv\, \frac{i}{2}\; \oint_{|s|=s_0} \frac{ds}{s_0}\; \omega (s) \left\{\Pi_{\cJ}(s)-\Pi^{\mathrm{OPE}}_{\cJ}(s)\right\}
\, =\, -\pi\int_{s_0}^\infty \frac{ds}{s_0}\;\omega(s)\;\Delta\rho_\cJ^{\mathrm{DV}}(s)
\, ,
\end{equation}
with
\begin{equation}
\Delta\rho_\cJ^{\mathrm{DV}}(s)\,\equiv\, \rho_\cJ(s) - \rho_\cJ^{\mathrm{OPE}}(s)
\end{equation}
the difference between the physical spectral function and its OPE expression. For large-enough values of $s$, the OPE provides the correct average value of $ \rho_\cJ(s)$, while missing the hadronic resonance structures that generate oscillations around this mean value. These differences decrease very fast when $s$ increases, so that one may expect
$\Delta\rho_\cJ^{\mathrm{DV}}(s)\sim \mathrm{e}^{-\gamma s}$ asymptotically. Therefore, the DV correction on the right-hand-side of Eq.~(\ref{eq:DV_A}) is completely dominated by the region of $s$ values just slightly above $s_0$. In fact, the relatively large oscillations of the spectral function at $s_0\lesssim m_{\tau}^2$ have a very minor numerical role in the integrals $A_\cJ^\omega (s_0)$. Additionally, as it is well-known in the QCD literature \cite{Braaten:1991qm,LeDiberder:1992zhd,Dominguez:1998wy,Maltman:1998uzw,Gonzalez-Alonso:2010lvh,Pich:2021yll,Cirigliano:2021yto}, taking weight functions that vanish at $s_0$ (pinched weights), one is then further minimizing the numerical impact of the unwanted DV effects.

\section{\boldmath Different strategies to obtain $\alpha_s(m_\tau^2)$}
\label{sec:alphasResults}

\begin{figure}[tb]
\centerline{
\includegraphics[width=0.38\textwidth]{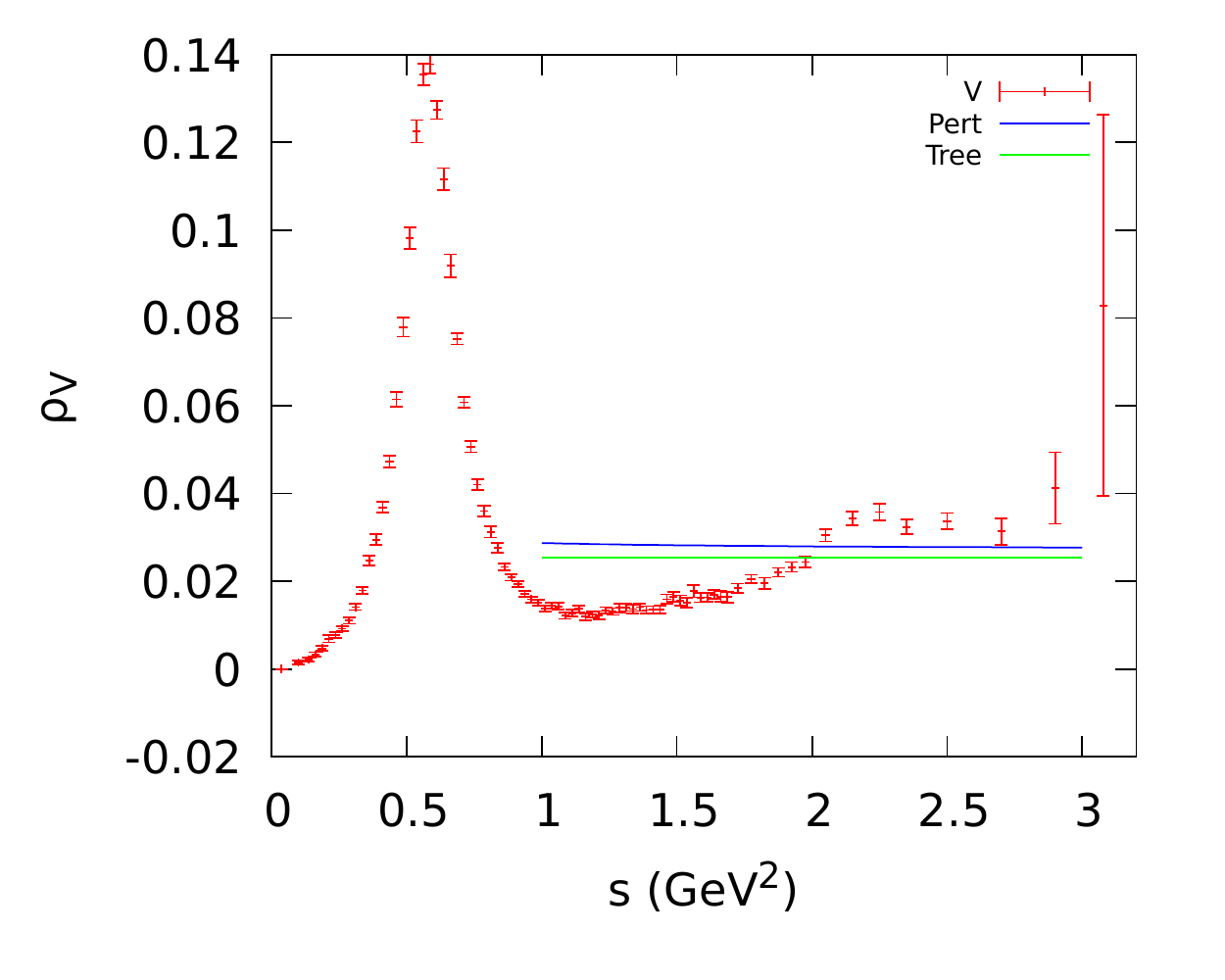}\hskip -.25cm
\includegraphics[width=0.38\textwidth]{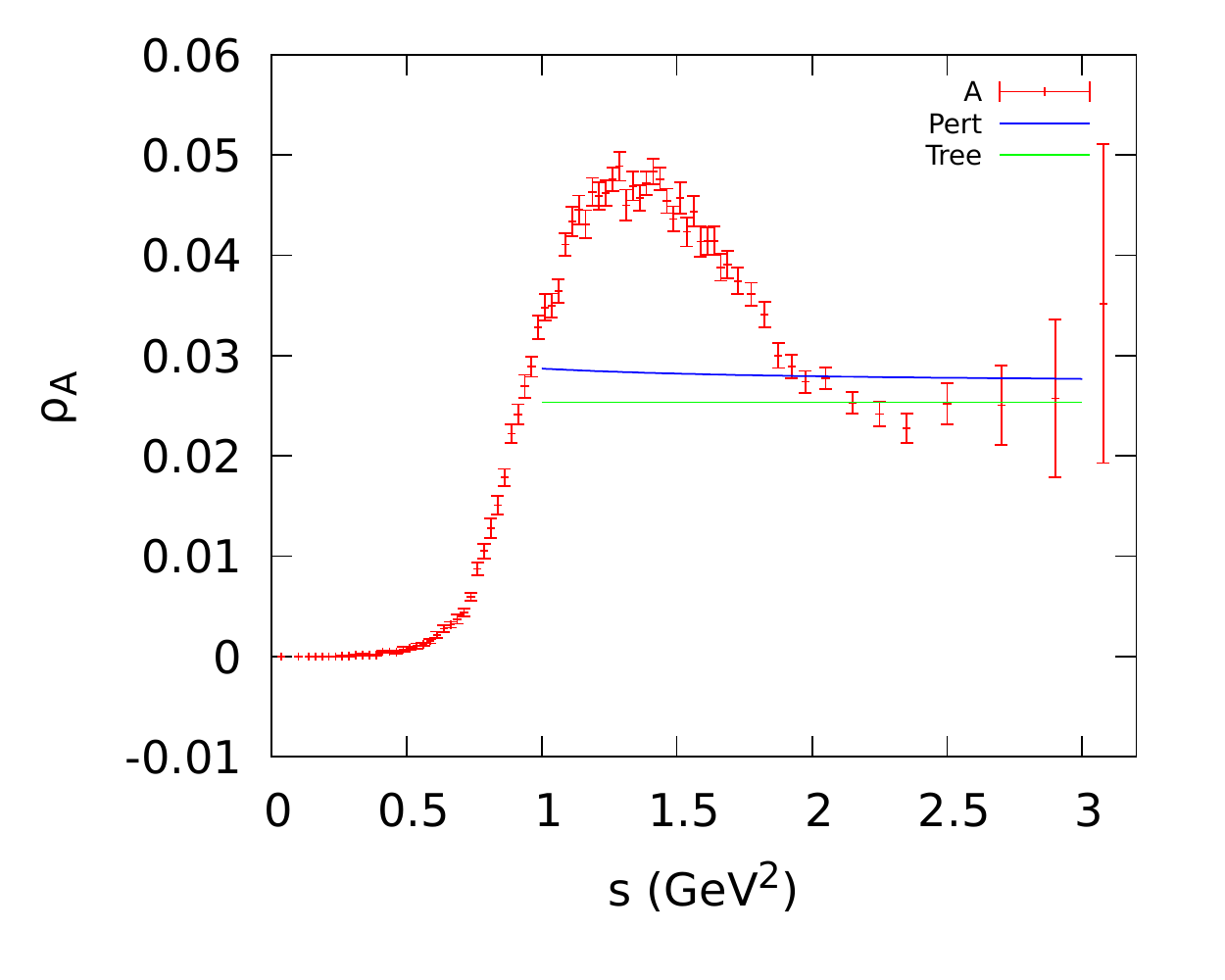}\hskip -.25cm
\includegraphics[width=0.38\textwidth]{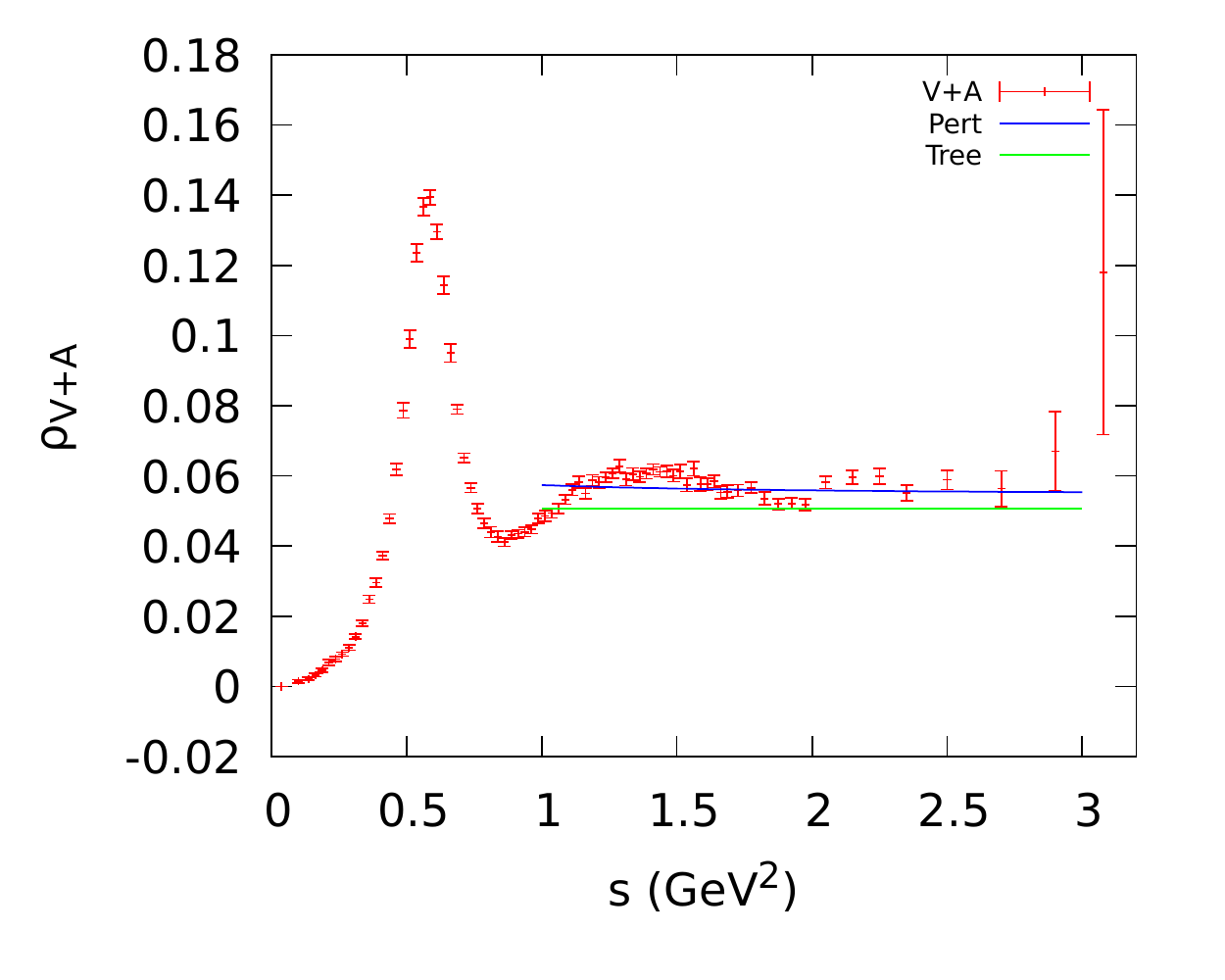}
}
\vskip -.3cm
\caption{\protect\label{fig:ALEPHsf}
Updated ALEPH spectral functions for the $V$, $A$ and $V+A$ channels \protect\cite{Davier:2013sfa}. The pion pole is not displayed.}
\end{figure}

From the measured invariant-mass distribution of the final hadrons in $\tau$ decays,
Ref.~\cite{Davier:2013sfa} extracted the spectral functions $ \rho_\cJ(s)$ shown in Fig.~\ref{fig:ALEPHsf}. 
Together with the experimental data points, the figure displays the naive parton-model expectations (horizontal green lines) and the predictions of (massless) perturbative QCD for $\alpha_s(m_\tau^2) = 0.329$ (blue lines). Resonance structures are clearly visible at low $s$ values, especially the prominent $\rho(2\pi)$ and $a_1(3\pi)$ resonance peaks, but as the invariant-mass increases they are soon diluted by the opening of high-multiplicity hadronic thresholds, leading to much smoother inclusive distributions, as expected from quark-hadron duality considerations \cite{Poggio:1975af}. The flattening of the spectral function is remarkably fast for the most inclusive $V+A$ channel, where perturbative QCD seems to work even at quite low values of $s\sim 1.2\;\mathrm{GeV}^2$.

An exhaustive re-analysis of the $\alpha_s(m_\tau^2)$ determination was performed in Ref.~\cite{Pich:2016bdg}. The aim was to carefully assess all significant sources of non-perturbative systematic uncertainty. Table~\ref{tab:AlphaTauSummary} summarizes the most reliable determinations, obtained with the total $V+A$ spectral function. Compatible results, although with larger uncertainties, can be extracted from the separate $V$ and $A$ distributions. The different rows in the table correspond to different choices of pinched weights, with very different sensitivities to non-perturbative effects:
\begin{align}
\omega_{kl}(x) & = (1-x)^{2+k}\, x^l \, (1+2x) \, ,
\qquad\qquad & (k,l) = \{(0,0),\, &(1,0), (1,1), (1,2), (1,3)\}\, ,
\nonumber\\
\hat\omega_{kl}(x) & = (1-x)^{2+k}\, x^l \, ,
\qquad\qquad  &(k,l) = \{(0,0),\, &(1,0), (1,1), (1,2), (1,3)\}\, ,
\nonumber\\
\omega^{(2,m)}(x) & = 
1-(m + 2)\, x^{m + 1} + (m + 1)\, x^{m + 2}\, ,
&&  1\le m\le 5\, ,
\nonumber\\
\omega_{a}^{(1,m)}(x)& = (1- x^{m+1})\, \mathrm{e}^{-ax}\, ,
&&  0\le m\le 6\, . \label{eq:weightsdet}
\end{align}
In this section we summarize the key points.

\begin{table}[t]
\centering
\begin{tabular}{|c|c|c|c|}
\hline 
\multirow{2}{*}{Method}
  & \multicolumn{3}{c|}{$\alpha_{s}^{(n_f=3)}(m_{\tau}^2)$}
\\[1.2pt] \cline{2-4}
& \raisebox{-2pt}{CIPT} & \raisebox{-2pt}{FOPT} & \raisebox{-1.5pt}{Average}
\\[1.2pt] \hline &&&\\[-8pt]
$\omega_{kl}(x)$ weights & $0.339 \,{}^{+\, 0.019}_{-\, 0.017}$ &
$0.319 \,{}^{+\, 0.017}_{-\, 0.015}$ & $0.329 \,{}^{+\, 0.020}_{-\, 0.018}$
\\[3pt]
$\hat\omega_{kl}(x)$ weights  & $0.338 \,{}^{+\, 0.014}_{-\, 0.012}$ &
$0.319 \,{}^{+\, 0.013}_{-\, 0.010}$ & $0.329 \,{}^{+\, 0.016}_{-\, 0.014}$
\\[3pt]
$\omega^{(2,m)}(x)$ weights  & $0.336 \,{}^{+\, 0.018}_{-\, 0.016}$ &
$0.317 \,{}^{+\, 0.015}_{-\, 0.013}$ & $0.326 \,{}^{+\, 0.018}_{-\, 0.016}$
\\[1.5pt]
$s_0$ dependence  & $0.335 \pm 0.014$ &
$0.323 \pm 0.012$ & $0.329 \pm 0.013$
\\[1.5pt]
$\omega_{a}^{(1,m)}(x)$ weights  & $0.328 \, {}^{+\, 0.014}_{-\, 0.013}$ &
$0.318 \, {}^{+\, 0.015}_{-\, 0.012}$ & $0.323 \, {}^{+\, 0.015}_{-\, 0.013}$
\\[2pt] \hline &&&\\[-9pt]
Average & $0.335 \pm 0.013$ & $0.320 \pm 0.012$ & $0.328 \pm 0.013$
\\[1pt] \hline
\end{tabular}
\caption{Determinations of $\alpha_{s}^{(n_f=3)}(m_{\tau}^2)$ from $\tau$ decay data, in the $V+A$ channel~\protect\cite{Pich:2016bdg}.}
\label{tab:AlphaTauSummary}
\end{table}

\subsection{ALEPH-like sets of weights}

The theoretical framework described in section~\ref{sec:formalism} implies that the weighted integrals $A^{(n)}_\cJ (s_0)$ depend on a large number of unknown parameters:
\begin{equation}\label{eq:parametric}
A^{(n)}_\cJ (s_0)[\alpha_s,K_{m\ge 5},\beta_{m\ge 6}, \mathcal{O}_{2n+2,\cJ}(\mu), \mathcal{P}_{D\neq 2n+2,\cJ},\Delta A^{(n)}_\cJ(s_0)] \, .
\end{equation}
If these parameters were allowed to take arbitrary values, without any physics justification, one could fit any given set of $A^{(n)}_\cJ (s_0)$ inputs, independently of whether they correspond to actual measurements or are just fake data. As in any power expansion, the series need to be truncated in order to have predictive power, and this entails some theoretical notion about the natural size of their coefficients.

Given the relatively good behaviour of the perturbative Adler series, we take a very conservative range $K_5=275\pm 400$ for the unknown fifth-order coefficient, and assume that higher-order corrections are encapsulated by this variation. Since the known fifth-order coefficient of the QCD $\beta$ function has already a negligible numerical impact on the results, we can safely disregard the unknown contributions from $\beta_{m\ge 6}$. In order to estimate the perturbative uncertainty, we supplement the $K_5$ variation
with the residual dependence on the renormalization scale within the interval $\mu^2/(-s)\in (0.5,2)$.

On the other hand, it is obvious that there is an energy regime where the most relevant power correction comes from the 
operator of lowest dimension, irrespectively from
whether it enters suppressed or not by short-distance QCD loops. The corresponding truncated prescription would correspond to keeping just the lowest-dimension contribution, disregarding whether it 
involves $\mathcal{O}_{D,\cJ}$, $\mathcal{P}_{D,\cJ}$ or both.

In the ALEPH-like fits one assumes that power corrections are small enough so that only the lowest-dimensional condensates $\mathcal{O}_{D,\cJ}$ can have some impact on the observables at $s_{0}=m_{\tau}^2$. Thus, one neglects 
all $\mathcal{P}_D$ factors and only the $\mathcal{O}_{D,\cJ}$ contributions with dimension smaller than $D_{\mathrm{cut}}$ are taken into account.
The original ALEPH fit adopts
the truncation prescription $D_{\mathrm{cut}}=10$, {\it i.e.}, the higher-dimensional corrections from 
$\mathcal{O}_{D \ge 10,\cJ}$ are neglected. Additionally, one assumes that DVs are negligible for double-pinched weight functions at the $\tau$ mass scale. In general, this is expected to be a safe assumption. The sizable fluctuations of the spectral functions observed in Figure~\ref{fig:ALEPHsf}, which are expected to go to zero exponentially at large values of $s$, already have a negligible numerical role for those integrated moments in a rather large $s_{0}$ interval.

The first row in Eq.~(\ref{eq:weightsdet}) shows the five weights employed in the ALEPH analysis. Although the resulting fit quality is good, there is some arbitrariness in this specific choice of weights and in the adopted truncation. Therefore, one must test the stability of the results under variations of the weight factors and analyze the uncertainties associated with the truncation of the OPE. The impact on $\alpha_s$ from neglected condensates of higher dimensions has been estimated including $\mathcal{O}_{10,\cJ}$ in the fit and taking the difference as an additional uncertainty.
As far as experimental errors do not increase too much, and barring accidental (or artificial) fine-tuning, the size of the $\alpha_s$ variation gives a good estimator of the systematic uncertainty due to truncation.
This leads to the determination of $\alpha_s(m_\tau^2)$ shown in the first row of Table~\ref{tab:AlphaTauSummary} for both perturbative prescriptions, FOPT and CIPT. 
The values obtained with the two prescriptions have been finally combined, adding quadratically half their difference as an additional systematic uncertainty.

The second and third rows in Table~\ref{tab:AlphaTauSummary} show the results obtained with the two alternative sets of weights $\hat{\omega}_{kl}$ and $\omega^{(2,m)}$, defined in the second and third rows of Eq.~(\ref{eq:weightsdet}). Apart from leading to further not redundant self-consistence tests for $\alpha_s$, each set of weights brings a different asset. The former eliminates the kinematic $(1+2x)$ factor of the ALEPH weights, nullifying any possible contribution of
$\mathcal{O}_{16,\cJ}$ and slightly reducing the potential impact of DVs. The second removes the contribution from $D=4$.
The three sets of weights give fits of excellent quality in the more inclusive $V+A$ channel. The fitted values for the power corrections are always small and the $\alpha_s$ determination is very stable (see Table~\ref{tab:AlphaTauSummary}). The very same value of the strong coupling is obtained from different combinations of weights, with very different sensitivities to the vacuum condensates.

\subsection{Complementary tests}

The observation made in the previous paragraph led us to make further tests in Ref.~\cite{Pich:2016bdg}. The role of power corrections appears to be rather marginal at $s_0\sim m_{\tau}^2$. This suggests that perturbation theory alone, {\it i.e.}, Eq.~(\ref{eq:parametric}) with all power corrections neglected, may give a good description of the data, so that similar $\alpha_s$ values would be obtained from different weights.
Table~\ref{tab:noOPE} shows the fitted values for $\alpha_s(m_\tau^2)$ obtained from a single moment, neglecting all non-perturbative contributions. The twelve different results correspond to twelve different choices of weights: $\omega^{(1,m)}(x) = 1-x^{m+1}$ and $\omega^{(2,m)}(x)$, with $0\le m\le 5$. While these numbers cannot be used in the final determination of the strong coupling, they do provide a useful assessment of the neglected corrections because each weight has a different sensitivity to these effects. The table exhibits an amazing stability of the results, which in all cases are well within the error ranges of our determinations in Table~\ref{tab:AlphaTauSummary}, suggesting that the missing non-perturbative contributions are most likely small.

\begin{table}[t]
\renewcommand\arraystretch{1.2}
\centering
\begin{tabular}{|c|c|c||c|c|c|}
\hline
Weight & \multicolumn{2}{|c||}{$\alpha_s(m_\tau^2)$} &
Weight & \multicolumn{2}{|c|}{$\alpha_s(m_\tau^2)$}
\\ \cline{2-3} \cline{5-6}  $(n,m)$ & FOPT    & CIPT   & $(n,m)$ & FOPT    & CIPT
\\ \hline
(1,0) & $0.315 \, {}^{+0.012}_{-0.007}$ & $0.327 \, {}^{+0.012}_{-0.009}$ &
(2,0) & $0.311 \, {}^{+0.015}_{-0.011}$ & $0.314 \, {}^{+0.013}_{-0.009}$
\\ \hline
(1,1) & $0.319 \, {}^{+0.010}_{-0.006}$ & $0.340 \, {}^{+0.011}_{-0.009}$ &
(2,1) & $0.311 \, {}^{+0.011}_{-0.006}$ & $0.333 \, {}^{+0.009}_{-0.007}$
\\ \hline
(1,2) & $0.322 \, {}^{+0.010}_{-0.008}$ & $0.343 \, {}^{+0.012}_{-0.010}$ &
(2,2) & $0.316 \, {}^{+0.010}_{-0.005}$ & $0.336 \, {}^{+0.011}_{-0.009}$
\\ \hline
(1,3) & $0.324 \, {}^{+0.011}_{-0.010}$ & $0.345 \, {}^{+0.013}_{-0.011}$ &
(2,3) & $0.318 \, {}^{+0.010}_{-0.006}$ & $0.339 \, {}^{+0.011}_{-0.008}$
\\ \hline
(1,4) & $0.326 \, {}^{+0.011}_{-0.011}$ & $0.347 \, {}^{+0.013}_{-0.012}$ &
(2,4) & $0.319 \, {}^{+0.009}_{-0.007}$ & $0.340 \, {}^{+0.011}_{-0.009}$
\\ \hline
(1,5) & $0.327 \, {}^{+0.015}_{-0.013}$ & $0.348 \, {}^{+0.014}_{-0.012}$ &
(2,5) & $0.320 \, {}^{+0.010}_{-0.008}$ & $0.341 \, {}^{+0.011}_{-0.009}$
\\ \hline
\end{tabular}
\caption{\label{tab:noOPE} Values of the strong coupling extracted from a single $A^\omega_{V+A}(s_0)$ moment with weights $\omega^{(1,m)}(x)$  or $\omega^{(2,m)}(x)$, $0\le m\le 5$, at $s_0 = 2.8\;\mathrm{GeV}^2$ and neglecting all non-perturbative corrections \cite{Pich:2016bdg}.}
\end{table}

\begin{figure}[tb]
\centerline{
\includegraphics[width=0.498\textwidth]{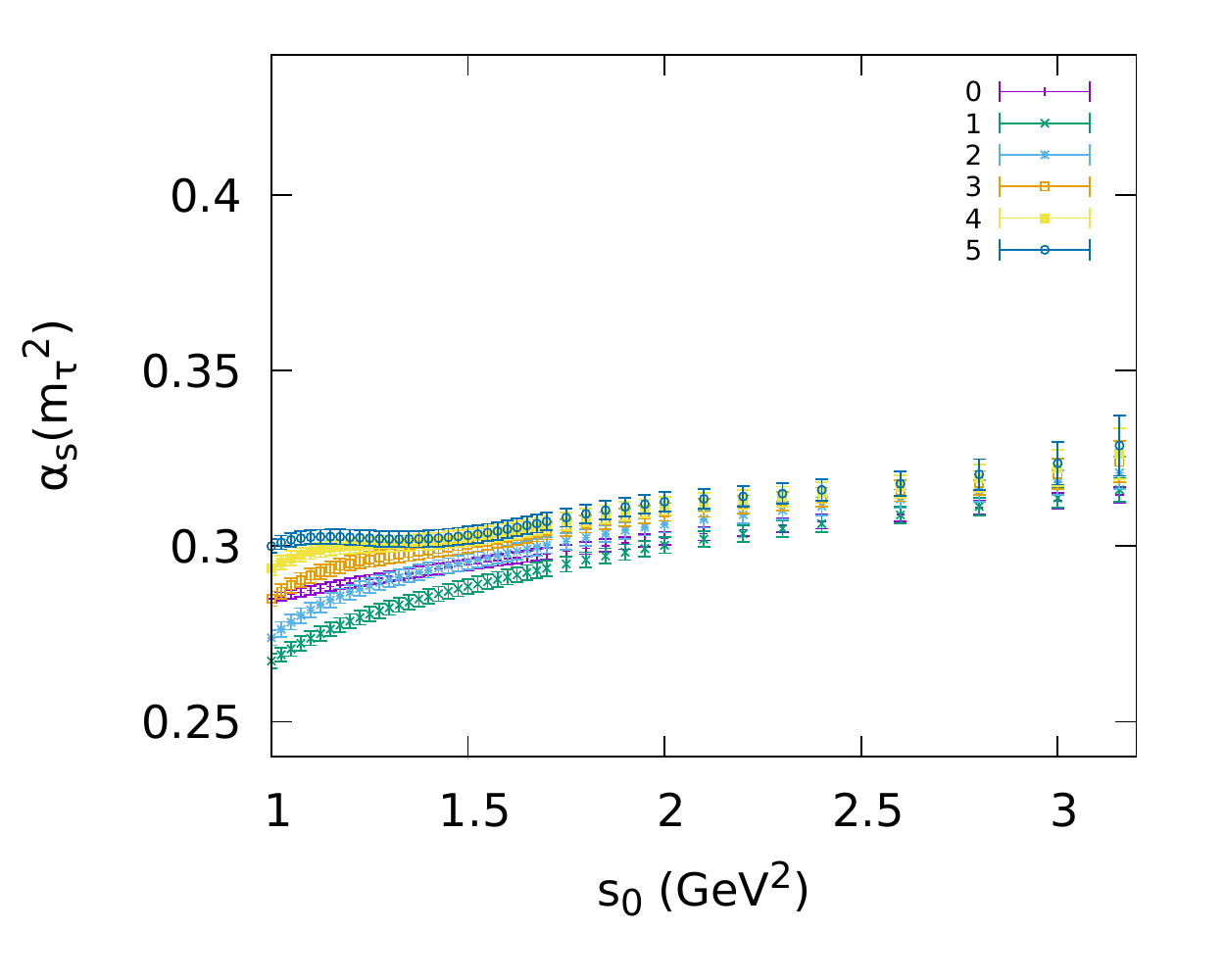}\hskip .2cm
\includegraphics[width=0.498\textwidth]{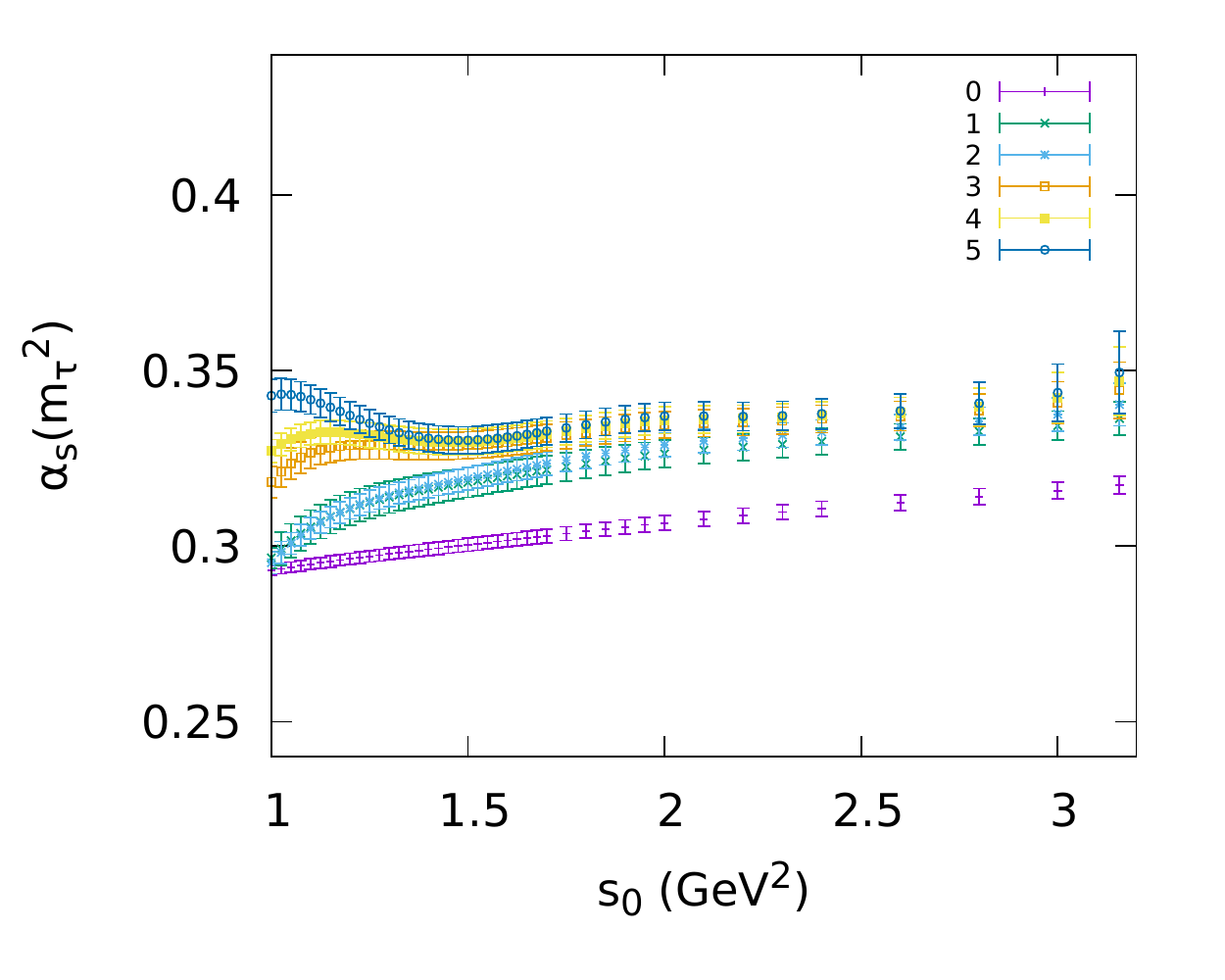}
}
\vskip -.4cm
\caption{\label{fig:todoa0}
$\alpha_{s}(m_{\tau}^{2})$ determinations with FOPT (left) and CIPT (right),
extracted from a single $A^\omega_{V+A}(s_0)$ moment, with different weights $\omega^{(2,m)}(x)$ ($0\le m\le 5$), at different values of $s_0$, ignoring all non-perturbative effects \protect\cite{Pich:2016bdg}.
Only experimental uncertainties are displayed.}
\end{figure}

Figure~\ref{fig:todoa0} displays how these results vary when the same exercise is performed at different values of $s_0$. The six weights $\omega^{(2,m)}(x)$ lead to fully compatible results in practically the whole range of $s_0$ analyzed. Notice that only the experimental errors are shown. The small observed fluctuations remain always within the larger perturbative uncertainties, which increase as $s_0$ decreases. 
The missing non-perturbative corrections to these moments are very different, spanning a large variety of inverse powers of $s_0$ that do not show up in the figure. The very similar $s_0$-dependence of the six curves provides a clear confirmation of the small size of power corrections. Similar results have been obtained with seven  $\omega^{(1,m)}(x)$ weights ($0\le m\le 6$) \cite{Pich:2016bdg}.

\begin{figure}[bht]
\centerline{
\includegraphics[width=0.498\textwidth]{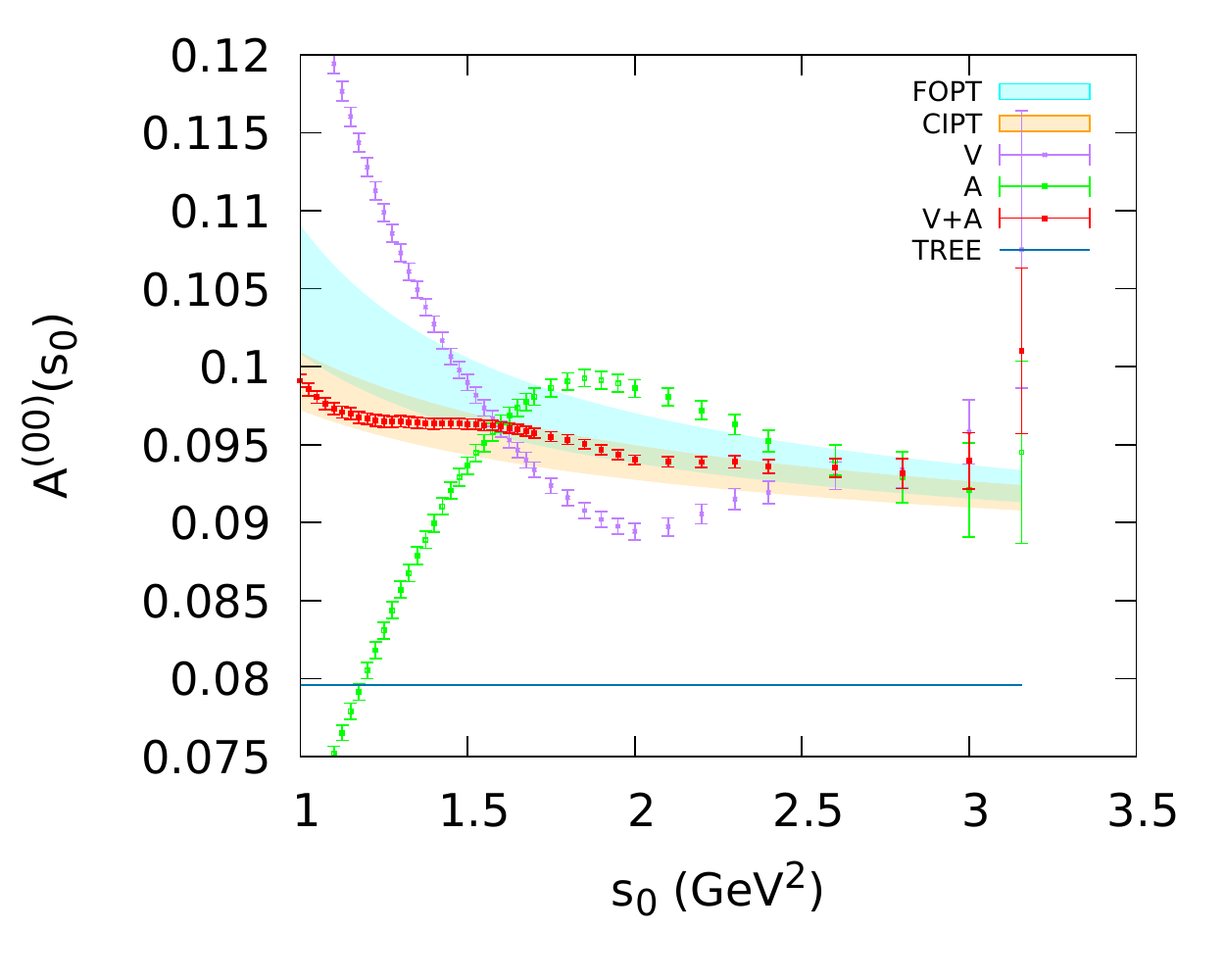}\hskip .15cm
\includegraphics[width=0.498\textwidth]{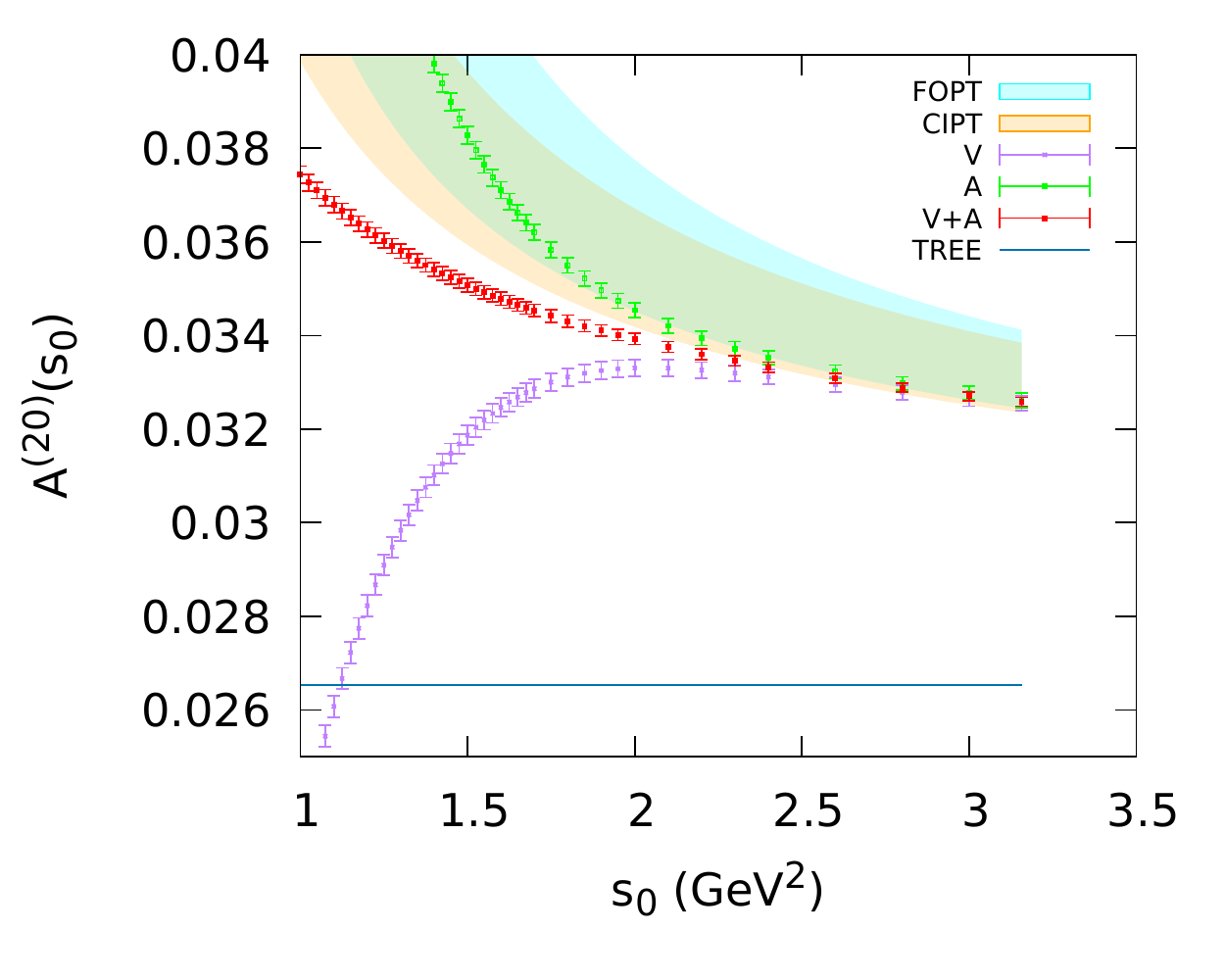}
}
\vskip -.4cm
\caption{\label{fig:pinchs}
Dependence on $s_0$ of the experimental moments $A^\omega_\cJ (s_{0})$ for the $V$ (purple), $A$ (green) and $\frac{1}{2}\, (V+A)$ (red) channels. The left plot corresponds to the weight $\omega^{(0,0)}(x) =1$, and the right one to
$\omega^{(2,0)}(x) =(1-x)^2$. The orange and light-blue regions are the CIPT and FOPT perturbative predictions for
$\alpha_s(m_\tau^2) = 0.329 \,{}^{+\, 0.020}_{-\, 0.018}$ \protect\cite{Pich:2016bdg}.
The blue horizontal lines at the bottom indicate the parton-model prediction.}
\end{figure}

Figure~\ref{fig:pinchs} compares two experimental moments, for the vector, axial-vector and $\frac{1}{2}\, (V+A)$ distributions, with their perturbative predictions, ignoring all non-perturbative contributions. Perturbation theory gives an identical prediction for the three distributions; its variation within the range $\alpha_s(m_\tau^2) = 0.329 \,{}^{+\, 0.020}_{-\, 0.018}$, in FOPT and CIPT, is indicated by the coloured bands. The left plot corresponds to the weight $\omega^{(0,0)}(x) =1$, {\it i.e.}, a direct integration of the measured spectral function without any weight. This moment does not receive any leading-order OPE power correction, but it is more exposed to violations of quark-hadron duality. The experimental curves show indeed a beautiful signal of duality violations: a clear oscillation of the $V$ and $A$ curves in opposite directions that cancels to a rather large extent in the total $V+A$ moment. The $V+A$ curve exhibits a surprisingly smooth behaviour, remaining within the $1\sigma$ CIPT band even at low values of $s_0\sim 1$~GeV. The $V$, $A$ and $\frac{1}{2}\, (V+A)$ experimental moments nicely join above $2.5~\mathrm{GeV}^2$, so that one can no-longer identify any duality-violation signal.

The right plot in Figure~\ref{fig:pinchs} corresponds to the weight $\omega^{(2,0)}(x) =(1-x)^2$. It clearly shows that the double-pinch factor has eliminated the visible signal of duality violations. Wiggles are no-longer present in any of the three curves. At the same time, it exhibits the presence of a clear ($D=6$) power correction, 
with opposite signs in the $V$ and $A$ moments, which matches the behaviour expected from the OPE. However, this correction seems to be tiny at $s_0\sim m_\tau^2$ because the $V$, $A$ and $\frac{1}{2}\, (V+A)$ experimental curves join  above $2.2~\mathrm{GeV}^2$ and, moreover, remain within the $1\sigma$ perturbative bands. In the higher energy bins, the numerical size of DVs and power corrections gets then masked by the much larger perturbative uncertainties.

\subsection[Determinations based on the $s_0$ dependence]{\boldmath Determinations based on the $s_0$ dependence}

Fitting the $s_0$ dependence of a single $A^{(2,m)} (s_{0})$ moment, above some $\hat s_0 \ge 2.0\;\mathrm{GeV}^2$, one can also extract the values of $\alpha_s(m_\tau^2)$, $\mathcal{O}_{2(m+2),\cJ}$ and $\mathcal{O}_{2(m+3),\cJ}$. The sensitivity to power corrections is poor, as expected, but one finds a surprising stability in the extracted values of $\alpha_s(m_\tau^2)$ at different $\hat s_0$. The fourth line of Table~\ref{tab:AlphaTauSummary} combines the information from three different moments ($m=0,1,2$), adding as an additional theoretical error the fluctuations with the number of fitted bins. 
Notice that this determination of the strong coupling is much more sensitive to violations of quark-hadron duality because the $s_0$ dependence of consecutive bins feels the local structure of the spectral function. The agreement with the other determinations shown in the table confirms the small size of duality violations in the $V+A$ distribution above $\hat s_0$.\footnote{Instead of fitting all energy points at the same time and inflate uncertainties based on the fluctuations in $\alpha_s$, we could have opted for taking a set of points with larger energy separation, removing to some extent the sensitivity to those DV fluctuations. However, the result would be  essentially equivalent, since then we would have eventually averaged over the arbitrary selection of energy points, using finally the same amount of experimental information.}

Weights with an extra exponential suppression $\mathrm{e}^{-ax}$, with $a>0$, are also interesting for determining $\alpha_s$. As shown in Eq.~(\ref{eq:DV_A}), they clearly reduce DVs. Moreover, for small values of $a \lesssim 0.5$, their induced power corrections are suppressed by a numerical factor $\frac{a^{D/2}}{(D/2)!}$ and, therefore, are
not going to be larger than the previously neglected $\mathcal{P}_{D,\cJ}$ contributions, leading in principle to a free gain with respect to the $a=0$ case.\footnote{In practice the further suppression of $\mathcal{P}_{4,\cJ}$ suggests taking prefactors that nullify $\mathcal{O}_{4,\cJ}$, such as $\omega(x) =(1+ax)\, e^{-ax}$.}
Taking into account that power corrections appeared to have a marginal role at $s_0 \sim m_{\tau}^2$, in Ref.~\cite{Pich:2016bdg} we opted for taking the weights $\omega_{a}^{(1,m)}$, defined in the last row of Eq.~(\ref{eq:weightsdet}). They provide a completely different sensitivity to non-perturbative corrections because their exponential suppression nullifies the higher s region, strongly reducing the violations of quark-hadron duality, at the price of being more exposed to OPE contributions of arbitrary dimensionality. Performing a pure perturbative analysis, the neglected power corrections should manifest as large instabilities of $\alpha_s$ under variations of $s_0$ and $a\not=0$; however, stable results are found for a broad range of values of $s_0$ and $a$, which again indicates small power corrections. The last line in Table~\ref{tab:AlphaTauSummary} combines the information extracted from seven different moments  with $0\le m\le 6$.

The excellent agreement among all determinations shown in Table~\ref{tab:AlphaTauSummary}, obtained with a broad variety of approaches that have very different sensitivities to non-perturbative corrections, demonstrates the small numerical impact of these contributions.

\section{Duality-violation approach to the strong coupling}
\label{sec:DVmethod}

We have now all the ingredients needed to analyze the strategy advocated in 
Refs.~\cite{Cata:2008ye,Cata:2008ru,Boito:2011qt,Boito:2012cr,Boito:2014sta} and assess its advantages and weaknesses.
The basic quantity being investigated is the integral
\begin{equation}
A^{\omega_0}_\cJ (s_0) \,\equiv\, A^{\omega^{(0,0)}}_\cJ (s_0) \, = \, \pi\int_{s_{\mathrm{th}}}^{s_0}\frac{ds}{s_0}\;\rho_\cJ(s)\, ,
\end{equation}
constructed with the simplest weight factor $\omega_0(s)\equiv \omega^{(0,0)}(s) = 1$. Since there is no weight, the leading power corrections $\mathcal{O}_{D,\cJ}$ do not contribute to this particular moment. However, it does receive contributions from the subleading $\mathcal{P}_{D,\cJ}$ terms in the OPE and, moreover, it is not protected against duality violations.

If one neglects all $\mathcal{P}_{D,\cJ}$ contributions, the moment $A^{\omega_0}_\cJ (s_0)$ only depends on the strong coupling and the DV correction. A sensible approach would be going to the most inclusive channel, $V+A$, and use the fact that, practically in the whole range where perturbation theory makes sense, the local DV fluctuations have a very subleading role in the integral, as explicitly shown by the red data points in Fig.~\ref{fig:pinchs}. The left panel in this figure exhibits indeed a very smooth $s_0$ dependence of the moment  $A^{\omega_0}_{V+A} (s_0)$.
Since the fluctuations are expected to go to zero very fast when $s_0$ increases, their size in a large-enough interval can give a conservative assessment of DVs. Essentially this corresponds to the determination given before in the fourth row of Table~\ref{tab:AlphaTauSummary} that nicely agrees with the 
determinations based on pinched weights, which are also shown in the table.

One may further insist in analyzing the separate $V$ and $A$ channels where sizeable oscillations are visible in Fig.~\ref{fig:pinchs}. In that case, given the flattening of the purple and green curves in the higher energy bins, it would still be possible to assume that near the $\tau$ mass, before experimental uncertainties become too large, DVs are small. Remarkably, it is obvious from the figure that one would obtain a value of the strong coupling very close to the $V+A$ one. 
However, the robustness of this isolated assumption is weaker. 

In order to have a better control on DV effects, one may assume a functional form for the spectral function, to be fitted from data, and use Eq.~(\ref{eq:DV_A}) to measure the size of the duality-violation contribution to $A^{\omega_0}_\cJ (s_0)$.
Let us then assume that the functional ansatz 
\begin{equation}
\label{eq:DVparam}
\Delta\rho^{\mathrm{DV}}_{\cJ}(s)\; =\; \mathcal{G}_\cJ(s) \; e^{-(\delta_{\cJ}+\gamma_{\cJ}s)}\;\sin{(\alpha_{\cJ}+\beta_{\cJ}s)}\, ,
\qquad\qquad s>
\hat s_0\, .
\end{equation}
provides a reasonable description of $\Delta\rho^{\mathrm{DV}}_{\cJ}(s)$ above some invariant mass $\hat s_0$. The combination of an oscillatory function with a damping exponential is assumed to describe the fall-off of duality violations at very high energies  \cite{Shifman:2000jv}. The ansatz adopted in~\cite{Cata:2008ye,Cata:2008ru,Boito:2011qt,Boito:2012cr,Boito:2014sta} corresponds to the choice
\begin{equation}\label{eq:default}
\left.\mathcal{G}_\cJ(s)\right|_{\mathrm{Default}} \, =\, 1\, ,
\qquad\qquad\qquad
\left. \hat{s}_{0}\right|_{\mathrm{Default}}=1.55 \, \mathrm{GeV}^2 \, .
\end{equation}
This four-parameter functional form is theoretically well motivated, but it
cannot be derived from first principles and nobody really knows above which
value of $\hat s_0$ it could start to be a good approximation. We have added the global factor $\mathcal{G}_\cJ(s)$ in order to assess later the stability of the results under slight variations of the assumed parametrization.

With all these assumptions, the DV ansatz parameters and the strong coupling can be extracted from a fit to the $s_0$ dependence of the experimental moment  $A_\cJ^{\omega_0}(s_0)_{\mathrm{exp}}$.
The algorithmic procedure
involves essentially the following simple steps:
\begin{enumerate}
\item The ansatz parameters are fitted, bin by bin, to the $s_0$ dependence of $A^{\omega_0}_\cJ (s_0)_{\mathrm{exp}}$, in the interval $\hat{s}_0 < s_0\le m_\tau^2$. This is mathematically equivalent to a direct fit of $\rho_\cJ(s)$ (the derivative of the integral of the spectral function),\footnote{Let us note that $\hat{s}_0$ is typically chosen in such a way that the fit quality is good. Having several free parameters, there is always an interval where this is going to occur, independently on whether the ansatz or the rest of assumptions are correct or not outside that interval, which is where we want to assess DVs. 
This procedure cannot be tested with $A^{(n)}_\cJ(s_0)$ moments because of the supplementary free parameters needed to fit them.}
as demonstrated in Eq.~(\ref{eq:ImPifrommom}) and appendix~\ref{subsubsec:s0-dependence}.

\item Eq.~(\ref{eq:DV_A}) is then used to compute $\Delta A^{\omega_0}_\cJ (\hat s_0)$, by integrating the fitted ansatz from the chosen $\hat s_0$ to infinity. The parametric errors of the fit are the only uncertainties considered. No error whatsoever is included to account for the arbitrary choice of a particular ansatz.

\item Since the strong coupling is largely insensitive to the local spectral function and the small correlation of $A_{\mathcal{J}}^{\omega_0}(\hat{s}_0)$ and $\rho_{\mathcal{J}}(s)$ at $s>\hat{s}_0$ plays a very marginal role, $\alpha_s$ is mostly extracted from $A^{\omega_0}_\cJ (\hat s_0)_{\mathrm{pert}} \approx A^{\omega_0}_\cJ (\hat s_0)_{\mathrm{exp}}-\Delta A^{\omega_0}_\cJ (\hat s_0)$, assuming that this difference is well described by perturbative QCD at the scale $\hat s_0$.
\end{enumerate}
The dangers of this prescription are quite obvious. Since one needs to employ enough energy bins to fit the ansatz parameters, the strong coupling is finally fixed at a very low scale $\hat s_0 \sim (1.2~\mathrm{GeV})^2$ where theoretical errors are large and perturbative QCD is suspect.
 Moreover, the subtracted duality-violation contribution is a rather sizeable integral that has been estimated in a model-dependent way, without any study of its possible variation under reasonable modifications of the ansatz. Thus, $A^{\omega_0}_\cJ (\hat s_0)_{\mathrm{pert}}$ is determined from a difference of two numbers and its resulting precision is limited by the actual theoretical control on $\Delta A^{\omega_0}_\cJ (\hat s_0)$. 

Once the ansatz parameters and $\alpha_s(\hat s_0)$ have been determined, one can use different weights to determine the OPE vacuum condensates from the moments $A^{(n)}_\cJ (\hat s_0)_{\mathrm{exp}}$. This is the only additional information available because the whole experimental data above $\hat s_0$ has been already used in the previous fit. Again, the duality-violation corrections $\Delta A^{(n)}_\cJ (\hat s_0)$ are first computed with the fitted ansatz, and the wanted condensates are approximately extracted from the differences $A^{(n)}_\cJ (\hat s_0)_{\mathrm{exp}}-\Delta A^{(n)}_\cJ (\hat s_0)$. The accuracy of this procedure can easily deteriorate in a very fast way. A slight modification of the assumed ansatz can generate non-negligible changes of the duality corrections, which get amplified when subtracting them from the experimental moments, 
implying slightly different values of $\alpha_s(\hat s_0)$ and corresponding modifications of all power corrections to compensate for the resulting differences in $A^{(n)}_\cJ (\hat s_0)_{\mathrm{pert}}$.

\subsection{Truncated versus non-truncated OPE}

A rather surprising argument, based on rejecting the truncation of the OPE, has been put forward in Refs.~\cite{Boito:2016oam,Boito:2019iwh,Boito:2020xli}, aiming to criticize the more standard determination of $\alpha_s$ discussed in section~\ref{sec:alphasResults} and 
to advocate the alternative use of a specific model of duality violations in non-protected moments with the algorithmic DV procedure described above.

The starting point consists in assuming a too small value for the strong coupling, obtained from a very unstable DV fit to
$A^{\omega_0}_V (\hat s_0)$ with the default ansatz in Eqs. (\ref{eq:DVparam}) and (\ref{eq:default}). A too small (or too large) strong coupling leads to very poor perturbative predictions for all the moments, when directly comparing them with data. An ad-hoc way of curing it, without correcting the input value of $\alpha_s$, is by adding as many arbitrary model parameters as observables.

In moments with pinched weight functions, duality violations are found to be very suppressed, independently of their exact shape. Therefore, $\Delta A^{\omega}_\cJ (\hat s_0)$ cannot compensate an incorrect value assumed for $\alpha_s$. The proposed solution advocated in Refs.~\cite{Boito:2016oam,Boito:2019iwh,Boito:2020xli} consists then in keeping all (an infinite number) higher-dimension $\mathcal{O}_{D,\cJ}$ coefficients, arguing that they are very large (divergent series), while at the same time all $\mathcal{P}_{D,\cJ}$ corrections are neglected. Clearly, the first statement is incompatible with the second approximation.

In any phenomenological application of the OPE, one needs to assume from the very beginning that there is an energy regime where the inverse-power expansion makes sense. Otherwise, the theoretical OPE description would be meaningless, irrespective of whether one truncates or not the series. 
For a given dimension $D$, neglecting the $\mathcal{P}_{D,\cJ}$ correction with respect to the corresponding $\mathcal{O}_{D,\cJ}$ contribution is reasonable because $\mathcal{P}_{D,\cJ}$ carries an additional $\alpha_s$ suppression, so that typically $|\mathcal{P}_{D,\cJ}|\sim 0.2 \, |\mathcal{O}_{D,\cJ}|$. However, this suppression is largely compensated by the much stronger power suppression of the higher-dimension condensate contributions. The validity of the OPE requires that
\begin{equation}
 \frac{|\mathcal{O}_{D+2k,\cJ}|}{s_{0}^k} \, \lesssim\,  |\mathcal{O}_{D,\cJ}|  \, .
\end{equation}
Thus, it looks quite implausible finding an energy regime where
$|\mathcal{P}_{D,\cJ}| \sim 0.2 \, |\mathcal{O}_{D,\cJ}| \ll |\mathcal{O}_{D+2k,\cJ}|/s_{0}^k$ could be satisfied. For this to be the case, one would need the following fine-tuned condition on $s_0$, for all positive values of $k$:
\begin{equation}
\left|\frac{\mathcal{O}_{D+2k,\cJ}}{\mathcal{O}_{D,\cJ}}\right|^{1/(2k)}< \;\sqrt{s_0}\; \ll\; \eta_k\,\left|\frac{\mathcal{O}_{D+2k,\cJ}}{\mathcal{O}_{D,\cJ}}\right|^{1/(2k)} \, ,
\end{equation}
where $\eta_k\sim 0.2^{-1/(2k)}\approx 1$
($\eta_4\sim 1.22$, $\eta_6\sim 1.14$, $\eta_8\sim 1.11$, $\eta_{10}\sim 1.08$, \ldots). Obviously, the assumption $|\mathcal{P}_{D,\cJ}| \ll |\mathcal{O}_{D+2k,\cJ}|/s_{0}^k$ does not make any sense.

In fact one of the many assumptions made in Ref.~\cite{Boito:2014sta} to obtain their advocated values of the strong coupling and the corresponding condensates consists in neglecting all $\mathcal{P}_{D,\cJ}$ contributions not at $s_0=m_{\tau}^2$, but at the much lower scale $\hat s_0<\frac{1}{2}\, m_{\tau}^2$, while arguing that 
the $\mathcal{O}_{D\ge 10}$ corrections
are too large to be neglected at $s_0=m_{\tau}^2$.
The impact of neglecting the former with respect to the latter scales as $0.2 \cdot 2^{D/2}$.  
A more explicit calculation shows that numerical pre-factors slightly damp this effect, but not nearly enough. Using Eq.~(\ref{eq:suleadingPowers}),  the neglected $\mathcal{P}_{D,\cJ}$ contributions to the moments with representative weights $\omega_0=1$ and $\omega_\tau$ are:
\begin{align}
\left. A^{\omega_0}_\cJ(s_0)\right|_{\mathcal{P}_{D,\cJ}}&=\,\pi\left(
-\frac{\mathcal{P}_{6,\cJ}}{2s_0^3}+\frac{\mathcal{P}_{8,\cJ}}{3s_0^4}
-\frac{\mathcal{P}_{10,\cJ}}{4s_0^5}+\frac{\mathcal{P}_{12,\cJ}}{5s_0^6}-\frac{\mathcal{P}_{14,\cJ}}{6s_0^7}+\frac{\mathcal{P}_{16,\cJ}}{7s_0^8}+\cdots\right) , \\
\left. A^{\omega_{\tau}}_{\cJ}(s_0)\right|_{\mathcal{P}_{D,\cJ}} &=\,\pi\left( 
\frac{3}{2}\frac{\mathcal{P}_{6,\cJ}}{s_0^3}
-\frac{8}{3}\frac{\mathcal{P}_{8,\cJ}}{s_0^4}
-\frac{3}{4}\frac{\mathcal{P}_{10,\cJ}}{s_0^5}
+\frac{1}{5}\frac{\mathcal{P}_{12,\cJ}}{s_0^6}
-\frac{1}{12}\frac{\mathcal{P}_{14,\cJ}}{s_0^7}
+\frac{3}{70}\frac{\mathcal{P}_{16,\cJ}}{s_0^8}+\cdots\right) \, .
\end{align}
Taking $|\mathcal{P}_{D,\cJ}|\sim 0.2\, |\mathcal{O}_{D,\cJ}|$
and the tree-level values of $\mathcal{O}_{D,\cJ}$ advocated
in Refs.~\cite{Boito:2014sta,Boito:2016oam} would completely spoil any precise determination of the strong coupling at $\hat{s}_{0}=1.55\, \mathrm{GeV}^2$, as illustrated by the estimated size of the individual corrections displayed in Table~\ref{tab:Pcont}.
Independently of any consideration about duality violations, the large values claimed for the condensates make the whole procedure inconsistent.

\begin{table}[tb]
\centering
{\renewcommand{\arraystretch}{1.3}
\begin{tabular}{|c|cccccc|c|}\hline 
Moment  & $\mathcal{P}_{6,V+A}$   & $\mathcal{P}_{8,V+A}$ &
$\mathcal{P}_{10,V+A}$   & $\mathcal{P}_{12,V+A}$
& $\mathcal{P}_{14,V+A}$& $\mathcal{P}_{16,V+A}$ & $20\%$ partonic
\\ \hline
$A^{\omega_0}_{V+A}(\hat{s}_0)$ &  
$0.0011$ & $0.0013$ &
$0.0015$ &  $0.0015$ & $0.0008$ & $0.0015$ & $0.032$
\\ \hline
$A^{\omega_{\tau}}_{V+A}(\hat{s}_0)$ &  
$0.0032$ & $0.0101$ &
$0.0044$ & $0.0015$ & $0.0004$ &  $0.0004$  & $0.016$ \\
\hline
\end{tabular}}
\caption{\label{tab:Pcont} Estimated size of the neglected $|\mathcal{P}_{D,V+A}|$ power corrections at $\hat{s}_0=1.55\, \mathrm{GeV}^2$, for the hypothetical divergent condensates advocated in Refs.~\cite{Boito:2014sta,Boito:2016oam}, compared with the size of the perturbative contribution used to extract $\alpha_s$. }
\end{table}

Let us note that, as shown in section~\ref{sec:alphasResults}, data do not indicate any signal of large condensate corrections at $s_0\sim m_\tau^2$. On the contrary, all tests performed there exhibit a smooth dependence on $s_0$, suggesting a very well-behaved OPE even at lower energy values around $s_0\sim 1.5~\mathrm{GeV}^2$.

\subsection{Modelling duality violations with the default ansatz}

In order to assess the actual uncertainties of the DV procedure described before, we will first perform several fits with the default ansatz of Ref.~\cite{Boito:2014sta}, using the same ALEPH data \cite{Davier:2013sfa}. Afterwards, we will investigate the robustness of these results by exploring how much they can change with small modifications of the assumed ansatz.
For simplicity, we adopt here the FOPT prescription when evaluating the perturbative series. Very similar conclusions can be obtained with CIPT. 

Fitting the $A_V^{\omega_0}(s_0)_{\mathrm{exp}}$ moment
with the ansatz in Eqs. (\ref{eq:DVparam}) and (\ref{eq:default}), one finds the results displayed in Table \ref{tab:tabla}.
Although the strong coupling is basically determined by the difference
$A_V^{\omega_0}(\hat s_0)_{\mathrm{exp}} - \Delta A_V^{\omega_0}(\hat s_0)$, at
$\hat{s}_0=1.55 \, \mathrm{GeV}^2$, its numerical value has been evolved to the usual reference scale $m_\tau^2$.

\begin{table}[th]\centering
\begin{tabular}{|c|cccc|c|}\hline &&&&&\\[-12pt]
 $\alpha_{s}^{(n_f=3)}(m_{\tau}^{2})$ & $\delta_V$      & $\gamma_V$      & $\alpha_V$       & $\beta_V$
& p-value (\% )
\\ \hline
 $0.298 \pm 0.010$          & $3.6 \pm 0.5$ & $0.6 \pm 0.3$ & $-2.3 \pm 0.9$ & $4.3 \pm 0.5$
& 5.3
\\ \hline
\end{tabular}
\caption{Fitted values of $\alpha_s^{(n_f=3)}(m_\tau^2)$, in FOPT, and the spectral ansatz parameters in Eq.~(\ref{eq:DVparam}\label{tab:tabla})
with the default choice (\ref{eq:default}).}
\label{tab:models}
\end{table}

\begin{figure}[tbh]
\centerline{
\includegraphics[width=0.498\textwidth]{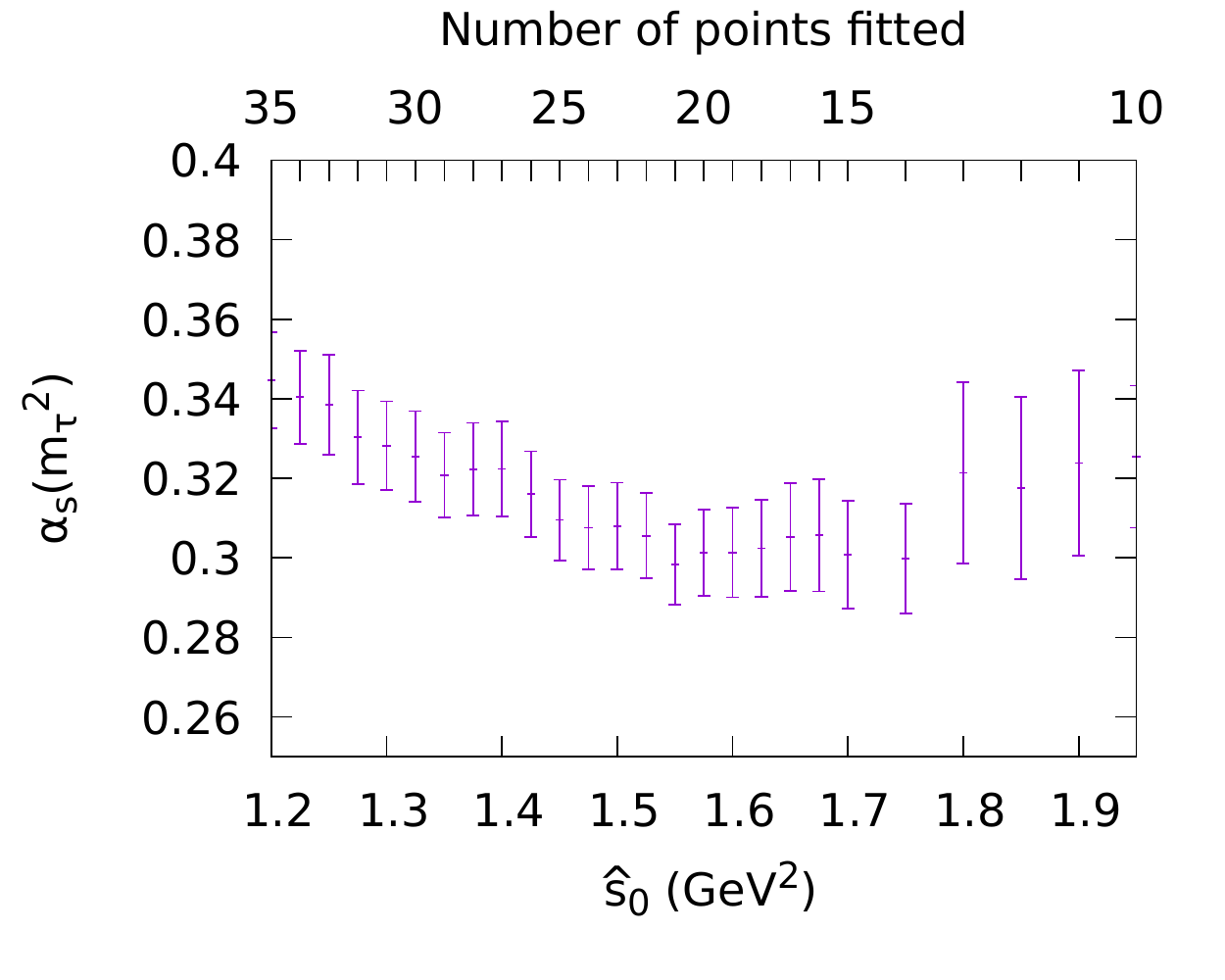}\hskip .15cm
\includegraphics[width=0.498\textwidth]{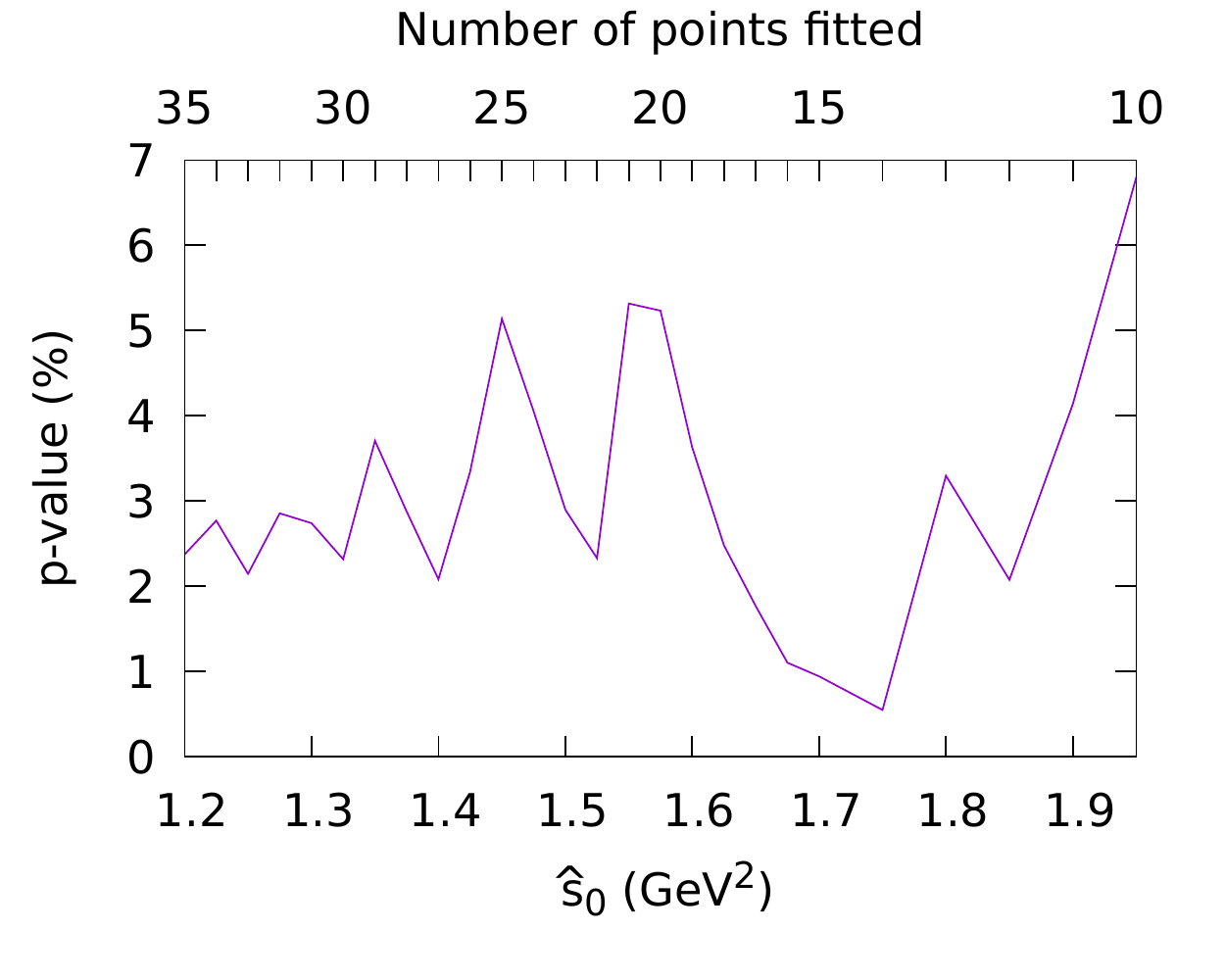}}
\vskip -.01cm
\centerline{\includegraphics[width=0.498\textwidth]{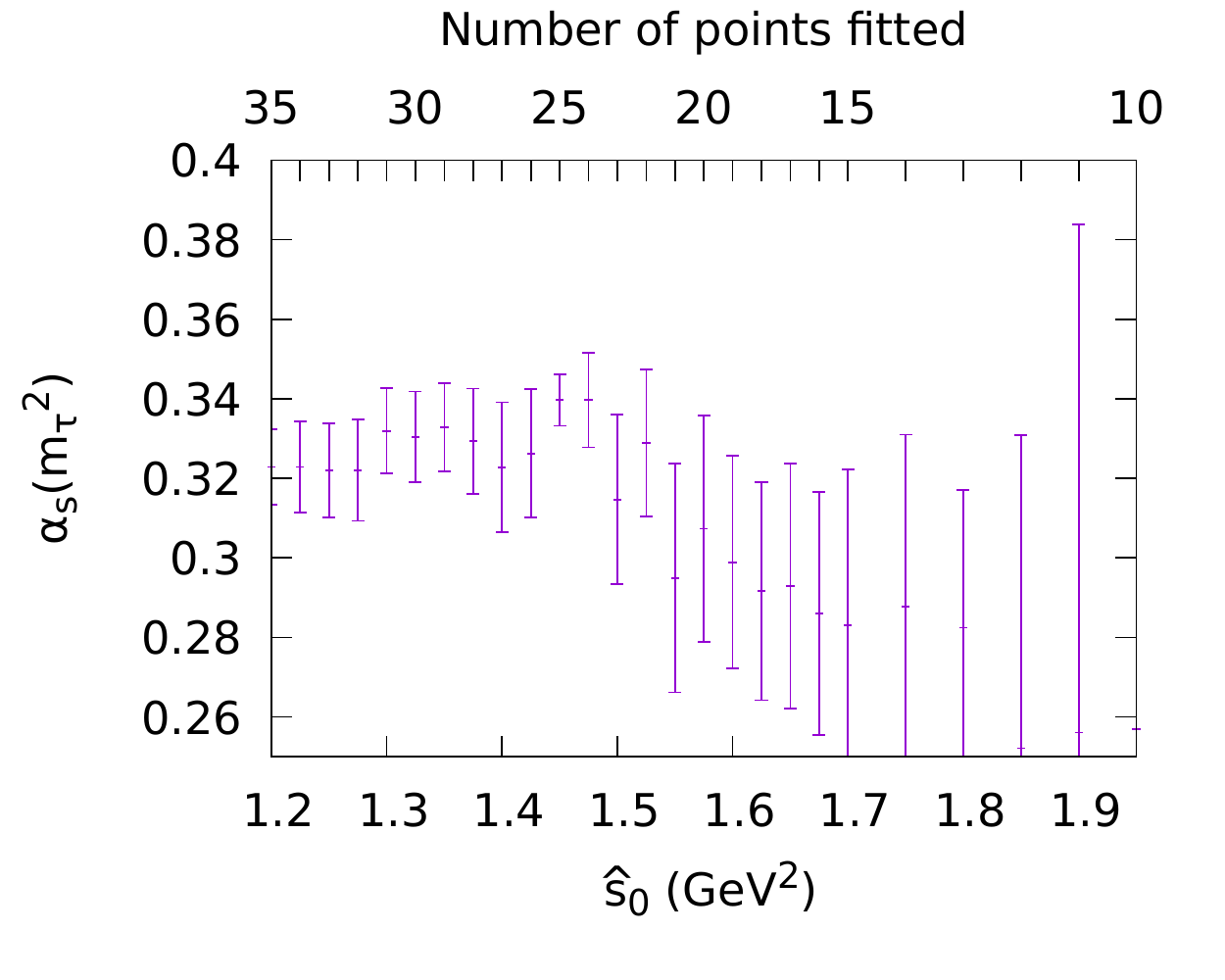}\hskip .15cm
\includegraphics[width=0.498\textwidth]{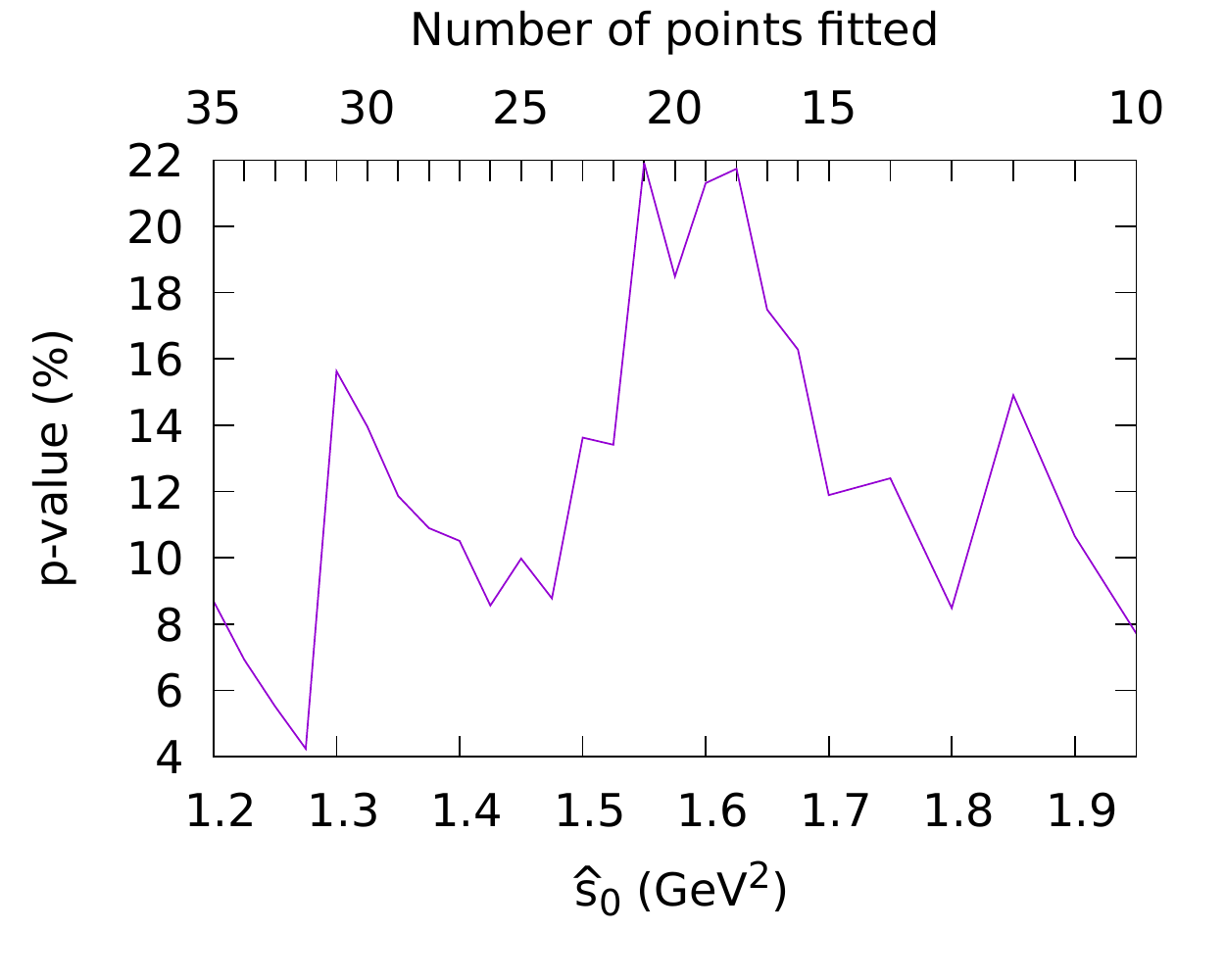}}
\vskip -.4cm
\caption{\label{fig:s0scan}
Fitted values of $\alpha_{s}$ in FOPT (left) for different choices of $\hat{s}_0$, with the default ansatz, and associated p-values (right). The top (bottom) figures correspond to the vector (axial) channels~\protect\cite{Pich:2016bdg}.}
\end{figure}

Once $\alpha_s$ and the ansatz parameters have been fixed, one can easily extract corresponding values for the OPE vacuum condensates from the experimental moments
$A^{(n)}_V(\hat{s}_{0})_{\mathrm{exp}}$, subtracting first the estimated DV contribution $\Delta A^{(n)}_V(\hat{s}_{0})$.\footnote{We refrain from making a full correlated statistical analysis to extract the condensates, because the systematic errors that we are going to discuss later are much larger than the statistical ones.}
The fitted central values are given in Table \ref{tab:models2a}.
This table shows also an approximate estimate of the corresponding axial condensates, which is good enough for our test purposes. 
At the chosen default value $\hat{s}_0=1.55\, \mathrm{GeV}^2$, no competitive information about $\alpha_s$ is obtained in the axial channel with the DV approach, even accepting all the assumptions (see Fig.~\ref{fig:s0scan}). In a combined fit, $\alpha_s$ is then going to be fixed by the vector channel. 
Thus, 
we have speeded up the numerical algorithm by taking the central value of $\alpha_s$ obtained in the vector channel to then fit the axial parameters. 
This suffices to obtain the corresponding axial condensates following the same procedure as for the vector one. The values obtained for both the DV parameters and the condensates are a good approximation of the corresponding ones given in Refs.~\cite{Boito:2014sta,Boito:2016oam}.

\begin{table}[h]\centering
{\begin{tabular}{|c|ccccccc|}\hline &&&&&&&\\[-12pt] Channel&$\mathcal{O}_{4,\cJ}$ & $\mathcal{O}_{6,\cJ}$ & $\mathcal{O}_{8,\cJ}$   & $\mathcal{O}_{10,\cJ}$   & $\mathcal{O}_{12,\cJ}$ & $\mathcal{O}_{14,\cJ}$   & $\mathcal{O}_{16,\cJ}$\\
\hline
$\cJ=V$ & $\phantom{-}0.0016$ & $-0.0082$ & $0.014$ & $-0.019$ &  $0.023$ &$-0.028$& $\phantom{-}0.037$ \\ \hline
$\cJ=A$ & $\phantom{-}0.0006$ & $-0.0016$ & $0.016$ & $-0.052$ &  $0.11\phantom{0}$ &$-0.11\phantom{0}$& $-0.28\phantom{0}$ \\ \hline
\end{tabular}}
\caption{Fitted values of the OPE condensates (in GeV units), with FOPT and $\hat s_0= 1.55\;\mathrm{GeV}^2$, using the default ansatz in Eq.~(\ref{eq:default}).}
\label{tab:models2a}
\end{table}

In order to have a better feeling on the numerical role of the DV corrections, we give in Table \ref{tab:sep_contrib0} the separate contributions of the OPE and the DV term to different moments of the vector distribution with weights $x^n$ and $1-x^n$. The table shows also the corresponding contributions at ${s}_{0}=2.8  \, \mathrm{GeV}^2$, using the same fitted parameters.
Within this model set-up, the DV contributions are found to be very suppressed when using pinched weight functions, as expected. 
Rather strikingly, even for the more unprotected $x^n$ weights, the predicted DV effects turn out to be only at the level of experimental uncertainties at $s_{0}=2.8  \, \mathrm{GeV}^2$.

\begin{table}[t]\centering
{\begin{tabular}{|c|c|ccc|ccc|}\hline 
\cline{3-8}
&& \multicolumn{3}{c|}{$\hat s_0 = 1.55\;\mathrm{GeV}^2$} &
\multicolumn{3}{c|}{$s_0 = 2.8\;\mathrm{GeV}^2$}
\\ \cline{3-8}
Weight & $n$  & OPE & DV & Exp  & OPE & DV & Exp
\\ \hline
\multirow{3}{*}{$x^n$} &
$0$ & $0.09453$ & $\phantom{-}0.00288$ & $0.09735\; (51)$ & $0.09116$ & $\phantom{-}0.00142$  & $0.09340\; (114)$
\\
& 1 & $0.04220$ & $\phantom{-}0.00228$  & $0.04447\; (34)$ & $0.04334$  & $\phantom{-}0.00143$ & $0.04529\; (100)$
\\
& 4 & $0.01081$ & $-0.00032$ & $0.01049\; (20)$ & $0.01762$  & $\phantom{-}0.00133$ & $0.01896\; \phantom{1}(77)$
\\ \hline
\multirow{3}{*}{$1-x^n$} &
1 & $0.05233$ & $\phantom{-}0.00061$ & $0.05288\; (24)$ & $0.04782$ & $-0.00001$  & $0.04810\; \phantom{1}(24)$
\\
& 2 & $0.07223$ & $\phantom{-}0.00139$  & $0.07356\; (36)$ & $0.06320$  & $\phantom{-}0.00000$ & $0.06342\; \phantom{1}(36)$
\\
& 4 & $0.08372$ & $\phantom{-}0.00320$ & $0.08686\; (52)$ & $0.07409$  & $\phantom{-}0.00008$ & $0.07444\; \phantom{1}(51)$
\\ \hline
\end{tabular}}
\caption{Separate values of the OPE and DV contributions to $A^{\omega_n}_V(s_0)$, together with the fitted experimental moments, for the relevant weights $\omega_n (x) = x^n$ and $1-x^n$, using the default ansatz. The splitting is shown at the scales $\hat s_0= 1.55\;\mathrm{GeV}^2$ and  $s_0= 2.8\;\mathrm{GeV}^2$.}
\label{tab:sep_contrib0}
\end{table}

Instead of fitting the default ansatz with the assumed default value of $\hat{s}_0$, we can change the latter, since its default choice is a priori not justified. This exercise, already done in Ref.~\cite{Pich:2016bdg} for the vector channel, is displayed both for the vector and axial distributions in Fig.~\ref{fig:s0scan}. 
%
%
The left panels show, as a function of $\hat s_0$, the value of $\alpha_s(m_\tau^2)$ extracted from a fit to all $s_0$ bins with $s_0\ge \hat s_0$, in the vector (top) and axial (bottom) channels. The right panels display the associated p-values of the different fits, which indicate a rather low statistical quality and a strong $\hat{s}_0$ dependence, specially in the vector case where the largest p-value is around 5\% only.\footnote{Better p-values and a (model-dependent) central value for $\alpha_s$ halfway between \cite{Pich:2016bdg} and \cite{Boito:2014sta} (within a given perturbative prescription) have been recently obtained with an ``improved vector-isovector spectral function'', which incorporates experimental information from $e^+e^-$ annihilation \cite{Boito:2020xli}. A public release of the associated correlation matrix could be of interest for the community.}
The default choice $\hat s_0= 1.55\;\mathrm{GeV}^2$ corresponds to the lowest value of $\alpha_s$. This ad-hoc choice was adopted in Ref.~\cite{Boito:2014sta} with the argument that it has the largest p-value, but it is difficult to justify from the behaviour observed in the figure.
Applying the same somewhat arbitrary criteria in the axial channel, this is, finding a local maximum in the $p$-value (which is larger in the axial channel) with not-too-large uncertainties,
one obtains:
\begin{align}\label{eq:DefVsol}
\hat{s}_0^{V}=1.55\, \mathrm{GeV}^{2}, \qquad\qquad \alpha_{s}^{V}(m_{\tau}^2)_{\mathrm{FOPT}}&=0.298 \pm 0.010 \, ,\\ \label{eq:DefAsol}
\hat{s}_0^{A}=1.30\, \mathrm{GeV}^{2},\qquad\qquad \alpha_{s}^{A}(m_{\tau}^2)_{\mathrm{FOPT}}&=0.332 \pm 0.011 \, .
\end{align}
One may argue that it makes more sense to pick the solution at $\hat s_0= 1.55\;\mathrm{GeV}^2$, away from the $\rho$ and $a_1$ peaks, as more likely, but looking at the fit results, we do not really have a very strong justification to prefer the former solution over the latter, as assumed in Ref.~\cite{Boito:2014sta} without noting the possible axial solution.

\subsection{Sensitivity to the assumed ansatz}
\label{subsec:DVansatz}

In Ref.~\cite{Pich:2016bdg} we already showed that there is a strong dependence of the fitted results with the assumed functional form of the ansatz. Inserting in Eq.~(\ref{eq:DVparam}) a multiplicative factor $\mathcal{G}_\cJ(s) = s^{\lambda_\cJ}$ ($\mathrm{GeV}^2$ units) and repeating the fit to the vector distribution with the same $\hat s_0= 1.55\;\mathrm{GeV}^2$ and different values of the power $\lambda_V$ in the interval $\lambda_V\in [0,8]$, we observed a very significant correlation between the input value of $\lambda_V$ and the fitted result for $\alpha_s$. Moreover, the outcomes from these fits show a pattern that can be summarized through the following properties:
\begin{enumerate}
\item The fit quality, as measured by the p-value, increases with the power $\lambda_V$.
\item The fitted value of $\alpha_s$ increases when the fit quality ($\lambda_V$) increases, approaching the
results in Table~\ref{tab:AlphaTauSummary}.
\item All models reproduce well $\rho_V(s)$ in the fitted region. However, the default value $\lambda_V=0$ implies a spectral function that strongly deviates from data at $s < \hat s_0$. As $\lambda_V$ increases, the ansatz slightly approaches the data below the fitted region.
\item The size of the DV correction $\Delta A^{\omega_0}_V(\hat s_0)$ decreases as $\lambda_V$ increases.
\item The fitted values of the vacuum condensates decrease in a very significant way when the fit quality ($\lambda_V$) increases.
\end{enumerate}
For completeness, we compile the details of this analysis in appendix~\ref{app:lambdaV_Fit}. 

The chosen functional form
$\mathcal{G}_\cJ(s) = s^{\lambda_\cJ}$ is of course completely ad-hoc, as it was the original default choice $\mathcal{G}_\cJ(s) = 1$, but it demonstrates that the fitted results strongly depend on the assumed spectral-function model and, therefore, are unreliable.
The orthodox DV practitioners could still argue \cite{Boito:2016oam} that the power dependence $s^{\lambda_\cJ}$ does not seem to comply with their expectations for the asymptotic behaviour of the spectral function at very large values of the hadronic invariant mass \cite{Boito:2017cnp}. However, the fitted region of $s_0$ values is not really asymptotic. One could use instead a functional form
$\mathcal{G}_\cJ(s) = 1 - a_\cJ/s$, which incorporates the expected leading inverse-power correction at $s\to\infty$, with very similar results. For any set of fitted parameters 
$\{\lambda_V,\alpha_s,\delta_V, \gamma_V, \alpha_V,\beta_V\}$ one can easily find an alternative set $\{a_V,\alpha_s,\delta_V, \gamma_V, \alpha_V,\beta_V\}$ that provides an equally good fit to the spectral function in the fitted region and exhibits the same strong correlation between the fitted value of the strong coupling and the remaining ad-hoc parameters.

A quick scan of possible ansatz variations reveals many possible solutions with very different behaviours. Let us just pick the following four illustrative examples (in $\mathrm{GeV}^2$ units): 
\begin{enumerate}
\item $\mathcal{G}_{V}(s)=s^8 ,\qquad \hat{s}_{0}=1.55$.
\item $\displaystyle\mathcal{G}_{V}(s)=1- \frac{1.35}{s}, \qquad  \hat{s}_{0}=1.55$.
\item $\displaystyle\mathcal{G}_{V}(s)=1-\frac{2}{s} , \qquad  \hat{s}_{0}=1.55$.
\item $\mathcal{G}_{V}(s)=1 ,\qquad \hat{s}_{0}=2 , \qquad \alpha_{s}=0.320$.
\end{enumerate}
The first one gives the fit with the highest p-value among the $s^{\lambda_V}$ models analyzed in Ref.~\cite{Pich:2016bdg}. The second and third examples correspond to ansatz modifications giving, respectively, particularly larger and smaller values of the strong coupling, while having a functional form that satisfies the asymptotic behaviour assumed in  Ref.~\cite{Boito:2017cnp}.
The last example exhibits another possible solution, fully allowed by data, with the default choice $\mathcal{G}_{V}(s)=1$ but selecting a higher value for $\hat{s}_0$. While at larger $\hat{s}_{0}$ one may not be able to give a unique precise solution for the other ansatz parameters, this fact does not make them any less likely, as shown by giving the corresponding $p$-value. Let us note that if one assumed the default $V$ solution in Eq.~(\ref{eq:DefVsol}) to be the correct one, this would actually be the case for the axial channel, where one would need to rule out the apparent local solution at $\hat{s}_{0}^{A}=1.30\, \mathrm{GeV}^2$ and argue that the physical one must be at larger $\hat{s}_{0}^{A}$, where data does not shed much light about the corresponding value of $\alpha_s$. 

\begin{figure}[bht]
\begin{center}
\includegraphics[width=0.6\textwidth]{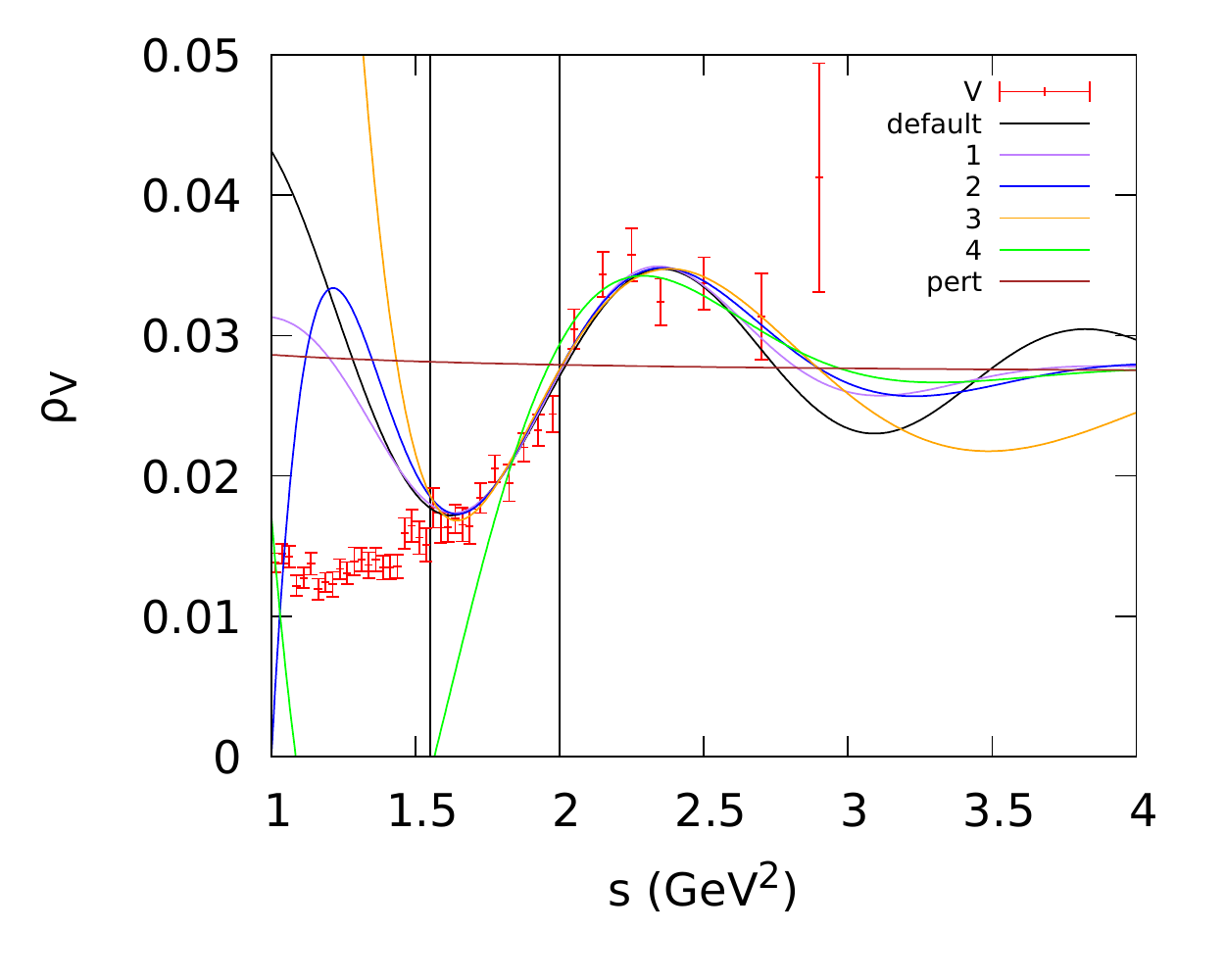}
\vskip -.4cm
\caption{\label{fig:DVmodels1} Comparison of the measured $\rho_V(s)$ (red data points) with the hypothetical vector spectral functions, obtained with the different model variations.}
 \end{center}
\end{figure}

Let us start by giving in Table~\ref{tab:ansatzesparams} the central values of the fitted DV parameters for the four different ansatz modifications, together with the reference values found before with the default choice. The four selected examples result in higher p-values than the default fit.
These model variations imply changes of about $\pm 10\%$ in the fitted value of $\alpha_s$, with respect to the default set-up, which may serve as an approximate assessment of the uncertainty inherent to a specific model choice. 
\begin{table}[t]\centering
\begin{tabular}{|c|ccccc|c|}\hline &&&&&&\\[-12pt]
Variation  & $\alpha_{s}^{(n_f=3)}(m_{\tau}^{2})$ & $\delta_V$      & $\gamma_V$      & $\alpha_V$       & $\beta_V$
& p-value (\% )
\\ \hline
Default & $0.298$          & $3.6$ & $0.6$ & $-2.3$ & $4.3$
& 5.3
\\ 
$1$ & $0.314$          & $1.0$ & $4.6$ & $-1.5$ & $3.9$ & $7.7$
\\ 
$2$ & $0.319$          & $-0.19$ & $1.8$ & $-0.8$ & $3.5$ & $7.8$
\\ 
$3$ & $0.260$          & $0.23$ & $1.2$ & $3.2$ & $2.1$ & $6.4$
\\ 
$4$ &  $0.320$          & $0.56$ & $1.9$ & $0.15$ & $3.1$ & $6.9$ \\ \hline
\end{tabular}
\caption{Fitted values of the spectral function ansatz parameters and $\alpha_s^{(n_f=3)}(m_\tau^2)$, in FOPT, for the four different modifications of the ansatz, compared with the default choice in Eq.~(\ref{eq:default}). 
\label{tab:ansatzesparams}}
\end{table}
%
The corresponding predictions for the vector spectral function
are compared with the experimental data in Fig.~\ref{fig:DVmodels1}, which exhibits their completely different behaviour outside the fitted region. This explains the sizeable splitting of their associated $\alpha_s$ determinations.
One can also observe that below the assumed $\hat{s}_0$ points the convergence of the data to the DV models is actually worse than the convergence of the data to the OPE itself at the alternative reference point $s_0 = m_{\tau}^2$.

The corresponding vacuum condensates, calculated in the same way as before,\footnote{For the sake of simplicity, we choose the default set-up for the corresponding axial channel and make the same approximation as before, that is, the value of $\alpha_s$ extracted from the vector channel is used in the fit to the axial data. }
are given in Tables~\ref{tab:condensatesv} and~\ref{tab:condensatesa}, for the vector and axial channels, respectively. 
From the tables, we observe how minor modifications in the ansatz can change the condensate values by several orders of magnitude. This instability is very easy to understand because, once $\alpha_s$ and the ansatz parameters get fixed, the vacuum condensates are enforced to re-absorb all the perturbative (through a modified $\alpha_s$) and DV deformations introduced by the different models, in order to approach the data. 

\begin{table}[tbh]\centering
{\begin{tabular}{|c|ccccccc|c|}\hline &&&&&&&\\[-12pt] Variation& $\mathcal{O}_{4,V}$ & $\mathcal{O}_{6,V}$ & $\mathcal{O}_{8,V}$   & $\mathcal{O}_{10,V}$   & $\mathcal{O}_{12,V}$ & $\mathcal{O}_{14,V}$   & $\mathcal{O}_{16,V}$ & $\alpha_s$\\
\hline
Default & $\phantom{-}0.0016$ & $-0.0082$ & $\phantom{-}0.014$ & $-0.019$ &  $\phantom{-}0.023$ &$-0.028$& $\phantom{-}0.037$ &$0.298$\\ \hline
1 & $-0.0003$ & $-0.0037$ & $\phantom{-}0.004$ & $\phantom{-}0.001$ &  $-0.010$ & $\phantom{-}0.006$& $\phantom{-}0.079$ & $0.314$ \\ \hline
2 & $-0.0009$ & $-0.0023$ & $\phantom{-}0.001$ & $\phantom{-}0.008$ &  $-0.026$ &$\phantom{-}0.046$& $-0.032$& $0.319$ \\ \hline
3 & $\phantom{-}0.0079$ & $-0.0318$ & $\phantom{-}0.091$ & $-0.25\phantom{0}$ &  $\phantom{-}0.59\phantom{0}$ &$-1.04\phantom{0}$& $\phantom{-}0.24\phantom{0}$ & $0.260$ \\ \hline
4 & $-0.0009$ & $-0.0012$ & $-0.003$ & $\phantom{-}0.021$ &  $-0.06\phantom{0}$ &$\phantom{-}0.13\phantom{0}$& $-0.20\phantom{0}$& $0.320$ \\ \hline
\end{tabular}}
\caption{Fitted values of the OPE vector condensates, in GeV units, obtained with the different modifications of the ansatz and FOPT.} 
\label{tab:condensatesv}
\end{table}


\begin{table}[tbh]\centering
{\begin{tabular}{|c|ccccccc|c|}\hline &&&&&&&\\[-12pt] Variation& $\mathcal{O}_{4,A}$ & $\mathcal{O}_{6,A}$ & $\mathcal{O}_{8,A}$   & $\mathcal{O}_{10,A}$   & $\mathcal{O}_{12,A}$ & $\mathcal{O}_{14,A}$   & $\mathcal{O}_{16,A}$ & $\alpha_s$\\
\hline
Default & $\phantom{-}0.0006$ & $-0.0016$ & $\phantom{-}0.016\phantom{0}$ & $-0.052$ &  $\phantom{-}0.11\phantom{0}$ &$-0.11\phantom{0}$& $-0.28\phantom{0}$ &$0.298$\\ \hline
1 & $-0.0015$ & $\phantom{-}0.0039$ & $\phantom{-}0.0024$ & $-0.023$ &  $\phantom{-}0.060$ &$-0.084$& $-0.037$ & $0.314$ \\ \hline
2 & $-0.0021$ & $\phantom{-}0.0055$ & $-0.0012$ & $-0.015$ &  $\phantom{-}0.047$ &$-0.074$& $-0.000$ & $0.319$ \\ \hline
3 & $\phantom{-}0.0054$ &  $-0.017\phantom{0}$ & $\phantom{-}0.058\phantom{0}$ & $-0.16\phantom{0}$ &  $\phantom{-}0.31\phantom{0}$ &$-0.16\phantom{0}$& $-2.15\phantom{0}$ & $0.260$ \\ \hline
4 & $-0.0014$ & $\phantom{-}0.0048$ & $-0.0005$ & $-0.016$ &  $\phantom{-}0.048$ &$-0.08\phantom{0}$& $\phantom{-}0.028$& $0.320$ \\ \hline
\end{tabular}}
\caption{Fitted values of the OPE axial condensates, in GeV units, obtained with the different modifications of the ansatz and FOPT.}\label{tab:condensatesa}
\end{table}


Tables \ref{tab:sep_contrib} and \ref{tab:DVpinched} show the separate OPE and DV contributions to the moments with weight functions $\omega_N(x) = x^N$ and $\hat\omega_N(x) = 1-x^N$, respectively, for the four different ansatz variations and the default set-up, in the vector channel. The splitting of the two contributions is given both at $\hat s_0$ and at $s_0 = 2.8~\mathrm{GeV}^2$. 
At $\hat s_0$, the relative size of the DV corrections strongly depends on the assumed ansatz parametrization, specially for the unprotected $x^N$ weights, which gets translated into the large variations observed before in the fitted values of the strong coupling and vacuum condensates. However, in all models, one observes again that pinched weights indeed suppress DVs at $s_0\sim m_{\tau}^2$, and that one pinch ({\it i.e.} one single zero at $s_0$) is enough to put them below the experimental uncertainties in most cases.

\begin{table}[tbh]\centering
\centerline{\begin{tabular}{|c|c|ccc|ccc|}\hline 
& & \multicolumn{6}{c|}{$A^{\omega_N}_V(s_0)$}
\\ \cline{3-8}
 && \multicolumn{3}{c|}{$\hat s_0$} &
\multicolumn{3}{c|}{$s_0 = 2.8\;\mathrm{GeV}^2$}
\\ \cline{3-8}
Ansatz & $N$  & OPE & DV & Exp  & OPE & DV & Exp
\\ \hline
\multirow{3}{*}{Default} &
0 & $\phantom{-}0.09453$ & $\phantom{-}0.00288$ & $\phantom{-}0.09735\; (51)$ & $\phantom{-}0.09116$ & $\phantom{-}0.00142$  & $\phantom{-}0.09340\; (114)$
\\
& 1 & $\phantom{-}0.04220$ & $\phantom{-}0.00228$  & $\phantom{-}0.04447\; (34)$ & $\phantom{-}0.04334$  & $\phantom{-}0.00143$ & $\phantom{-}0.04529\; (100)$
\\
& 4 & $\phantom{-}0.01081$ & $-0.00032$ & $\phantom{-}0.01049\; (20)$ & $\phantom{-}0.01762$  & $\phantom{-}0.00133$ & $\phantom{-}0.01896\; \phantom{1}(77)$
\\ \hline
\multirow{3}{*}{1} &
0 & $\phantom{-}0.09606$ & $\phantom{-}0.00132$ & $\phantom{-}0.09735\; (51)$ & $\phantom{-}0.09204$  & $\phantom{-}0.00088$ & $\phantom{-}0.09340\; (114)$
\\
& 1 & $\phantom{-}0.04466$ & $-0.00019$ & $\phantom{-}0.04447\; (34)$ &  $\phantom{-}0.04424$ & $\phantom{-}0.00096$ & $\phantom{-}0.04529\; (100)$
\\
& 4 & $\phantom{-}0.01776$ & $-0.00726$ & $\phantom{-}0.01049\; (20)$ & $\phantom{-}0.01746$  & $\phantom{-}0.00119$ & $\phantom{-}0.01896\; \phantom{1}(77)$
\\ \hline
\multirow{3}{*}{2} &
0 & $\phantom{-}0.09655$ & $\phantom{-}0.00082$ & $\phantom{-}0.09735\; (51)$ & $\phantom{-}0.09232$  & $\phantom{-}0.00088$ & $\phantom{-}0.09340\; (114)$
\\
& 1 & $\phantom{-}0.04542$ & $-0.00095$ & $\phantom{-}0.04447\; (34)$ &  $\phantom{-}0.04552$ & $\phantom{-}0.00098$ & $\phantom{-}0.04529\; (100)$
\\
& 4 & $\phantom{-}0.02020$ & $-0.00970$ & $\phantom{-}0.01049\; (20)$ & $\phantom{-}0.01760$  & $\phantom{-}0.00128$ & $\phantom{-}0.01896\; \phantom{1}(77)$
\\ \hline
\multirow{3}{*}{3} &
0 & $\phantom{-}0.09124$ & $\phantom{-}0.00611$ & $\phantom{-}0.09735\; (51)$ & $\phantom{-}0.08911$  & $\phantom{-}0.00429$ & $\phantom{-}0.09340\; (114)$
\\
& 1 & $\phantom{-}0.03366$ & $\phantom{-}0.01081$ & $\phantom{-}0.04447\; (34)$ &  $\phantom{-}0.04038$ & $\phantom{-}0.00500$ & $\phantom{-}0.04529\; (100)$
\\
& 4 & $-0.06903$ & $\phantom{-}0.07952$ & $\phantom{-}0.01049\; (20)$ & $\phantom{-}0.01280$  & $\phantom{-}0.00620$ & $\phantom{-}0.01896\; \phantom{1}(77)$
\\ \hline
\multirow{3}{*}{4} &
0 & $\phantom{-}0.09454$ & $-0.00511$ & $\phantom{-}0.08944\; (52)$ & $\phantom{-}0.09241$ & $\phantom{-}0.00042$  & $\phantom{-}0.09340\; (114)$
\\ &
1 & $\phantom{-}0.04500$ & $-0.00575$ & $\phantom{-}0.03926\; (40)$ & $\phantom{-}0.04452$ & $\phantom{-}0.00051$  & $\phantom{-}0.04529\; (100)$
\\
& 4 & $\phantom{-}0.01949$ & $-0.00725$ & $\phantom{-}0.01224\; (27)$ & $\phantom{-}0.01783$  & $\phantom{-}0.00081$ & $\phantom{-}0.01896\; \phantom{1}(77)$
\\ \hline
\end{tabular}}
\caption{Separate values of the OPE and DV contributions to the moments of the vector distribution $A^{\omega_N}_V(s_0)$, obtained with the different ansatz variations, together with their experimental values, for the relevant weights $\omega_N (x) = x^N$.}
\label{tab:sep_contrib}
\end{table}

\begin{table}[tbh]\centering
\centerline{\begin{tabular}{|c|c|ccc|ccc|}\hline 
& & \multicolumn{6}{c|}{$A^{\hat\omega_N}_V(s_0)$}
\\ \cline{3-8}
 && \multicolumn{3}{c|}{$\hat s_0$} &
\multicolumn{3}{c|}{$\hat s_0 = 2.8\;\mathrm{GeV}^2$}
\\ \cline{3-8}
Ansatz & $N$  & OPE & DV & Exp  & OPE & DV & Exp
\\ \hline
\multirow{3}{*}{Default} &
1 & $\phantom{-}0.05233$ & $\phantom{-}0.00061$ & $\phantom{-}0.05288\; (24)$ & $\phantom{-}0.04782$ & $-0.00001$  & $\phantom{-}0.04810\; (24)$
\\
& 2 & $\phantom{-}0.07223$ & $\phantom{-}0.00139$  & $\phantom{-}0.07356\; (36)$ & $\phantom{-}0.06320$  & $\phantom{-}0.00000$ & $\phantom{-}0.06342\; (36)$
\\
& 4 & $\phantom{-}0.08372$ & $\phantom{-}0.00320$ & $\phantom{-}0.08686\; (52)$ & $\phantom{-}0.07409$  & $\phantom{-}0.00008$ & $\phantom{-}0.07444\; (51)$
\\ \hline
\multirow{3}{*}{1} &
1 & $\phantom{-}0.05140$ & $\phantom{-}0.00151$ & $\phantom{-}0.05288\; (24)$ &  $\phantom{-}0.04781$ & $-0.00008$ & $\phantom{-}0.04810\; (24)$
\\
& 2 & $\phantom{-}0.07002$ & $\phantom{-}0.00358$ & $\phantom{-}0.07356\; (36)$ & $\phantom{-}0.06338$  & $-0.00016$ & $\phantom{-}0.06342\; (36)$
\\
& 4 & $\phantom{-}0.07830$ & $\phantom{-}0.00859$ & $\phantom{-}0.08686\; (52)$ & $\phantom{-}0.07458$  & $-0.00031$ & $\phantom{-}0.07444\; (51)$
\\ \hline
\multirow{3}{*}{2} &
1 & $\phantom{-}0.05113$ & $\phantom{-}0.00177$ & $\phantom{-}0.05288\; (24)$ &  $\phantom{-}0.04780$ & $-0.00011$ & $\phantom{-}0.04810\; (24)$
\\
& 2 & $\phantom{-}0.06937$ & $\phantom{-}0.00421$ & $\phantom{-}0.07356\; (36)$ & $\phantom{-}0.06344$  & $-0.00021$ & $\phantom{-}0.06342\; (36)$
\\
& 4 & $\phantom{-}0.07635$ & $\phantom{-}0.01053$ & $\phantom{-}0.08686\; (52)$ & $\phantom{-}0.07472$  & $-0.00040$ & $\phantom{-}0.07444\; (51)$
\\ \hline
\multirow{3}{*}{3} &
1 & $\phantom{-}0.05758$ & $-0.00470$ & $\phantom{-}0.05288\; (24)$ &  $\phantom{-}0.04873$ & $-0.00071$ & $\phantom{-}0.04810\; (24)$
\\
& 2 & $\phantom{-}0.08889$ & $\phantom{-}0.00358$ & $\phantom{-}0.07356\; (36)$ & $\phantom{-}0.06475$  & $-0.00137$ & $\phantom{-}0.06342\; (36)$
\\
& 4 & $\phantom{-}0.160268$ & $-0.073411$ & $\phantom{-}0.08686\; (52)$ & $\phantom{-}0.07631$  & $-0.00191$ & $\phantom{-}0.07444\; (51)$
\\ \hline
\multirow{3}{*}{4} &
1 & $\phantom{-}0.04953$ & $\phantom{-}0.00064$ & $\phantom{-}0.05019\; (24)$ & $\phantom{-}0.04789$  & $-0.00009$ & $\phantom{-}0.04810\; (24)$
\\
& 2 & $\phantom{-}0.06579$ & $\phantom{-}0.00126$ & $\phantom{-}0.06707\; (32)$ & $\phantom{-}0.06337$ & $-0.00018$   & $\phantom{-}0.06342\; (36)$
\\
& 4 & $\phantom{-}0.07505$ & $\phantom{-}0.00214$ & $\phantom{-}0.07720\; (38)$ &  $\phantom{-}0.07457$ & $-0.00039$ & $\phantom{-}0.07444\; (51)$
\\ \hline
\end{tabular}}
\caption{Separate values of the OPE and DV contributions to the moments of the vector distribution $A^{\hat\omega_N}_V(s_0)$, obtained with the different ansatz variations, together with their experimental values, for the pinched weights $\hat\omega_N (x) = 1-x^N$.}
\label{tab:DVpinched}
\end{table}

\section{\boldmath Assessing the size of DV uncertainties in the $V+A$ channel}\label{sec:assessment}

Any determination of the strong coupling is affected by systematic uncertainties, originating in those effects which are not yet under full theoretical control, such as continuous extrapolation (in discretized computations), truncation of perturbation theory and/or the OPE, hadronization, duality violations, etc. They need to be estimated in a proper way, 
trying to avoid both naive underestimates and pessimistic overestimates.

Given a deviated strong coupling value as input, one can test how much one would need to inflate the initially assigned systematic errors to accommodate such a deviation. If a systematic uncertainty is well-estimated one should expect that the inflation leads to improbable scenarios, such as effective parameters acquiring values orders of magnitude off or crazy bumps in otherwise expected smooth functions. Otherwise the suggested inflation may be justified.

In our case, we can take the values of the strong coupling,
$\alpha^{\mathrm{FOPT}}_s(m_{\tau}^2)=(0.26-0.32)$,
emerging from the different (vector) DV scenarios discussed in the previous section, together with their corresponding modelling of the spectral function, and check whether those values which deviate from the ones given in Table~\ref{tab:AlphaTauSummary} lead  indeed to solutions that do not make much sense. 

\begin{figure}[t]
\begin{center}
\includegraphics[width=0.6\textwidth]{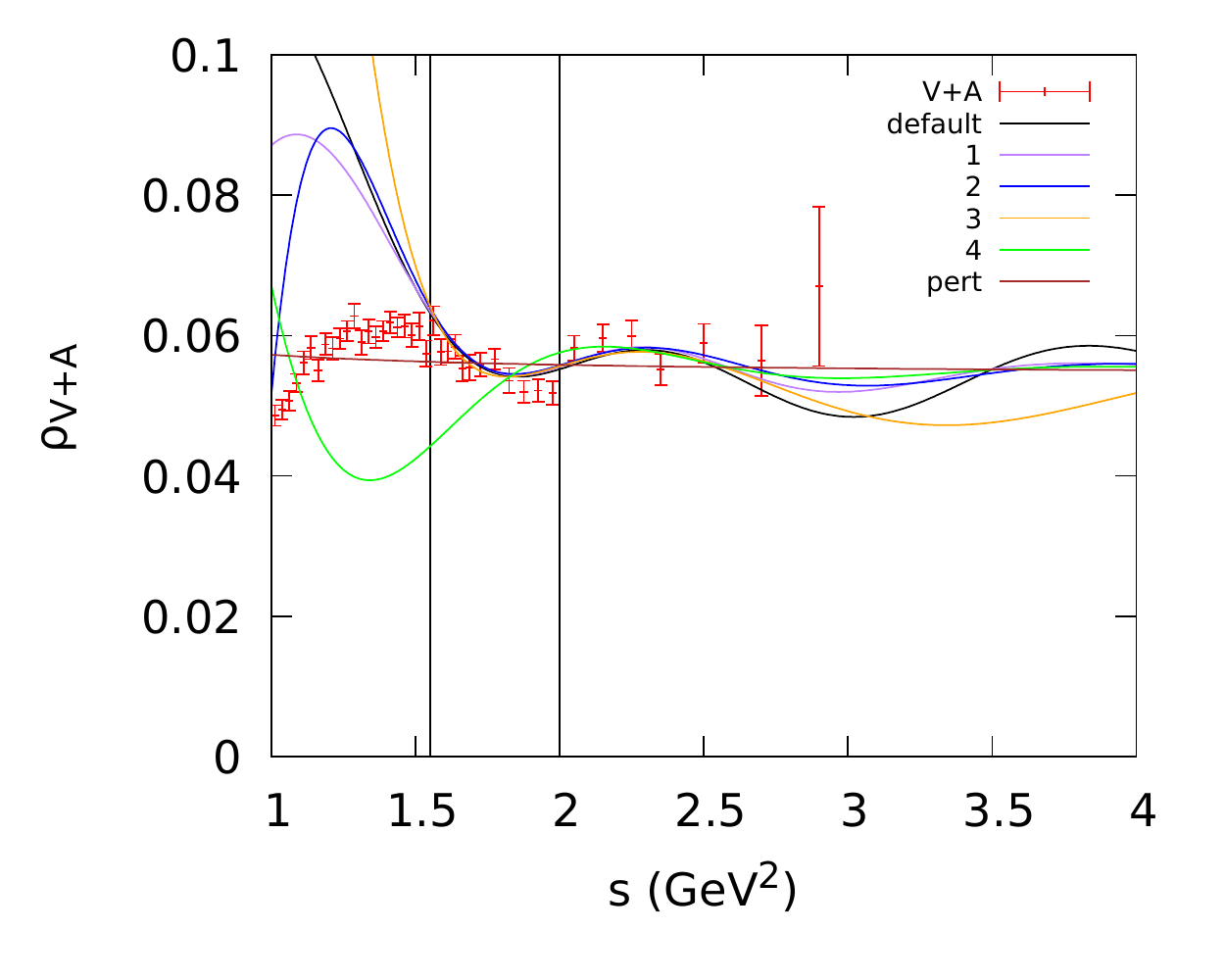}\vskip -.5cm
\caption{\label{fig:DVmodelsv+a}Hypothetical $V+A$ spectral functions obtained with the different model variations, compared with the experimental data.}
 \end{center}
\end{figure}

The $V+A$ spectral functions predicted by the different ansatz variations are compared with the data in Fig.~\ref{fig:DVmodelsv+a}. A discouraging feature for the use of all these models becomes evident. Their convergence to the data below the assumed point $\hat{s}_0$ is actually much worse than the convergence of the OPE itself around the reference value $m_{\tau}^2$ (and actually at any point within the plot region).
Since the lack of an exact convergence of the OPE to the data was the original motivation to introduce DV corrections, one may wonder whether the poor behaviour exhibited by the assumed ansatzs justifies at all this modelling of duality violations.
The same caveat can be observed with the extrapolation of the DV ansatzs at higher values of the hadronic invariant mass. In fact, both the default model and the variation $3$ imply a rather implausible shape, with local DVs above $m_{\tau}^2$ considerably larger than even the one corresponding to the peak of the first axial resonance $a_1(1260)$. 
Taking into account the behaviour of the experimental spectral function in practically all the measured energy range and the large number of hadronic channels already opened at this energy, the additional bumps/dips predicted by these two models look rather unlikely. Looking back into Table~\ref{tab:ansatzesparams}, we realize that these are precisely the two models leading to too low values of the strong coupling.

\begin{figure}[t]\centering
\includegraphics[width=0.45\textwidth]{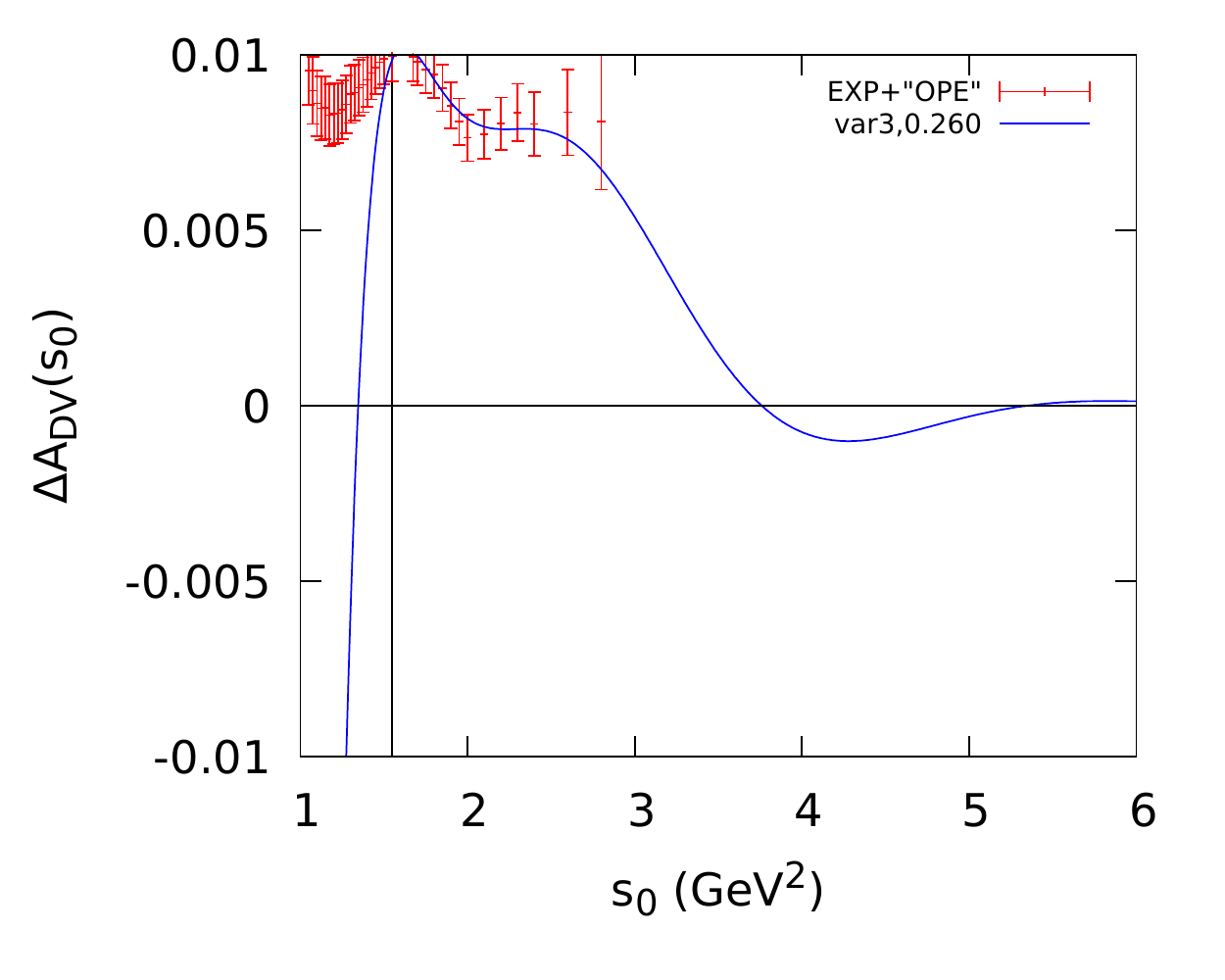}\hskip .25cm
\includegraphics[width=0.45\textwidth]{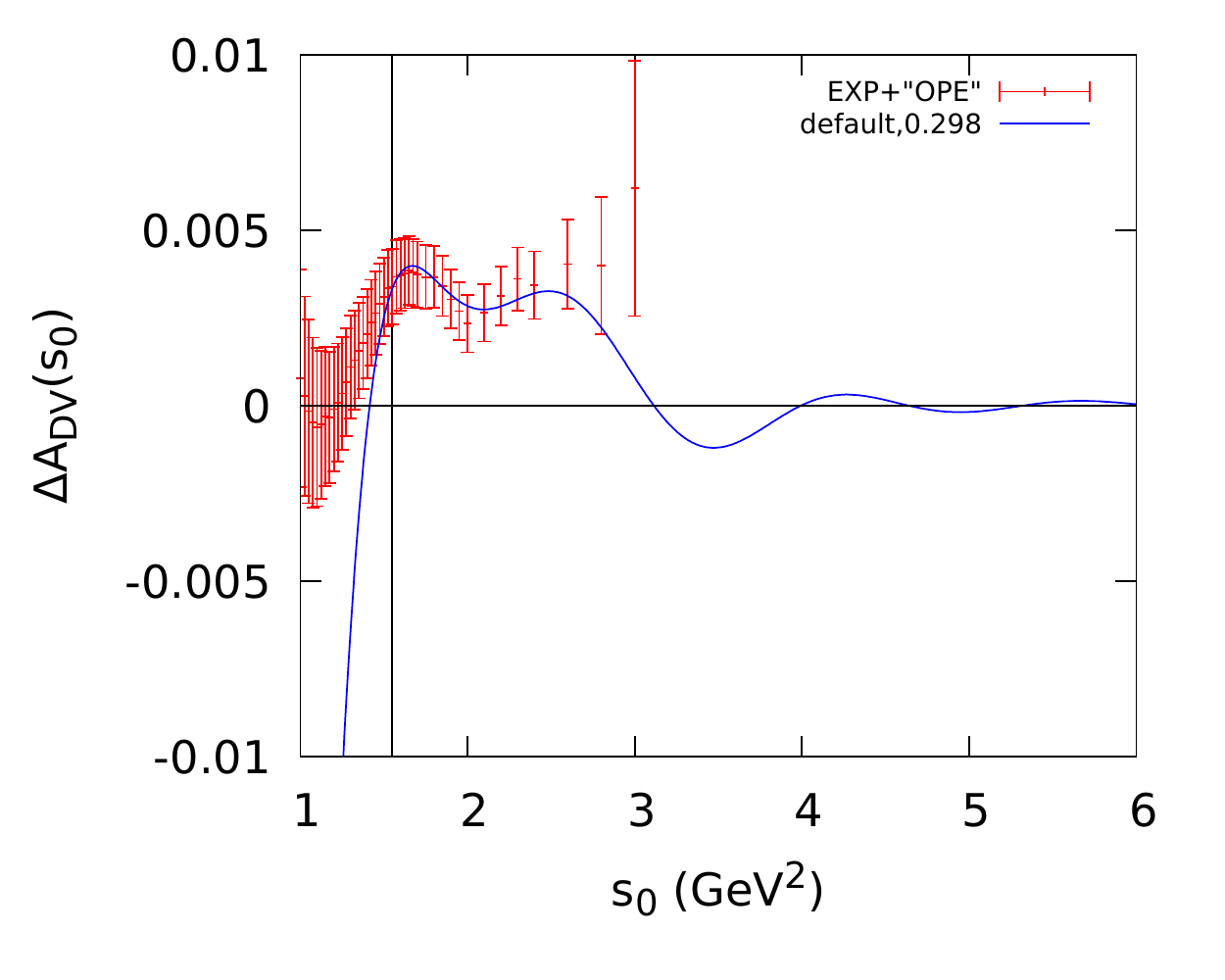}
\\[-5pt]
\includegraphics[width=0.45\textwidth]{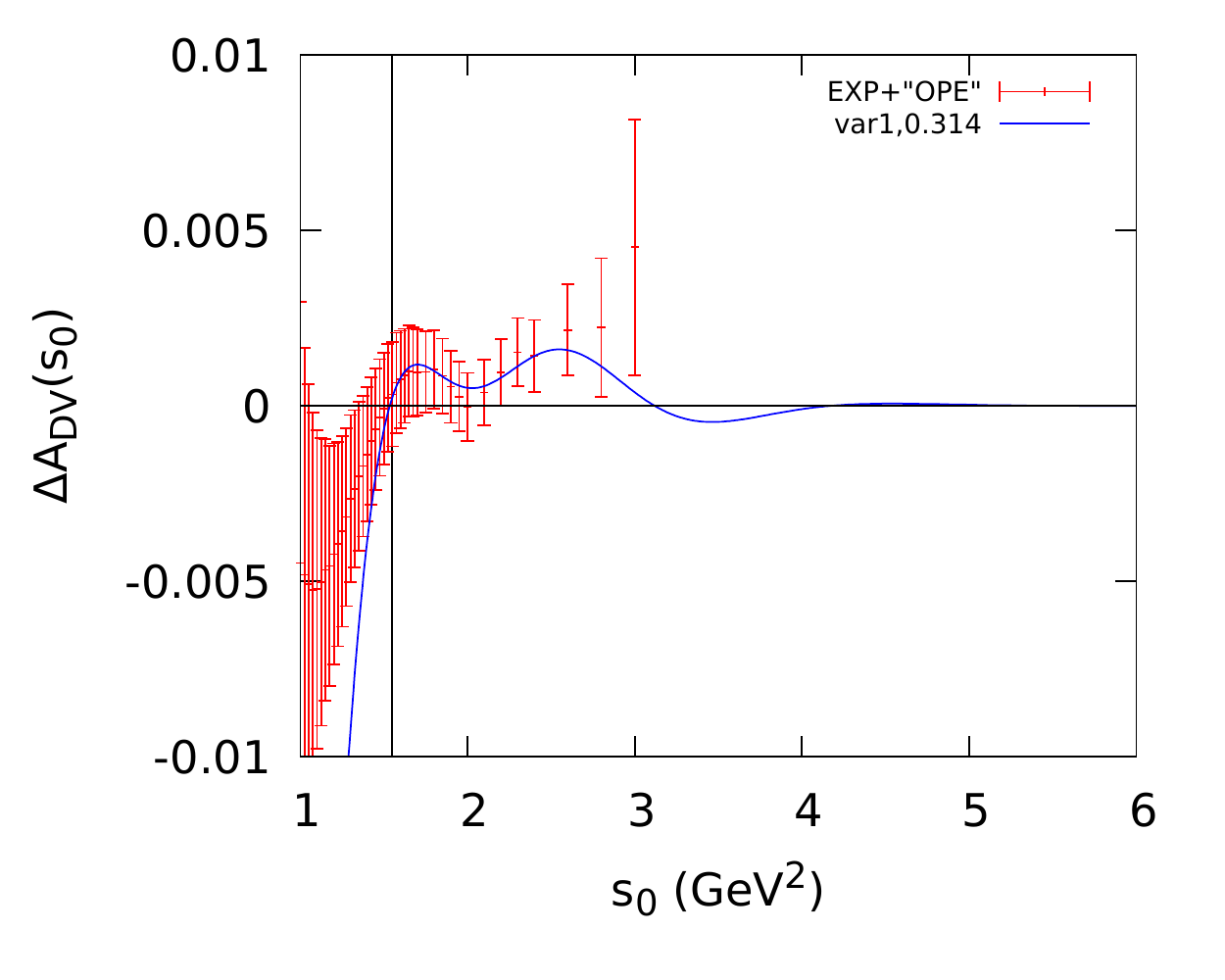}\hskip .25cm
\includegraphics[width=0.45\textwidth]{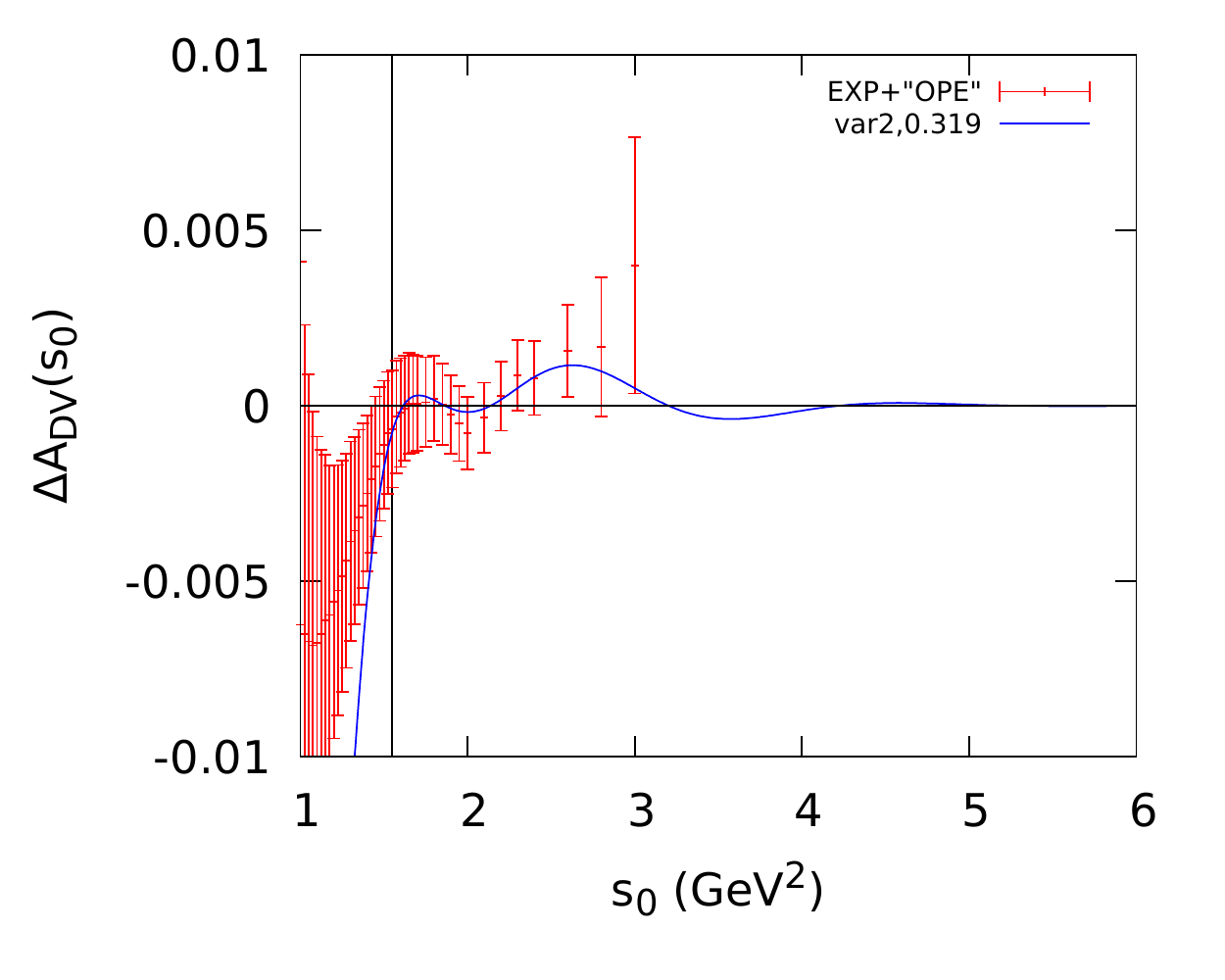} 
\\[-5pt]
\includegraphics[width=0.45\textwidth]{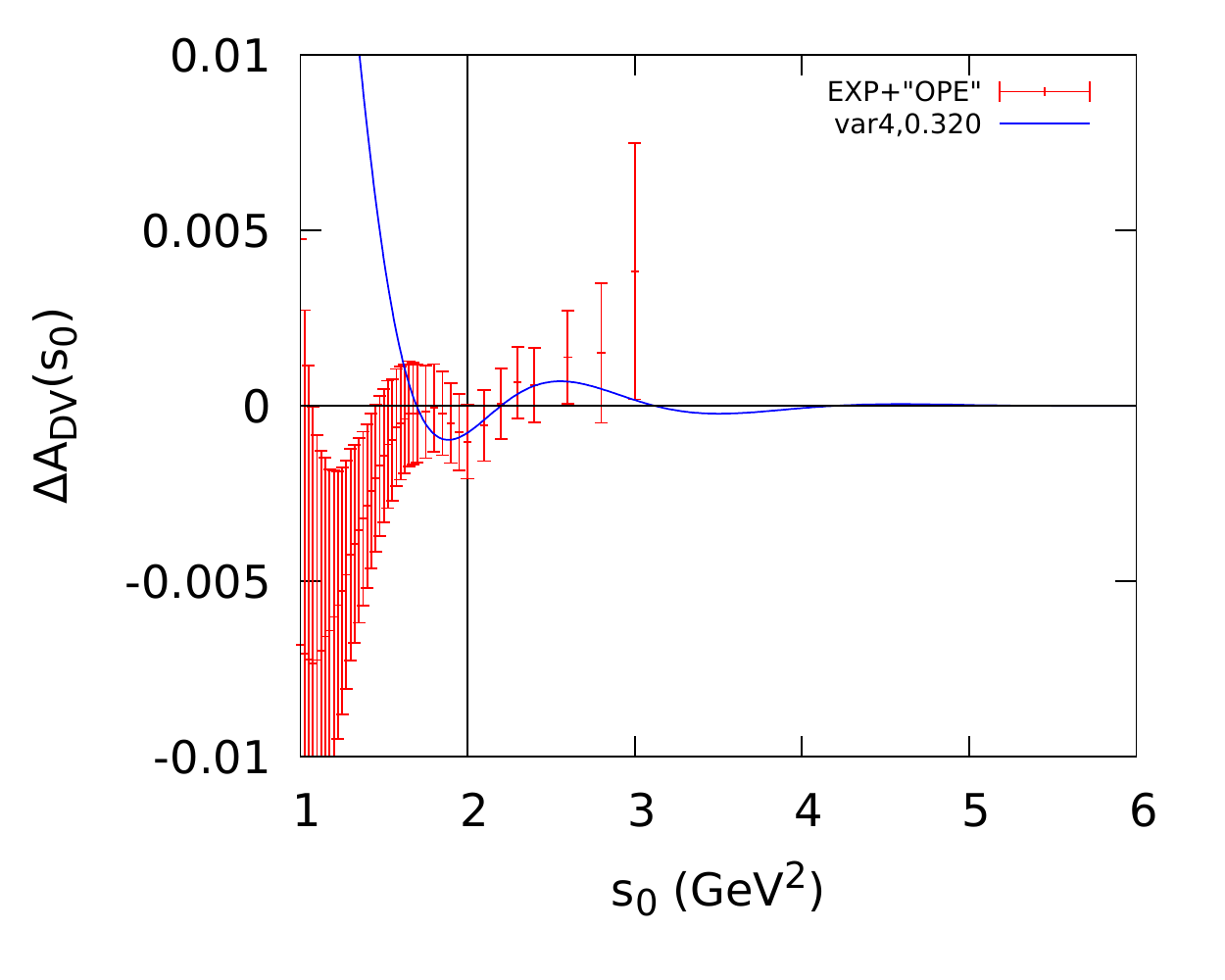}
\vskip -.4cm
\caption{\label{fig:DVmodelsplot}
Hypothetical values of $\Delta A_{V+A}^{\omega_0}(s_0)$,
obtained with the different ansatz variations, ordered from larger to smaller deviation from the reference value of $\alpha_s^{V+A}(m_\tau^2)_{\mathrm{FOPT}}$ in Table~\ref{tab:AlphaTauSummary}.}
\end{figure}

This unphysical behaviour becomes more evident when we display the corresponding values of $\Delta A_{V+A}^{\omega_0}(s_0)$. 
This is done in Fig.~\ref{fig:DVmodelsplot}, where the predicted DV contributions of the different models are compared with the corresponding ``experimental'' shapes of these quantities, {\it i.e.}, with $A_{V+A}^{\omega_0}(s_0)_{\mathrm{exp}}-A_{V+A}^{\omega_0}(s_0)_{\mathrm{OPE}}$, the OPE contribution being computed with the fitted value of $\alpha_s$ within the given model. By construction, for all models the two curves are in good agreement inside the fitted range $s_0\in [\hat s_0, m_\tau^2]$. However, a very different behaviour is observed outside this region. To better visualize the implied patterns, we have ordered the different panels attending to the corresponding deviation of the strong coupling from the results given in section~\ref{sec:alphasResults} (from larger to smaller deviation). The more $\alpha_s$ deviates from the quoted uncertainties in Table~\ref{tab:AlphaTauSummary}, the more absurd is the shape displayed by the function $\Delta A_{V+A}^{\omega_0}(s_0)$.
Obviously, the two Heaviside-like scenarios at the top of the figure (variation 3 and default ansatz) are very unlikely. They would imply a huge DV at $m_\tau^2$, not required by any experimental fact, that needs to fall down abruptly to zero in order to be consistent with asymptotic freedom. We find natural to leave them outside the quoted uncertainties without any need of guessing what is the exact shape of the spectral function, which is beyond theoretical control.

However, a Heaviside-like convergence of $\Delta A_{V+A}^{\omega_0}(s_0)$ would not be enough to take $\alpha_s$ outside our determination, since this kind of DV behaviour becomes very suppressed when using pinched weight functions. 
In this case, a deviated value of $\alpha_s$ can only be compensated with huge fine-tuned nonperturbative condensates, as shown in Table~\ref{tab:DVpinched11} for two representative double pinched moments with the five ansatz set-ups. Neglecting the $\mathcal{P}_\cJ$ corrections, these  
$A^{\omega^{(2,n)}}_{V+A}(s_0)$ moments\footnote{Let us note that $A^{\omega^{(2,1)}}_{V+A}(s_0)$ is the moment associated to the tau decay width, $R_{\tau}$.} only receive OPE
contributions from $\mathcal{O}_{2(n+2),V+A}$ and $\mathcal{O}_{2(n+3),V+A}$. Both the default ansatz and the variation 3 need to incorporate huge OPE corrections at $s_0=m_\tau^2$ in order to restore agreement with the data. As we argued above, this OPE scenario has no regime of validity or physical meaning and should be discarded. 
Indeed, for these two ansatzs the corresponding fine-tuning becomes totally unreliable if we go to the lower scale $\hat{s}_0$, where the OPE was assumed to be valid. This is demonstrated in Table~\ref{tab:DVpinched22}, which exhibits a completely crazy behaviour with individual OPE contributions much larger than the total perturbative correction. The OPE does not make any sense in these two scenarios.

\begin{table}[t]\centering
{\begin{tabular}{|c|c|ccccc|}\hline 
Weight & variation  & Pert & $\mathcal{O}_{2(n+2),V+A}$ & $\mathcal{O}_{2(n+3),V+A}$ & DV & Exp
\\ \hline
\multirow{5}{*}{$A^{\omega^{(2,1)}}_{V+A}(m_\tau^2)$} &
Default & $0.0938\; (5)$  &  $\phantom{-} 0.0029$  &  $ -0.0019$  &  $ -0.0001$  &  $ 0.0954\; (3)$ 
\\
& 1 & $0.0952\; (7)$  &  $ -0.0001$  &  $ -0.0004$  &  $ -0.0000$  &  $ 0.0954\; (3)$ 
\\
& 2 & $0.0957\; (8)$  &  $ -0.0010$  &  $\phantom{-} 0.0000$  &  $ -0.0000$  &  $ 0.0954\; (3)$ 
\\
& 3 & $0.0908\; (2)$  &  $\phantom{-} 0.0145$  &  $ -0.0095$  &  $ -0.0007$  &  $ 0.0954\; (3)$ 
\\
& 4 & $0.0958\; (8)$ & $ -0.0011$ & $ -0.0005$ & $-0.0000$  &  $ 0.0954\; (3)$ 
\\ \hline
\multirow{5}{*}{$A^{\omega^{(2,4)}}_{V+A}(m_\tau^2)$} &
Default & $0.1316\; (4)$  &  $\phantom{-} 0.0025$  &  $ -0.0007$  &  $ \phantom{-}0.0001$  &  $ 0.1344\; (8)$ 
\\
& 1 & $0.1331\; (5)$  &  $\phantom{-} 0.0009$  &  $ -0.0004$  &  $ \phantom{-}0.0001$  &  $ 0.1344\; (8)$
\\
& 2 & $0.1336\; (5)$  &  $\phantom{-} 0.0004$  &  $ -0.0001$  &  $ \phantom{-}0.0000$  &  $ 0.1344\; (8)$ 
\\
& 3 & $0.1282\; (2)$  &  $\phantom{-} 0.0171$  &  $ -0.0061$  &  $ -0.0056$  &  $ 0.1344\; (8)$ 
\\
& 4 & $0.1337\; (5)$  &  $  -0.0002$  &  $\phantom{-}  0.0002$  &  $  \phantom{-}0.0001$  &  $ 0.1344\; (8)$ 
\\ \hline
\end{tabular}}
\caption{Separate contributions to the moments $A^{\omega^{(2,n)}}_{V+A}(m_\tau^2)$ with $n=1,4$, together with their experimental values, at $s_0=m_{\tau}^2$.}
\label{tab:DVpinched11}
\end{table}

\begin{table}[tbh]\centering
{\begin{tabular}{|c|c|ccccc|}\hline 
Weight & variation  & Pert & $\mathcal{O}_{2(n+2),V+A}$ & $\mathcal{O}_{2(n+3),V+A}$ & DV & Exp
\\ \hline
\multirow{5}{*}{$A^{\omega^{(2,1)}}_{V+A}(\hat s_0)$} &
Default & $0.1010\; (18)$  &  $\phantom{-} 0.0248$  &  $ -0.0326$  &  $ \phantom{-}0.0062$  &  $ 0.0994\; (4)$ 
\\
& 1 & $0.1043\; (28)$  &  $ -0.0006$  &  $ -0.0071$  &  $\phantom{-} 0.0028$  &  $ 0.0994\; (4)$ 
\\
& 2 & $0.1054\; (32)$  &  $ -0.0081$  &  $\phantom{-} 0.0003$  &  $ \phantom{-}0.0018$  &  $ 0.0994\; (4)$ 
\\
& 3 & $0.0948\; (06)$  &  $\phantom{-} 0.1221$  &  $ -0.1629$  &  $ \phantom{-}0.0452$  &  $ 0.0994\; (4)$  
\\
& $4$ & $0.1010\; (18)$  &  $-0.0042$  &  $ \phantom{-}0.0015$  &  $-0.0001$  &  $0.0980\;(3)$ 
\\ \hline
\multirow{5}{*}{$A^{\omega^{(2,4)}}_{V+A}(\hat s_0)$} &
Default & $0.1391\; (10)$  &  $\phantom{-} 0.1808$  &  $ -0.1012$  &  $ -0.0787$  &  $ 0.1401\; (5)$ 
\\
& 1 & $0.1424\; (14)$  &  $\phantom{-} 0.0676$  &  $ -0.0572$  &  $ -0.0128$  &  $ 0.1401\; (5)$
\\
& 2 & $0.1434\; (16)$  &  $\phantom{-} 0.0281$  &  $ -0.0203$  &  $ -0.0112$  &  $ 0.1401\; (5)$ 
\\
& 3 & $0.1327\; (05)$  &  $\phantom{-} 1.2216$  &  $ -0.8833$  &  $ -0.3309$  &  $ 0.1401\; (5)$ 
\\
& $4$ & $0.1392\; (11)$  &  $-0.0036$  &  $\phantom{-}0.0058$  &  $ -0.0034$  &  $ 0.1378\; (4)$ 
\\ \hline
\end{tabular}}
\caption{Separate contributions to the moments $A^{\omega^{(2,n)}}_{V+A}(\hat s_0)$ with $n=1,4$, together with their experimental values, at $\hat s_0$.}
\label{tab:DVpinched22}
\end{table}

The other three DV ansatzs (variations 1, 2 and 4) do not exhibit any of these pathologies. They show an acceptable $s_0$ behaviour in Fig.~\ref{fig:DVmodelsplot}, falling down smoothly at large values of the hadronic invariant mass, as expected. Moreover, their corresponding OPE contributions in Tables~\ref{tab:DVpinched11} and \ref{tab:DVpinched22} have a reasonable size, consistent with the implicit assumption of negligible $\mathcal{P}_\cJ$ corrections near the $\tau$ mass. Not surprisingly, the values of $\alpha_s$ implied by these three scenarios are in excellent agreement with our more solid determinations with pinched weights presented in section~\ref{sec:alphasResults}.
In the three cases, $\alpha_s$ lies within our estimated $1\sigma$ interval, showing that systematic uncertainties were indeed correctly assessed in Ref.~\cite{Pich:2016bdg}.

\section{Summary}
\label{sec:Summary}

We have addressed in a quantitative way the role of violations of quark-hadron duality in low-energy determinations of the strong coupling. This type of effects are unavoidably present in any hadronic observable, preventing an exact (infinite accuracy) theoretical description. Assuming confinement, inclusive observables provide the best possible playground to make precision physics in QCD, using the powerful OPE techniques. However, even there, small DV corrections show up, owing to the different threshold behaviour (multi-hadron versus multi-parton) of the two dual descriptions of the QCD spectrum.

In the absence of a rigorous understanding of confinement, one usually tries to minimize the DV contributions in order to achieve the best possible phenomenological accuracy. This can be done by working at large-enough energies and/or by smearing the observable cross sections over a suitable energy range. This second approach is compulsory at low and intermediate energies, where precise QCD predictions can only be made for integrated moments of the spectral hadronic distributions.

The vector and axial-vector spectral functions extracted from the invariant-mass distribution of the final hadrons in $\tau$ decays have made possible to perform accurate determinations of the strong coupling with a N${}^3$LO accuracy.
We have reviewed the present status in sections \ref{sec:formalism} and \ref{sec:alphasResults}, where the reasons why a high sensitivity to $\alpha_s$ is obtained have been explained in detail. An important aspect of this nowadays classical determination is the strong suppression of DV contributions in spectral moments with pinched weights. This was actually one of the very first considerations made in the pioneering papers suggesting to extract $\alpha_s$ from the observable $R_\tau$ and related pinched moments \cite{Narison:1988ni,Braaten:1991qm,LeDiberder:1992zhd}.

In recent years, a different strategy has been suggested, advocating to model the oscillations observed in the experimental spectral functions with phenomenological ansatzs, and use them to quantify the DV corrections to non-protected (not pinched) moments. These ansatzs are elegantly motivated, but one should keep in mind that they correspond to particular hadronization models. An obvious question then arises, concerning how much the value of the strong coupling obtained with this procedure depends on the specific functional form assumed for the adopted ansatz. The clarification of this important question has been the main motivation of this work.

An exhaustive analysis of the DV-ansatz approach to $\alpha_s$ has been presented in section~\ref{sec:DVmethod}. We have anatomized the employed algorithm, in order to make its implicit assumptions as transparent as possible. This has allowed us to show that all experimental data in the interval $( \hat s_0, m_\tau^2 ]$ are actually used to fit the parameters modelling the spectral function, while the wanted QCD information is mostly obtained from moments computed at the lowest energy point $\hat s_0$. After subtracting the corresponding DV corrections, the strong coupling is finally extracted from $A^{\omega_0}_\cJ(\hat s_0)$ and the moments $A^{(n)}_\cJ(\hat s_0)$ provide the power corrections. This introduces two evident caveats: 1) the subtracted DV contributions are model dependent, since they have been computed with the particular ansatz that has been assumed, and 2) at $\hat s_0 \sim (1.2\;\mathrm{GeV})^2$ the perturbative uncertainties are quite large and the unknown power corrections are unavoidably enhanced. 

Using different functional forms for the hadronic ansatz, we have exhibited the very large sensitivity of the fitted value of $\alpha_s$ on the assumed parametrization. To simplify the discussion, we have focused on four particular model variations, all of them having a better statistical quality (p-value) than the default model originally adopted in Ref.~\cite{Boito:2014sta}.
As shown in Table~\ref{tab:ansatzesparams}, these model variations imply changes of up to $\pm 10\%$ in the fitted value of $\alpha_s$.\footnote{We have analyzed many other possible ansatzs, finding a larger spread of fitted results.}
Within any given model, the power corrections need to be adjusted to compensate a slightly incorrect value of $\alpha_s$ (plus the model-dependent DV contributions), in order to reproduce the corresponding experimental moment $A^{(n)}_\cJ(\hat s_0)$. Therefore, the spread of $\alpha_s$ values enforces a much larger spread of fitted OPE corrections,
shown in Tables~\ref{tab:condensatesv} and \ref{tab:condensatesa}, indicating a dangerous loss of theoretical control.
The strong correlation between the assumed functional form of the hadronic ansatz and the fitted values of $\alpha_s$ and the different condensates shows that the resulting parameters constitute at best an effective model description with unclear relation to QCD.

Figures~\ref{fig:DVmodels1} and \ref{fig:DVmodelsv+a} provide some enlightenment on what is actually happening with these fits. Although all analyzed models describe well the hadronic spectral function in the fitted region $( \hat s_0, m_\tau^2 ]$, they exhibit a quite different behaviour outside it. Below $\hat s_0$ the models deviate dramatically from data. Above $m_\tau^2$, those models giving too low values of $\alpha_s$ (and huge condensates) generate very large oscillations that seem unphysical. This is better appreciated in the $V+A$ distribution, given in Fig.~\ref{fig:DVmodelsv+a}, where one can observe the implausible shape of these bumps/dips with local oscillations above $m_\tau^2$ that are larger in amplitude than the $a_1(1260)$ resonance. As explicitly shown in Fig.~\ref{fig:DVmodelsplot}, these models generate a quite pathological $s_0$ dependence of the DV contribution $\Delta A^{\omega_0}_{V+A}(s_0)$, since this correction needs to be very large at $s_{0}\sim m_{\tau}^2$ to accommodate the associated $\alpha_{s}$ value and, at the same time, must fall down very abrutly to match the expected asymptotic behaviour. The DV ansatzs that do not exhibit these pathologies turn out to generate fitted condensates of more reasonable size and values of $\alpha_s$ in excellent agreement with the standard determination in Table~\ref{tab:AlphaTauSummary}.

The gained understanding on violations of quark-hadron duality has allowed us to go one step further and analyze the possible impact of this type of corrections in the standard determination of $\alpha_s(m_\tau^2)$ with pinched moments. Tables~\ref{tab:DVpinched} and \ref{tab:DVpinched11} compare the predicted size of the DV contributions to different pinched moments with the corresponding OPE contributions and with the experimental values, around the scale $m_\tau^2$. At this scale the estimated DV contributions turn out to be tiny in all cases, being much smaller than the OPE uncertainties, and always remaining below the experimental errors (except for the pathological variation 3). For the default ansatz assumed in Ref.~\cite{Boito:2014sta}, the obtained DV effects are completely negligible.

Taking all this into account, we conclude that systematic uncertainties have been correctly assessed in the standard determination of $\alpha_s(m_\tau^2)$ reviewed in section~\ref{sec:alphasResults}. 
The dominant errors originate in perturbation theory itself. Therefore, at present, the $\tau$ decay data imply the values of the strong coupling summarized in Table~\ref{tab:AlphaTauSummary} for the two alternative perturbative prescriptions:
\begin{equation}
    \alpha_s^{(n_f=3)}(m_\tau^2) \, =\, \left\{
    \begin{array}{cc}
     0.335\pm 0.013 \qquad   &  \mathrm{(CIPT)}\\
     0.320\pm 0.012  \qquad  &  \mathrm{(FOPT)} 
    \end{array}
    \right.\, .
\end{equation}
Combining these two values and adding quadratically half their difference as an additional systematic uncertainty, one finally gets 
\begin{equation}
\alpha_{s}^{(n_f=3)}(m_{\tau}^2)= 0.328\pm 0.013\, .
\end{equation}
After evolution up to the scale $M_Z$, this value decreases to
\begin{equation}
\alpha_{s}^{(n_f=5)}(M_Z^2)= 0.1197\pm 0.0015\, ,
\end{equation}
in excellent agreement with the direct N${}^3$LO determination at the $Z$ peak.

\acknowledgments
We are grateful to Bogdan Malaescu for discussions and useful comments on the manuscript.
This work has been supported by 
MCIN/AEI/10.13039/501100011033, Grant No. PID2020-114473GB-I00, by the Generalitat Valenciana, Grant No. Prometeo/2021/071,
and by the Agence Nationale de la Recherche (ANR) under grant ANR-19-CE31-0012 (project MORA).

\appendix

\section{Relations between observables and equivalent fits}\label{appendix}

In this appendix, we give trivial relations between observables, aiming to expose some redundancies in previous analyses with the aim of sorting the amount of meaningful information that one can extract from different fits.

\subsection{Some generic fit properties}
\label{subsec:FitProperties}

In a fit we typically have a set of $n$ experimental points $p_i$ ($i=1,\cdots n$) with an associated covariance matrix $V_{ij}$. For every $p_i$ we have a theoretical prediction $t_i(\theta_j)$
that depends on $m$ parameters $\theta_j$. Essentially, the result of a fit is supposed to give us the $\theta_j$ values that best match the predictions $t_i$ to the measurements $p_i$. 

The fit method should be invariant under linear transformations of the data points. For any invertible known matrix $A$, the fit should give the same $\theta_j$ independently on whether we fit $(p_i,t_i)$ or $(\tilde{p}_i,\tilde{t}_i)\equiv A_{ij}(p_j,t_j)$, whose associated covariance matrix is $\tilde{V}=A V A^{T}$. This is trivially realized with the $\chi^2(\theta_j)$ function,
\begin{equation}
\chi^2(\vec{\theta})=(\vec{p}-\vec{t}\, )^T V^{-1}(\vec{p}-\vec{t}\, ) = (\vec{\tilde{p}}-\vec{\tilde{t}}\, )^T (A V A^T)^{-1}(\vec{\tilde{p}}-\vec{\tilde{t}}\, )=(\vec{\tilde{p}}-\vec{\tilde{t}}\, )
^T \tilde{V}^{-1}(\vec{\tilde{p}}-\vec{\tilde{t}}\, ) \, .
\end{equation}
Notice that the $\chi^2$ function is ill-defined when the covariance matrix is singular. This can occur when the same data point $\tilde{p}_i$ has been introduced twice. Since adding many times the same data point to a fit cannot change the fit result, the solution is straightforward: remove the redundancies.

Another condition that any fit should satisfy is the fact that adding a new data point $(t_{n+1},p_{n+1})$ dependent on  an extra unknown parameter $\theta_{m+1}$ does not give us any information on the previous $\theta_{i}$ or in the agreement of the theory with data. Indeed, when minimizing the $\chi^{2}$, 
the new parameter $\theta_{m+1}$ will simply adapt its value to exactly match $t_{n+1}$ with $p_{n+1}$,
leaving the $\chi^{2}$ unmodified.\footnote{If $p_{n+1}$ is correlated with the rest of $p_{i}$, it will 
instead adapt itself to exactly match the uncorrelated combination of $p_{n+1}$ with the original $\{p_{i}\}$ set.}

\subsection{Explicit redundancies in several approximations}

When dealing with experimental distributions, such as the ones from ALEPH \cite{Davier:2013sfa}, we work with  a discrete spectrum. For a set of consecutive energy-squared values $s_i$, which are the central energy points of bins with width $\hat{\Delta}_i$, ending at $\tilde{s}_i\equiv s_i + \frac{\hat{\Delta}_i}{2}$, we have the measured spectral function $\rho_i\equiv\rho(s_i)$. Correlations among the different data points can be large and need to be taken into account.

The associated discrete integrals of Eq. (\ref{eq:A(n)}) are simply given by
\begin{equation}\label{eq:Andiscrete}
A^{(n)}(\tilde{s}_j)=\pi \sum_{i}^{j}\left(\frac{s_i}{\tilde{s}_j}\right)^n \rho_i \, \Delta_i \, ,
\end{equation}
where $\Delta_i\equiv \hat{\Delta}_i/\tilde{s}_j$. If we stick to a single energy point $\tilde{s}_j$, we can fit $A^{(n)}(\tilde{s}_j)$ for several monomial functions or combinations of them.

\subsubsection{ALEPH-like fits}

In the ALEPH-like fits one assumes that power corrections are small enough so that only the lowest-dimensional condensates $\mathcal{O}_D$ have any impact on the observables.
Thus, one neglects both the lower-dimensional $\mathcal{P}_D$ factors and the higher-dimensional $\mathcal{O}_{D}$. The original ALEPH fit takes
\begin{equation}\label{kl}
\omega_{kl}(s)\; =\;\left(1-\frac{s}{m_{\tau}^{2}}\right)^{2+k}\left( \frac{s}{m^{2}_{\tau}} \right)^{l}\left( 1+\frac{2s}{m_{\tau}^{2}}   \right) \, ,
\end{equation}
with $(k,l)=\{(0,0), (1,0), (1,1), (1,2), (1,3)\}$. The truncation choice consists in neglecting all $\mathcal{P}_{D}$ and $\mathcal{O}_{D>8}$, so that:
\begin{eqnarray}
A_{00,V/A}^{^{\mathrm{ALEPH}}} &=& A_{00,V/A}^{^{\mathrm{ALEPH}}}(a_{s},\mathcal{O}_{6\, V/A},\mathcal{O}_{8\, V/A}) \, ,
\nonumber\\
A_{10,V/A}^{^{\mathrm{ALEPH}}} &=& A_{10,V/A}^{^{\mathrm{ALEPH}}}(a_{s},\langle a_{s}GG\rangle,\mathcal{O}_{6\, V/A},\mathcal{O}_{8\, V/A}) \, ,
\nonumber\\
A_{11,V/A}^{^{\mathrm{ALEPH}}} &=& A_{11,V/A}^{^{\mathrm{ALEPH}}}(a_{s},\langle a_{s}GG\rangle,\mathcal{O}_{6\, V/A},\mathcal{O}_{8\, V/A}) \, ,
\nonumber\\
A_{12,V/A}^{^{\mathrm{ALEPH}}} &=& A_{12,V/A}^{^{\mathrm{ALEPH}}} (a_{s},\mathcal{O}_{6\, V/A},\mathcal{O}_{8\, V/A}) \, ,
\nonumber\\
A_{13,V/A}^{^{\mathrm{ALEPH}}} &=& A_{13,V/A}^{^{\mathrm{ALEPH}}} (a_{s},\mathcal{O}_{8\, V/A}) \, .
\label{eq:ALEPHmom}
\end{eqnarray}
The small values  obtained for the condensates give some illuminating information: the deviations from the purely perturbative predictions are very small for all moments. Nonetheless, if we are only interested in the value of $\alpha_s$, we can take into account the 
previous discussion in subsection~\ref{subsec:FitProperties}
and isolate the two independent linear combinations that, with our truncation choice, only depend on $\alpha_s$:
\begin{align}
\omega_{1}(x)&=(1-x)^2\, (1+2x)\, (1+3x^2-2x^3+9x^4) \, ,\\
\omega_{2}(x)&=(1-x)^2\, (1+2x)\, x^4 \, ,
\end{align}
with $x=\frac{s}{m_{\tau}^2}$. By construction, the fitted value of $\alpha_s$ will be exactly the same as in the full fit, since the three additional points in (\ref{eq:ALEPHmom}) depend on three completely unknown parameters and, as remarked before,
adding as many free parameters as data points does not give us any information about the fit quality or the previous parameter, $\alpha_s$. Analogously, including $\mathcal{O}_{10}$ in the fit is equivalent to only using $\omega_{1}$ for the $\alpha_s$ determination. Taking the difference of both results, which is related to the quality of the fit, is a good assessment of the neglected higher-order power corrections, which appear enhanced by large prefactors for the relevant weight functions.

The corresponding weights for the fit without the kinematic factor $(1+2x)$ are
\begin{align}
\hat\omega_{1}(x) &=(1-x)^2\, (1+2x+3x^2+4x^3+5x^4) \, ,\\
\hat\omega_{2}(x) &=(1-x)^2\, x^4 \, .
\end{align}
Finally, for the $A^{(2,m)}$ moments, one has the analogous combinations
\begin{align}
\tilde\omega_{1}(x) &=(1-x)^2(1+2x+3x^2+4x^3+5x^4+6x^5) \, ,\\
\tilde\omega_{2}(x) &=(1-x)^2 x^5 \, .
\end{align}
While they are not uncorrelated, it is rather clear that these tests are not redundant, as one can explicitly check by observing the larger instabilities of the fits in the separate $V$ and $A$ channels \cite{Pich:2016bdg}.

\subsubsection[Fit to the $s_0$ dependence]{\boldmath Fit to the $s_0$ dependence}
\label{subsubsec:s0-dependence}

On the other hand, if we stick to a monomial function and make a fit to the $s_{0}$ dependence of the moment $A^{(n)}(s_0)$, 
it corresponds in the discrete version to fitting
\begin{equation}
\left\{ \pi\sum_{i}^{j_{\mathrm{in}}}\left(\frac{s_i}{\tilde{s}_{j_{\mathrm{in}}}}\right)^n \rho_i \, \Delta_i\, ,\, \pi\sum_{i}^{j_{\mathrm{in}}+1}\left(\frac{s_i}{\tilde{s}_{j_{\mathrm{in}}+1}}\right)^n \rho_i \, \Delta_i\, , \,\cdots\, ,\, \pi\sum_{i}^{j_{\mathrm{end}}}\left(\frac{s_i}{\tilde{s}_{j_{\mathrm{end}}}}\right)^n \rho_i \, \Delta_i  \right\}\, .
\end{equation}
As explicitly discussed before, the fit must be invariant under linear transformations of the data points. It is then trivial that the previous fit is necessarily equivalent to a fit to
\begin{equation}
\left\{ \sum_{i}^{j_{\mathrm{in}}}s_i^n \rho_i \, \Delta_i\, ,\, \sum_{i}^{j_{\mathrm{in}}+1}s_i^n \rho_i \, \Delta_i\, ,\, \cdots\, ,\, \sum_{i}^{j_{\mathrm{end}}} s_i^n \rho_i \, \Delta_i  \right\}\, ,
\end{equation}
or
\begin{equation}\label{eq:boitofit}
\left\{ A^{(n)}(\tilde{s}_{j_\mathrm{in}})\, ,\, \rho_{j_{\mathrm{in}}+1}\, ,\,\cdots\, ,\, \rho_{j_{\mathrm{end}}}  \right\}\, .
\end{equation}
Fitting the $s_{0}$-dependence of $A^{(n)}(s_0)$ is exactly the same as fitting $A^{(n)}(s_0)$ in the initial point, plus the spectral function: one can trivially reproduce one set of data points from the other without \emph{any} theoretical input. Note how the continuum version of this equivalency is simply given by Eq.~(\ref{eq:ImPifrommom}). The slope of the moments is given by the spectral function.

It is at this stage where we can demonstrate that the five different fits to the $s_0$ dependence of the vector channel made in Refs.~\cite{Boito:2014sta,Boito:2020xli} trivially reduce to Eq.~(\ref{eq:boitofit}) for $n=0$, not giving any new information on $\alpha_s$ or in the validity of their theory assumptions, as incorrectly claimed in those references. Defining $x\equiv s/s_0$, let us consider each fit separately: 
\begin{itemize}
\item Fit 1: $\omega_0(x)=1$. This corresponds to the equivalence we have just shown, for $n=0$:
\begin{equation}
\left\{ A^{(0)}(\tilde{s}_{j_\mathrm{in}}) , ,\rho_{j_{\mathrm{in}}+1} ,\dots, \rho_{j_{\mathrm{end}}}\right\}\, .
\end{equation}
\item Fit 2: $\omega_1(s)=1$, $\omega_2(s)=1-x^2$, $\omega_3(s)=(1-x^2)(1+2x)$, One has:
\begin{equation}
\left\{ A^{\omega_1}(\tilde{s}_{j_\mathrm{in}}) , \dots, A^{\omega_2}(\tilde{s}_{j_\mathrm{in}}) , \dots ,A^{\omega_3}(\tilde{s}_{j_\mathrm{in}}) ,\dots\right\}\, .
\end{equation}
A trivial linear transformation gives 
\begin{equation}
\left\{ A^{(0)}(\tilde{s}_{j_\mathrm{in}}) , \dots, A^{(2)}(\tilde{s}_{j_\mathrm{in}}) , \dots ,A^{(3)}(\tilde{s}_{j_\mathrm{in}}) ,\dots\right\}\, ,
\end{equation}
which again reduces to:
\begin{equation}
\left\{ A^{(0)}(\tilde{s}_{j_\mathrm{in}}) , \rho_{j_{\mathrm{in}}+1} ,\cdots, \rho_{j_{\mathrm{end}}}, A^{(2)}(\tilde{s}_{j_\mathrm{in}}) , \rho_{j_{\mathrm{in}}+1} ,\cdots, \rho_{j_{\mathrm{end}}} ,A^{(3)}(\tilde{s}_{j_\mathrm{in}}) ,\rho_{j_{\mathrm{in}}+1} ,\cdots, \rho_{j_{\mathrm{end}}}\right\}\, .
\end{equation}
Removing the repeated data points, which do not carry any information and can only distort the fit, one gets
\begin{equation}
\left\{ A^{(0)}(\tilde{s}_{j_\mathrm{in}}) , A^{(2)}(\tilde{s}_{j_\mathrm{in}})  ,A^{(3)}(\tilde{s}_{j_\mathrm{in}}) ,\rho_{j_{\mathrm{in}}+1} ,\dots, \rho_{j_{\mathrm{end}}}\right\}\, .
\end{equation}
Taking into account (see discussion above) that in the working condensate approximation $A^{(2)}$ and $A^{(3)}$ are adding as many data points ($2$) as completely unknown parameters, $\mathcal{O}_{6}$ and $\mathcal{O}_{8}$, the fit for the remaining parameters and for the test of the theory is exactly equivalent to a fit without those two points, leading to
\begin{equation}
\left\{ A^{(0)}(\tilde{s}_{j_\mathrm{in}})  ,\rho_{j_{\mathrm{in}}+1} ,\dots, \rho_{j_{\mathrm{end}}}\right\}\, .
\end{equation}

\item Fit 3: $\omega_1(s)=1$, $\omega_2(s)=1-x^2$. This is a trivial variation of the previous one.

\item Fit 4: $\omega_1(s)=1$, $\omega_2(s)=(1-x)^2(1+2x)$.  One has
\begin{equation}
\left\{ A^{\omega_1}(\tilde{s}_{j_\mathrm{in}}) , \dots, A^{\omega_2}(\tilde{s}_{j_\mathrm{in}}) , \dots\right\}\, ,
\end{equation}
which is equivalent to
\begin{equation}
\left\{ A^{(0)}(\tilde{s}_{j_\mathrm{in}}) , \rho_{j_{\mathrm{in}}+1} ,\dots, \rho_{j_{\mathrm{end}}}, 3A^{(2)}(\tilde{s}_{j_\mathrm{in}})-2A^{(3)}(\tilde{s}_{j_\mathrm{in}}), 3A^{(2)}(\tilde{s}_{j_\mathrm{in}+1})-2A^{(3)}(\tilde{s}_{j_\mathrm{in}+1}), \dots  \right\}\, .
\end{equation}
Finally, using the discrete version of Eq.~(\ref{eq:ImPifrommom}), 
arranging the appropriate linear combinations and removing repeated data points this is equivalent to 
\begin{equation}
\left\{ A^{(0)}(\tilde{s}_{j_\mathrm{in}}) , \rho_{j_{\mathrm{in}}+1} ,\dots, \rho_{j_{\mathrm{end}}}, A^{(2)}(\tilde{s}_{j_\mathrm{in}}), A^{(3)}(\tilde{s}_{j_\mathrm{in}})  \right\}\, ,
\end{equation}
which is the same as Fit 2.

\item Fit 5: $\omega_1(s)=1$, $\omega_2(s)=(1-x^2)^2$. The derivation is identical to the previous one.
\end{itemize}
Thus, the comparison among the results for $\alpha_s$ obtained from these five different fits constitute a tautological test.

\subsection{Other tautological tests}
\label{sec:tautologies}

The $s_0$-dependence of different moments has been claimed to provide an excellent consistency test of the DV-ansatz approach~\cite{Boito:2016oam}. However, once $A^{\omega}_\cJ(\hat s_0)$ and the experimental spectral function have been fitted with the parameters of the assumed hadronic model, all moments get determined in the fitted region. For instance, with the weights $\omega_n(x) = x^n$, one trivially has the exact mathematical identity
\begin{equation}\label{eq:tautology1}
 A^{(n)}_\cJ (s_0)\, =\, \left(\frac{\hat s_0}{s_0}\right)^{n+1}
 A^{(n)}_\cJ (\hat s_0) + \frac{\pi}{s_o^{n+1}}\int_{\hat s_0}^{s_0} ds\; s^n\,\rho_\cJ(s)\, ,
\end{equation}
valid for any value of $s_0 > \hat s_0$. 
The consistency plots shown in Ref.~\cite{Boito:2016oam} only display a range of $s_0$ values in the fitted region $[\hat s_0, m_\tau^2]$, where the agreement with data is guaranteed by Eq.~(\ref{eq:tautology1}), since both $A^{\omega}_\cJ(\hat s_0)$ and $\rho_\cJ(s)$ have been fitted to data. Any hadronic model would exhibit the same excellent agreement, provided that it fits well the data, independently of the numerical value of $\alpha_s$ emerging from it. Therefore, this type of plots are only testing the statistical quality of the multi-parameter fit to the spectral function, and do not provide any information about the actual relation of the assumed ansatz with QCD. 

In order to learn something about the ansatz itself, one should compare the model predictions with the data below the fitted region; however, such comparison is never shown. 
From Figs.~\ref{fig:DVmodels1}, \ref{fig:DVmodelsv+a} and \ref{fig:spectral_old}, it is evident that this exercise would exhibit a poor behaviour, instead of the claimed excellent performance of the DV model.
Let us stress once again that, in contrast with the assumed approximate convergence to the OPE  at $s_0= m_{\tau}^2$, used in the standard 
extraction of the strong coupling,
the DV-ansatz approach relies
on the (exact, since no uncertainty at all is assigned) validity of the hadronic model in the whole energy range from $\hat{s}_0=1.55 \, \mathrm{GeV}^2$ to
$s_{0}=m_{\tau}^2$.

This type of plots has also been used as a mean to demonstrate hypothetical failures of the OPE, by zooming scales and not displaying any error bars for the theoretical curves, when being compared to data points that have not been explicitly fitted. One usually displays differences such as $A_\cJ^\omega(m_\tau^2)-A_\cJ^\omega(s_0)$ or even double-differences, subtracting the corresponding experimental quantities, in order to magnify the claimed disagreement. As already discussed in section~\ref{sec:alphasResults}, in the standard determination of the strong coupling, the relevant power corrections turn out to be too small to be clearly identified at $s_0\sim m_\tau^2$ because they get masked by
the much larger noise of the perturbative uncertainties. Therefore, the fitted values of the condensates have rather large errors, but their impact on $\alpha_s$ is small and has been carefully (and conservatively) assessed. However, if one plots the $s_0$ dependence of $A^{(n)}_\cJ (s_0)$ without the corresponding power correction and taking away the theoretical error bars (see for instance Fig.~\ref{fig:pinchs}), a visible difference with the experimental data should certainly emerge, which does not have any meaningful impact in the standard determination of $\alpha_s$.

\section{\boldmath Results from DV fits with $\mathcal{G}_\cJ (s) = s^{\lambda_\cJ}$}
\label{app:lambdaV_Fit}

Ref.~\cite{Pich:2016bdg} already analyzed the sensitivity of the DV fits to the vector distribution with the ansatz (\ref{eq:DVparam}), using $\mathcal{G}_\cJ (s) = s^{\lambda_\cJ}$ ($\mathrm{GeV}^2$ units) and different values of $\lambda_V$ between zero and 8, while keeping the ad-hoc choice $\hat s_0 = 1.55~\mathrm{GeV}^2$. We reproduce in Table~\ref{tab:PowerModels} the fitted values of the strong coupling and the four ansatz parameters, together with the p-values of each fit. These results exhibit a very strong correlation between the input value assumed for $\lambda_V$ and the output value of $\alpha_s(m_\tau^2)$. The worse fit (p-value) corresponds to the default choice $\lambda_V=0$ and leads to the smallest $\alpha_s$. As $\lambda_V$ increases, the fit quality improves, while the strong coupling slowly approaches its reference value discussed in section~\ref{sec:alphasResults}.

\begin{table}[t]\centering
{\begin{tabular}{|c|c|cccc|c|}\hline &&&&&&\\[-12pt]
$\lambda_V$  & $\alpha_{s}^{(n_f=3)}(m_{\tau}^{2})$ & $\delta_V$      & $\gamma_V$      & $\alpha_V$       & $\beta_V$
& p-value (\% )
\\ \hline
0  & $0.298 \pm 0.010$          & $3.6 \pm 0.5$ & $0.6 \pm 0.3$ & $-2.3 \pm 0.9$ & $4.3 \pm 0.5$
& 5.3
\\ 
1  & $0.300 \pm 0.012$          & $3.3 \pm 0.5$ & $1.1 \pm 0.3$ & $-2.2 \pm 1.0$ & $4.2 \pm 0.5$
& 5.7
\\
2  & $0.302 \pm 0.011$          & $2.9 \pm 0.5$ & $1.6 \pm 0.3$ & $-2.2 \pm 0.9$ & $4.2 \pm 0.5$
& 6.0
\\ 
4  & $0.306 \pm 0.013$          & $2.3 \pm 0.5$ & $2.6 \pm 0.3$ & $-1.9 \pm 0.9$ & $4.1 \pm 0.5$
& 6.6
\\ 
8  & $0.314 \pm 0.015$          & $1.0 \pm 0.5$ & $4.6 \pm 0.3$ & $-1.5 \pm 1.1$ & $3.9 \pm 0.6$
& 7.7
\\ \hline
\end{tabular}}
\caption{Fitted values of $\alpha_s^{(n_f=3)}(m_\tau^2)$, in FOPT, and the spectral ansatz parameters in Eq.~(\ref{eq:DVparam}) with $\hat s_0= 1.55\;\mathrm{GeV}^2$, for different values of the power $\lambda_V$ \protect\cite{Pich:2016bdg}}
\label{tab:PowerModels}
\end{table}

\begin{figure}[t]
\centerline{\includegraphics[width=0.6\textwidth]{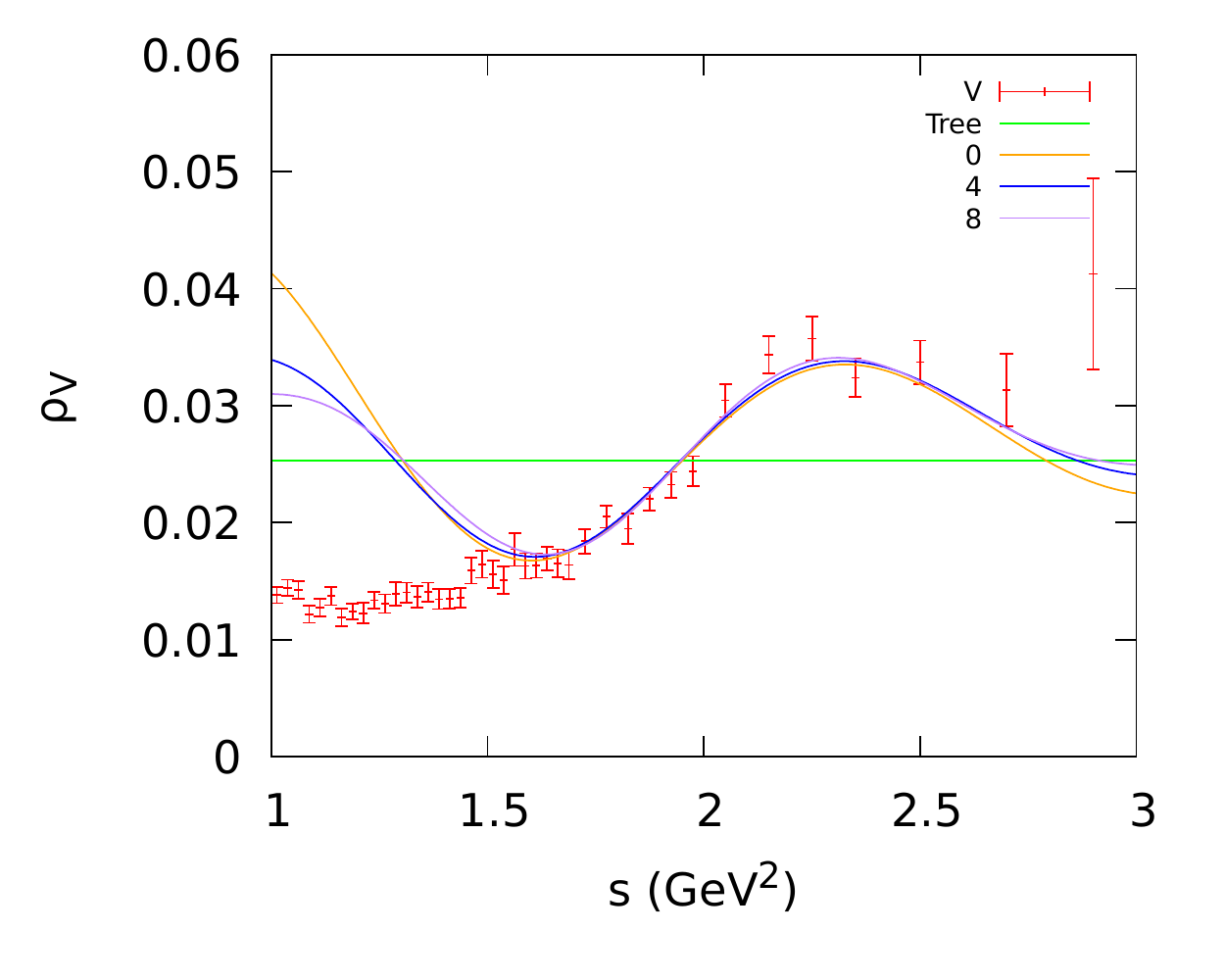}}
\vskip -.5cm
\caption{Vector spectral function $\rho_V^{\protect\phantom{()}}(s)$, fitted above 1.55~GeV${}^2$ with the ansatz (\ref{eq:DVparam}), for different values of $\lambda_V= 0, 4, 8$, compared with the data points \protect\cite{Pich:2016bdg}}
\label{fig:spectral_old}
\end{figure}

Fig.~\ref{fig:spectral_old} compares the vector spectral function predicted by the different fitted ansatzs with the experimental data.
%
%
All models reproduce well $\rho_V^{\protect\phantom{()}}(s)$ in the fitted region
of invariant masses  ($1.55\;\mathrm{GeV}^2\le s\le m_\tau^2 $), 
but they fail badly below it. The worse behaviour is obtained with the default model ($\lambda_V=0$). When $\lambda_V$ increases, the predicted spectral function slightly approaches the data below the fitted range, while the ansatz parameters adapt themselves to compensate the growing at high values of $s$ with the net result of a smaller duality-violation correction. 

The large variation in the output value of $\alpha_s$ obtained 
from the different fits gets obviously reflected in the fitted values of the power corrections that need to adapt themselves in order to reproduce the corresponding experimental moments $A^{(n)}_V(\hat s_0)$ with a different $\alpha_s$. This is shown in Table~\ref{tab:models2}, which compiles the values of the condensates $\mathcal{O}_{D\le 16,V}$ obtained with the different choices of $\lambda_V$. The observed changes are indeed very large, and even the signs get modified in some cases. The absolute size of the condensates decreases in a very sizable way when $\lambda_V$ (and $\alpha_s$) increases, except for $\mathcal{O}_{16,V}$. However the most important result from this exercise is the very strong model dependence of the fitted parameters, which are void of any physical meaning.

\begin{table}[t]\centering
{\begin{tabular}{|c|ccccccc|}\hline &&&&&&\\[-12pt]
$\lambda_V$  & $\mathcal{O}_{4,V}$ & $\mathcal{O}_{6,V}$ & $\mathcal{O}_{8,V}$   & $\mathcal{O}_{10,V}$   & $\mathcal{O}_{12,V}$
& $\mathcal{O}_{14,V}$ & $\mathcal{O}_{16,V}$
\\ \hline
0 & $\phantom{-}0.0016$ & $-0.0082$ & $0.014$ & $-0.019$ &  $\phantom{-}0.023$ & $-0.029$ & $0.037$
\\
1 & $\phantom{-}0.0014$ & $-0.0078$ & $0.013$ & $-0.018$ &  $\phantom{-}0.021$ & $-0.027$ & $0.036$
\\
2 & $\phantom{-}0.0012$ & $-0.0074$ & $0.012$ & $-0.016$ &  $\phantom{-}0.019$ & $-0.025$ & $0.037$
\\
4 & $\phantom{-}0.0008$ &  $-0.0064$ & $0.010$ & $-0.012$ &  $\phantom{-}0.014$ & $-0.022$ & $0.049$
\\
8 & $-0.0003$ &  $-0.0037$ & $0.004$ &  $\phantom{-}0.001$ &  $-0.0099$ &  $\phantom{-}0.006$  & $0.079$
\\ \hline
\end{tabular}}
\caption{Fitted values of the OPE condensates in GeV units, with FOPT and $\hat s_0= 1.55\;\mathrm{GeV}^2$, for different values of the power $\lambda_V$.}
\label{tab:models2}
\end{table}

\bibliographystyle{JHEP}
\bibliography{References-alphas}

\end{document}